\documentclass[11pt]{article}
\usepackage[a4paper, total={6in, 9in}]{geometry}
\usepackage[utf8]{inputenc}
\usepackage[english]{babel}
\usepackage{hyperref}
\usepackage{braket}
\usepackage{bbm}
\usepackage{graphicx}
\usepackage{subfigure}
\usepackage{bbold}
\usepackage[table]{xcolor}
\usepackage{rotating}
\usepackage{dsfont}
\usepackage[export]{adjustbox}
\usepackage{authblk}
\usepackage{amsmath,amssymb}
\usepackage{caption}
\numberwithin{equation}{section}
\numberwithin{figure}{section}
\numberwithin{table}{section}
\usepackage{float}
\allowdisplaybreaks

\newcommand{\F}[1]{{}^#1\!F}
\newcommand{\Fi}[1]{{}_#1\mspace{-1mu}F}
\newcommand{\mc}[1]{\mathcal{#1}}

\def\lact  {\,{\triangleright}\,}
\def\ract  {\,{\triangleleft}\,}

\newcommand{\tr}{\operatorname{Tr}}

\newcommand{\skippart}[1]{}

\begin{document}

\begin{center}{\Large \textbf{
Topological aspects of the critical three-state Potts model
}}\end{center}

\begin{center}
Robijn Vanhove\textsuperscript{1,a},
Laurens Lootens\textsuperscript{1,b},
Hong-Hao Tu\textsuperscript{2} and
Frank Verstraete\textsuperscript{1}
\end{center}

\begin{center}
{\bf 1} Department of Physics and Astronomy, Ghent University, \\
Krijgslaan 281, S9, B-9000 Ghent, Belgium
\\
{\bf 2} Institute of Theoretical Physics, Technische Universit\"at Dresden, \\
01062 Dresden, Germany
\\[1em]
\textsuperscript{a}robijn.vanhove@ugent.be \\
\textsuperscript{b}laurens.lootens@ugent.be
\end{center}

\begin{abstract}
\noindent
We explore the topological defects of the critical three-state Potts spin system on the torus, Klein bottle and cylinder. A complete characterization is obtained by breaking down the Fuchs-Runkel-Schweigert construction of 2d rational CFT to the lattice setting. This is done by applying the strange correlator prescription to the recently obtained tensor network descriptions of string-net ground states in terms of bimodule categories [Lootens, Fuchs, Haegeman, Schweigert, Verstraete, SciPost Phys. 10, 053 (2021)]. The symmetries are represented by matrix product operators (MPO), as well as intertwiners between the diagonal tetracritical Ising model and the non-diagonal three-state Potts model. Our categorical construction lifts the global transfer matrix symmetries and intertwiners, previously obtained by solving Yang-Baxter equations, to MPO symmetries and intertwiners that can be locally deformed, fused and split. This enables the extraction of conformal characters from partition functions and yields a comprehensive picture of all boundary conditions.
\end{abstract}
\tableofcontents
\section{Introduction} \label{Sec:Intro}

The intricate and beautiful connection between conformal field theory (CFT) and topological field theory (TFT) has been established on many levels throughout the last decades. The study of these connections was largely initiated in a seminal paper by Witten \cite{Witten1988}, where the equivalence between the state-space of 3d Chern-Simons theory and the conformal blocks of a 2d Wess-Zumino-Witten CFT were first understood. The main idea is that the CFT governing the gapless boundary of a system described by a TFT is largely determined by the TFT in the bulk. This holographic relation is observed in many systems in condensed matter physics: two examples being the fractional quantum Hall state (FQHS) (characterized on the edge by a CFT through the bulk-edge correspondence \cite{moore1991nonabelions}) and topological insulators \cite{kane2005z}, protected by time-reversal and charge-conjugation symmetries (exhibiting non-chiral edge modes). In 3d, this holographic duality between TFT and CFT culminated in the full classification of 2d rational conformal field theories \cite{Felder2000, Fuchs2002, Frohlih2004}. This was possible because the underlying mathematical framework for both 3d TFT and 2d CFT is a modular tensor category (MTC), which proved to be exactly the right language to express many topological features of 2d CFT such as partition functions, boundary conditions and certain aspects of correlation functions.\\

In the case of a non-chiral TFT, one can construct the partition function as a state-sum, meaning that one can discretize the underlying manifold and assign values to the building blocks of the tessellation such that the result depends only on the topology of the manifold. For 3d, this invariant is known as the Turaev-Viro state-sum \cite{Turaev1992, barrett1996invariants}, and in \cite{aasen2016topological, aasen2020topological} these constructions were used to construct critical lattice models described by CFTs in the continuum limit, reminiscent of the above holographic duality. The Turaev-Viro state-sums admit a (2+1)d Hamiltonian formulation known as the string-net models by Levin and Wen \cite{Levin2005}, the ground states of which can be interpreted as Turaev-Viro state-sums on a particular 3-manifold and correspond to the tensor network description of the string-net ground states in terms of projected entangled pair states (PEPS) \cite{luo2017structure,lootens2020matrix}. In the same way as \cite{aasen2016topological,aasen2020topological}, we used these PEPS descriptions of string-net ground states to construct critical lattice models \cite{vanhove2018mapping} using a \textit{strange correlator} (SC), first conceived for the detection of SPT phases \cite{You2014} and now generalized for the long-range entangled string-net wave functions The SC maps a (2+1)d PEPS with topological order to a classical 2d partition function by taking the overlap between the PEPS and an unentangled product state. The usefulness of this construction lies in the systematic description of the non-local symmetries of the emergent CFTs in terms of explicit matrix product operator (MPO) symmetries present in the topologically ordered PEPS. Anyonic excitations in the TFT are mapped to operators in the CFT \cite{Moore1989, kitaev2006anyons}, and the conformal blocks emerge from the topological sectors of the original string-net wave function. The map can be seen as a Euclidean counterpart to anyonic spin chain Hamiltonians \cite{PhysRevLett.98.160409, PhysRevLett.103.070401, Buican2017, ardonne2011microscopic}, which have successfully allowed for the computation of twisted CFT partition functions for e.g. the Fibonacci model \cite{PhysRevLett.98.160409}.\\

In this paper, we aim to further push the lattice understanding of critical spin systems described by a CFT by considering them on a number of surfaces: the torus, the Klein bottle and the cylinder. The holographic TFT construction of CFT described above provides a categorical description of the possible partition functions and boundary conditions required to construct a CFT on these surfaces. Although this construction is well understood, it is not obvious how these results translate to actual critical spin systems with Hamiltonians acting on a Hilbert space, or equivalently some critical statistical mechanics model with Boltzmann weights and local fluctuating degrees of freedom. It is our goal to show that this formulation is equally useful for studying lattice models by showing that many of the topological features of CFT are already present at finite size. We show this by applying the strange correlator to the recently generalized PEPS descriptions of string-net models, which turn out to exhibit exactly the same categorical structure present in 2d CFT, as briefly summarized in Section \ref{Sec::FRS}.\\

In Sections \ref{Sec:Torus}, \ref{Sec:TorusvsKlein} and \ref{Sec:cylinder} we will consider CFT partition functions on a torus, Klein bottle and cylinder respectively. We will emphasize the role of topological defects, as these provide a handle on the topological aspect of CFT in our finite-size lattice models constructed from a strange correlator, and show that these topological defects can be used to isolate characters in the partition function. In these sections, we will restrict to diagonal CFT (sometimes referred to as the Cardy case), a discussion which is instrumental in understanding non-diagonal theories such as the Potts model later. For the torus and cylinder partition functions this comprises essentially a review of previous work \cite{vanhove2018mapping,aasen2016topological,aasen2020topological}. The non-orientable Klein bottle case is newer and sits somewhere between the torus and cylinder partition functions due to the fact that it is a closed surface but nevertheless has a partition function that is linear in the characters, as is the case for the cylinder. In Ref. \cite{tu2017universal}, a CFT detection method was established on the Klein bottle for rational CFTs by calculation of the so-called Klein bottle entropy. This entropy depends on the quantum dimensions of the CFT primary states and its calculation restricts, at least partially, the possible CFTs. The Klein bottle entropy arises from performing modular transformations on the partition function (essentially probing the $S$-matrix of the theory) \cite{verlinde1988fusion} and is intimately analogous to the famous Affleck-Ludwig entropy for boundary CFTs \cite{affleck1991universal}, but this time on a closed manifold. \\

The discussion of the diagonal case serves mainly as a stepping-stone towards understanding general non-diagonal CFT partition functions, considered in Section \ref{Sec:Orbifold}, which is the main goal of this work. The construction is repeated on a torus, Klein bottle and cylinder for this more general case. An additional ingredient here is a type of defect that separates different realizations of the CFT which we will refer to as an intertwiner. This intertwiner can be used to map a non-diagonal partition function to the partition function of a diagonal model in the presence of topological defects, a procedure known as orbifolding, which in particular can also be used to explain the relation between boundary conditions of a non-diagonal model and a diagonal model. Notably, this procedure allows for a generalization of the previously found intertwiners using integrability theory \cite{pearce1993intertwiners}. However, putting the orbifolding procedure on a full categorical footing allows for the construction of intertwiners that can freely be moved through the lattice, offering more flexibility and the ability to study intersections of defects and intertwiners, a key ingredient in constructing boundary conditions in its full generality.\\

To illustrate this more concretely, we treat representative examples in the form of the Ising CFT and the non-unitary Yang-Lee CFT for the diagonal case. The non-diagonal case is represented by the three-state Potts CFT, which serves as the apex of this work. These models have been studied extensively, and this section serves to collect the relevant CFT data associated to them. In Section \ref{Sec:results}, we finally turn to explicit lattice realizations of these CFTs constructed as the strange correlator of the appropriate string-net model. We calculate the finite-size spectra of the three models on the torus, Klein bottle and cylinder and the Klein bottle entropy through exact diagonalization. This illustrates the strange correlator procedure by highlighting its application on three broad CFT classes: unitary CFTs with diagonal partition functions, non-unitary CFTs and CFTs with non-diagonal partition functions.

\section{Topological aspects of 2d rational CFT}
\label{Sec::FRS}

In this section, we aim to briefly review some relevant features of 2d CFT. Their structure can be divided into two parts:
\begin{itemize}
	\item
	A local geometric aspect, which among others concerns the construction of conformal blocks that serve as the building blocks of correlation functions. They are solutions to the conformal Ward identities, which decompose into left-moving and right-moving Virasoro Ward identities. For critical lattice models, this aspect differs from one model to another as it depends on the microscopic details of the model under consideration.
	\item
	A global, topological aspect that determines the appropriate correlation functions and boundary conditions on closed and open surfaces. This has to be done in a way that is compatible with the composition of these surfaces, which imposes conditions known as the sewing constraints. For critical lattice models, this aspect is essentially the same for every model in a given universality class.
\end{itemize}
For a rational full 2d CFT, a rigorous construction and classification of this topological aspect has been obtained by using a form of the holographic principle to construct these CFTs from a TFT \cite{Felder2000, Fuchs2002, Frohlih2004}. The Moore-Seiberg data concerning the representations of the chiral algebra is captured in this construction by an MTC $\mc{D}$ and is assumed to be given. To arrive at a full CFT with local correlation functions, an additional piece of data is required, since for a given MTC $\mc{D}$ there exist multiple distinct consistent CFTs. A simple example of this is the free boson with a $\mathfrak{u}(1)$ chiral algebra compactified on a circle of radius $R$. For each $R$, the full CFT has a different (modulo $T$-duality) torus partition function, so one might guess that the additional data required is the specification of the modular invariant on the torus. This is not quite right however, as there exists a plethora of examples of modular invariant bilinear character combinations that do not arise as the partition function of any CFT.\\

The correct additional piece of data turns out to be a right $\mc{D}$-module category $\mc{M}$, which in the original TFT construction is established from a particular type of algebra $A$ in $\mc{D}$, but we will not use that definition. The module category $\mc{M}$ determines the modular invariant on the torus, but more generally also allows the CFT to be defined on any closed surface in such a way that it is invariant under the mapping class group of that surface. On an open surface, the conformal boundary conditions are given by simple objects in $\mc{M}$. The topological defect lines of the CFT are then given by another tensor category $\mc{C}$, such that $\mc{M}$ is also a left $\mc{C}$-module category. The tensor category $\mc{C}$ of topological defects depends on $\mc{D}$ and $\mc{M}$ by the requirement that $\mc{M}$ is an invertible $(\mc{C},\mc{D})$-bimodule category; for a given $\mc{D}$ and $\mc{M}$, such a tensor category $\mc{C} = \mc{D}_\mc{M}^*$ is called the dual of $\mc{D}$ with respect to $\mc{M}$. For a basic review of the relevant category theory, we refer to Appendix \ref{App:Category}.\\

Recently, it was discovered that exactly the same categorical structure is present in generalized tensor network representations of ground states of string-net models \cite{lootens2020matrix}. Here, the string-net model is given by a (not necessarily modular) fusion category $\mc{D}$, for which an explicit PEPS representation of its ground state can be found by the module associator of some right $\mc{D}$-module category $\mc{M}$. The non-local MPO symmetries play the role of lattice topological defects and are indeed described by the dual fusion category $\mc{C} = \mc{D}_\mc{M}^*$. Explicit representations for all the relevant tensors can then be constructed from the associators of the invertible $(\mc{C},\mc{D})$-bimodule category. The ground states of these string-net model can be mapped to a critical statistical mechanics model by using the strange correlator, which is very reminiscent of the holographic construction of a CFT from a TFT. The generalized PEPS representations for string-net ground states and its features are reviewed in Appendix \ref{App:StringNets}. It is the aim of this paper to argue that the TFT formulation of CFT provides an equally useful language for studying spin systems. We will do this by considering critical lattice models on a number of surfaces, both open and closed, and show that many of the topological features of CFT are already present at finite size.\\

There is a particular choice of module category $\mc{M}$ that can always be made: one can always take $\mc{M} = \mc{D}$ as a right $\mc{D}$-module category over itself. This choice yields a diagonal partition function for the CFT and is sometimes called the Cardy case. In this case, the boundary conditions are given by simple objects in $\mc{D}$, which means that there is one boundary condition for each representation of the chiral algebra. The dual of $\mc{D}$ with respect to $\mc{D}$ is again $\mc{D}$, meaning that the topological defects are also in one-to-one correspondence with the representations of the chiral algebra, and in particular obey the same fusion rules. In the PEPS language, this choice of module category gives the representations that were first derived for the ground states of string-net models \cite{buerschaper2009explicit,gu2009tensor}. In our discussion, we will restrict to this case until Section \ref{Sec:Orbifold}, where we will allow $\mc{M}$ to be a generic right $\mc{D}$-module category.

\section{Torus partition functions with topological defects} \label{Sec:Torus}

\begin{figure}[h]
	\centering
	\begin{minipage}[b]{0.4\textwidth}
		\centering
		\includegraphics[clip,trim = 1cm 0 2cm 0,scale = 0.3]{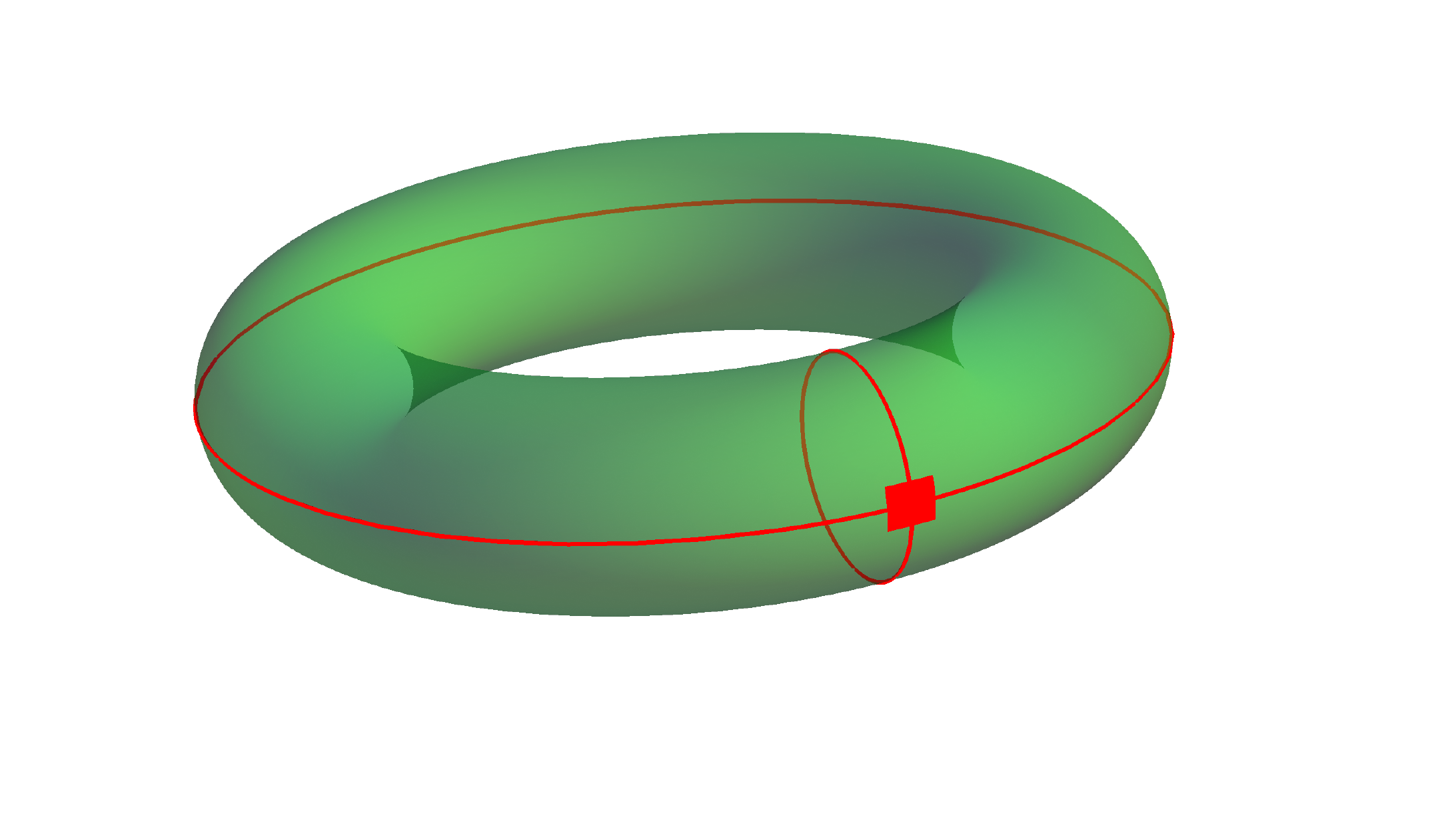}
	\end{minipage}\quad
	\begin{minipage}[b]{0.4\textwidth}
		\includegraphics[page=1]{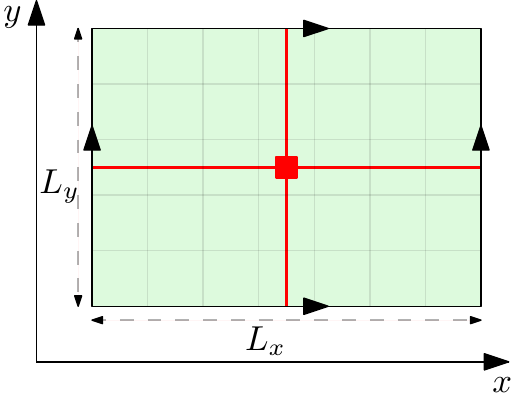}
	\end{minipage}
	\caption{The torus partition function with the insertion of topological defects in both x- and y-directions. The crossing of the defects is marked with a red square, indicating the insertion of a simple idempotent.}
	\label{Torus}
\end{figure}

As a start, let us review the identification of a critical statistical mechanics model with a CFT partition function on a torus. To this end, we make the identification between the transfer matrix of the model with periodic boundary conditions on $L_x$ sites, and the corresponding CFT Hamiltonian \cite{di1997senechal}:
\begin{align}
	T = e^{-H_{\text{ring}}} = \includegraphics[valign=c,page=9]{figures/kbtorus},
	\label{transferHam}
\end{align}

where we have introduced a tensor network representation of the transfer matrix as an MPO \cite{haegeman2017diagonalizing}. The single MPO tensors encode the Boltzmann weight of the statistical mechanics model between different sites.
The partition function of the classical 2D model on a torus of length $(L_x,L_y)$ is given by
\begin{align}
Z^{\text{tor}} &= \tr(T^{L_y}).
\end{align}
The finite-size Hamiltonian can be written as follows \cite{blote1986conformal,affleck1988universal}:

\begin{align}
H_{\text{ring}} = a + f_0L_x + vH_{\text{CFT}} + \mathcal{O}(\frac{1}{L_x^{\mu}}).
\label{Ham}
\end{align}

The shift $a$ and the rescaling factor (velocity) $v$ are non-universal numbers. $f_0$ is the free-energy density and $H_{\text{CFT}}$ the CFT Hamiltonian. Neglecting the terms decaying faster than $\frac{1}{L_x^{\mu}}$, we now focus on the universal part of the Hamiltonian. The Hamiltonian of the CFT ($H_{\text{CFT}}$) contains the sum of the right- and left moving Virasoro generators:

\begin{align}
H_{\text{CFT}} = \frac{2\pi}{L_x}\left(L_0+\overline{L}_0 - \frac{c}{12}\right),
\end{align}

with $c$ the central charge and $L_0+\overline{L}_0$ the generator of dilations on the plane and time translations on the cylinder. This is a CFT result and therefore a direct consequence of the conformal invariance. It is natural to work in the eigenbasis of these generators (the Hamiltonian) $\ket{\alpha,\overline{\alpha}}$, with the corresponding eigenvalues the characteristic conformal weights $h_{\alpha}$ and $h_{\overline{\alpha}}$. Introducing $q = e^{2\pi i \tau}$, with the modular parameter $\tau = iL_y/L_x$, let us define the character function

\begin{align}
\chi_{\alpha}(q) = \tr_{\alpha}(q^{L_0 - c/24}) = q^{h_{\alpha} - \frac{c}{24}} \sum_{n=0}^{\infty} c_{\alpha}(n) q^n,
\end{align}

with $c_{\alpha}(n)$ the CFT-specific degeneracy at level $n$. The universal CFT part of the partition function (keeping only $H_{\text{CFT}}$ in \ref{Ham}) becomes

\begin{align}
Z^\mathcal{\text{tor}} \simeq \tr(e^{-L_yH_{\text{CFT}}}) &\simeq \sum_{\alpha,\bar{\beta}}\chi_{\alpha}(q)M_{\alpha\bar{\beta}}\overline{\chi}_{\overline{\beta}}(\overline{q}) \\
&= q^{-\frac{c}{12}}\sum_{\alpha,\overline{\beta},n,m} M_{\alpha\bar{\beta}} c_{\alpha}(n) c_{\beta}(m) q^{h_{\alpha} + h_{\beta} + n + m}.
\label{partition}
\end{align}

Although the lattice partition function is extensive in both lengths, due to the free energy term in Eq. \ref{Ham}, the universal part of the partition function is only dependent on the ratio $L_y/L_x$. The partition function decomposes into a finite number (for rational CFT) of discrete Verma modules of right- and left- moving parts labeled by $\alpha$ and $\overline{\beta}$, combined by the integer matrix $M$. The effects of the lattice regularization - which differ for different statistical models in the same universality class - are entirely contained in the finite-size corrections to the character functions and they can be regarded as the ``geometric aspect" of the CFT. The matrix $M$ is purely topological and its form is dictated by requiring modular invariance on the torus. It is this formula that is used in practice to numerically identify the CFT (see section \ref{Sec:results}) as the degeneracies in the spectrum of the transfer matrix at some level $(n,m)$ are dictated by the character functions and the matrix $M$. For now, we restrict to diagonal partition functions, which have $M_{\alpha\bar{\beta}} = \delta_{\alpha\bar{\beta}}$. In this case the conformal scaling dimensions $\Delta_{\alpha} = h_{\alpha}+h_{\beta}$ appear in Eq. \eqref{partition}. We will explore non-diagonal partition functions in Section \ref{Sec:Orbifold}.\\

Let us now consider the case where we introduce a topological defect labeled by $\gamma$ in the lattice. These defects only have a non-trivial action on the partition function when wrapped around non-contractible cycles of the torus, i.e. they are locally invisible and can be freely deformed through the lattice:

\begin{equation*}
\includegraphics[valign=c,page=22]{figures/kbtorus}.
\end{equation*}

These defects themselves are also represented as MPOs, just like the transfer matrix. It was extensively discussed in \cite{csahinouglu2021characterizing, bultinck2017anyons, williamson2014matrix} how the local deformations, shown above, translate into local constraints on the MPO tensors of the topological defects (the so-called ``pulling-through" equations):
\begin{equation}
\includegraphics[valign=c,page=10]{figures/kbtorus}.
\label{pullingThrough}
\end{equation}
As the defects can be seen as ``symmetries" of the transfer matrix they are called MPO symmetries \footnote{Note that the word ``symmetry" is loosely used in this context. The MPO symmetries commute with the transfer matrix, but are not necessarily unitary representations of some symmetry group $G$. We use the word symmetry throughout the remainder of this work, including in particular what other authors would call dualities, i.e. objects that fuse to a sum of group-like objects.}. From this equation it is obvious that the product of two MPO symmetries is again an MPO symmetry, and if we assume a finite set of MPO symmetries they must be closed under multiplication and in general form representations of a fusion ring:
\begin{align*}
\includegraphics[page=18,valign=c]{figures/kbtorus} = \sum_{\gamma} N_{\alpha\beta}^{\gamma} \, \vcenter{\hbox{\includegraphics[page=19]{figures/kbtorus}}},
\end{align*}
with $N_{\alpha\beta}^{\gamma}$ non-negative integers. This equation implies the existence of fusion tensors that satisfy
\begin{equation}
\includegraphics[valign=c,page=20]{figures/kbtorus}.
\label{zipper}
\end{equation}
These MPO symmetries allow us to define a twisted partition function
\begin{equation}
Z_\gamma^\text{tor} = \tr(\widetilde{T}_{\gamma}^{L_y}), \quad \widetilde{T}_{\gamma} = \includegraphics[valign=c,page=11]{figures/kbtorus},
\end{equation}
where we have introduced the twisted transfer matrix $\widetilde{T}_{\gamma}$. The decomposition of such a twisted partition function on the torus in terms of characters was worked out in \cite{petkova2001generalised}:
\begin{align}
Z_{\gamma}^\text{tor} \simeq  \sum_{\alpha,\bar{\beta}}\chi_{\alpha}(q)\widetilde{M}^{\gamma}_{\alpha\bar{\beta}}\overline{\chi}_{\overline{\beta}}(\overline{q}).
\label{torusTwist}
\end{align}
The matrix $\widetilde{M}^{\gamma}$ again consists of non-negative integer entries, and we have $\widetilde{M}^{\mathbf{1}} = M$ where $\mathbf{1}$ labels the identity twist, which is of course equivalent to no twist at all.\\

This paper aims to explicitly show on the lattice how the matrix $\widetilde{M}$ (the ``topological aspect" of the CFT partition function) is obtained for general twists. At the heart of this discussion lies the ``pulling through" property (\ref{pullingThrough}). The MPOs, representing the lattice topological defects, are locally invisible and independent of the specific lattice geometry contained in the transfer matrix, consistent with our intuitive notion of ``topological".\\

When considering topological defects along both non-contractible cycles of the torus, we place the following object, which we will refer to as a tube, at the intersection of the defects:
\begin{equation}
\mathcal{T}_{\mu \nu \mu}^\gamma = \includegraphics[valign=c,page=21]{figures/kbtorus},
\end{equation}
which due to \eqref{pullingThrough} and \eqref{zipper} can also be freely moved through the lattice. In \cite{vanhove2018mapping} it was shown that the individual terms in the sum \eqref{torusTwist} can be isolated by inserting a projector
\begin{equation}
P_{\gamma,i}^{\alpha\bar{\beta}} = \includegraphics[valign=c,page=8]{figures/kbtorus}
\end{equation}
in the partition function to obtain
\begin{align}
(Z_\gamma^{\alpha\bar{\beta}})^{\text{tor}} = \tr(P_{\gamma,i}^{\alpha\bar{\beta}}\widetilde{T}^{L_y}_\gamma) \simeq \chi_{\alpha}(q)\overline{\chi}_{\overline{\beta}}(\overline{q}).
\label{torusProjected}
\end{align}

This is graphically depicted in Figure \ref{Torus}. The pulling-through property (\ref{pullingThrough}) also ensures that this projector commutes with the twisted transfer matrix. When constructing these partition functions as a strange correlator of some string-net model, these projectors are obtained as simple idempotents of the tube algebra, which naturally appear when considering the different possible ground states on the torus; for details, we refer to Appendix \ref{App:StringNets}. We will demonstrate this isolation explicitly on the lattice in section \ref{Sec:results}. \\

In the above formulas we have chosen the modular parameter $\tau$ to be purely imaginary. Choosing a general complex parameter corresponds to introducing momentum in the spectrum. The translation invariant transfer matrix commutes with the translation operator
\begin{equation}
\Gamma = e^{\frac{2\pi i}{L_x}P} = \includegraphics[valign=c,page=12]{figures/kbtorus},
\end{equation}
which shifts the lattice by one site. It is known from conformal invariance that the momentum operator $P = L_0-\overline{L}_0$. Choosing the modular parameter as $\tau = \frac{1}{L_x} + i\frac{L_y}{L_x}$, one can write down the partition function with a translation:

\begin{align}
 \tr(\Gamma T^{L_y}) \simeq q^{-\frac{c}{12}} \sum_{\alpha,\overline{\beta},n,m} M_{\alpha\bar{\beta}} c_{\alpha}(n) c_{\beta}(m) |q|^{h_{\alpha} + h_{\beta} +n+m} e^{\frac{2\pi i}{L_x}\left(h_{\alpha} - h_{\beta} +n-m \right)}.
 \label{translatedPartition}
 \end{align}

In the diagonal case ($M_{\alpha\bar{\beta}} = \delta_{\alpha\bar{\beta}}$), the conformal spin $s_{\alpha} = h_{\alpha}-h_{\alpha} = 0$ appears in the imaginary part of (\ref{translatedPartition}). The same can be done for twisted partition functions, but one has to be careful in defining the twisted translation operator:

\begin{equation}
\widetilde{\Gamma}_{\gamma} = \includegraphics[valign=c,page=13]{figures/kbtorus}.
\end{equation}

For the twisted translation operator $\widetilde{\Gamma}_{\gamma}^{L_x} \neq \mathbb{1}$; instead, we get a non-trivial operator called the Dehn twist:
\begin{equation}
\widetilde{\Gamma}_{\gamma}^{L_x} = D_{\gamma} = \includegraphics[valign=c,page=14]{figures/kbtorus}.
\end{equation}

The action of the Dehn twist consists of cutting the torus into a cylinder, fully twisting one end of the cylinder by $L_x$ sites and finally gluing the cylinder back to a torus. The result of the momentum labeling on the spectrum with a non-trivial horizontal twist is the appearance of topological corrections to the conformal spins ($h_{\alpha}-h_{\beta}$, the eigenvalues of the Dehn twist), which are no longer necessarily integers. The twist actually introduces an effective length $L_{\text{eff}}$ for which $\widetilde{\Gamma}_{\gamma}^{L_{\text{eff}}} = \mathbb{1}$. The result of the Dehn twist on the lattice partition function is trivial in the case of no twist. It is the modular transformation ($\mathcal{T}$) on the torus that implements the mapping $\tau \rightarrow \tau +1$. This is consistent with the fact that the untwisted (empty) partition function is modular invariant. \\

Besides the $\mathcal{T}$ transformation, the modular group is generated by an $\mathcal{S}$-transformation, which implements the mapping $\tau \rightarrow -\frac{1}{\tau}$. The effect of this transformation on the characters is given by the topological $S$-matrix:

\begin{align}
\chi_{\alpha}(q) = \sum_{\beta} S_{\alpha,\beta} \chi_{\beta}(\widetilde{q})
\label{Strans}
\end{align}

with $\widetilde{q} = e^{-\frac{2\pi i}{\tau}}$. The untwisted partition function is again invariant under this transformation, but single terms, obtained by the projection on the simple idempotents (\ref{torusProjected}), are not. The transformation on the idempotents is known and is implemented by the tensor product of the $S$-matrix with itself:

\begin{align}
P^{\alpha\bar{\beta}} \rightarrow \sum_{\delta,\bar{\epsilon}}S_{\alpha,\delta}\bar{S}_{\bar{\beta},\bar{\epsilon}} P^{\delta\bar{\epsilon}}.
\end{align}

Note that idempotents with a twist $\gamma$ in the $x$-direction are not mapped to idempotents with the same twist in the $x$-direction as the role of the $\mathcal{S}$ transformation is exactly to interchange the $x$ and $y$ direction. An important question is to find partition functions with twists in both directions of the torus such that the partition function is modular invariant (invariant under both $\mathcal{T}$ and $\mathcal{S}$). We will show that such a twisted modular invariant partition function can be mapped to an untwisted partition function of a different model; this procedure is called orbifolding and is explained in section \ref{Sec:Orbifold}. \\

A numerical shortcut to obtain both the scaling dimensions and the conformal spin for transfer matrices with a real spectrum is to directly diagonalize $\Gamma T$. The real part of the spectrum then contains the towers of $\Delta$ and the imaginary part the spins $s$. This is exactly how the conformal spectra are obtain in section \ref{Sec:results}. \\

We conclude this section by illustrating the torus partition function for the Ising CFT and the non-unitary Yang-Lee CFT. These results will be relevant for the numerical simulations performed in section \ref{Sec:results}. \\

The Ising CFT has central charge $c = 1/2$ and three primaries labeled by $\mathbf{1}$, $\sigma$ and $\psi$ with corresponding conformal weights $h_{\mathbf{1}} = 0$, $h_{\sigma} = 1/16$ and $h_{\psi} = 1/2$, represented in the \emph{Kac table} as
\begin{center}
	\begin{tabular}{c|c c c}
		\rule{0pt}{1.2em}%
		$\widetilde{\it{1/2}}$ & \cellcolor{blue!25}$1/2$ & $1/16$ & \cellcolor{blue!25}$0$ \\[0.1em]
		\rule{0pt}{1.2em}%
		$\widetilde{\it{0}}$ & $0$ & \cellcolor{blue!25}$1/16$ & $1/2$ \\[0.1em]
		\hline
		\rule{0pt}{1.2em}%
		&$\it{0}$&$\it{1/2}$&$\it{1}$
	\end{tabular}
\end{center}
where repeated fields are shaded blue. The corresponding MFC $\mc{D}_\text{Ising}$ has 3 simple objects with non-trivial fusion rules
\begin{equation}
\sigma \otimes \sigma = 1 + \psi, \quad \psi \otimes \sigma = \sigma, \quad \psi \otimes \psi = 1;
\end{equation}
the remaining data can be found in Appendix \ref{App:data}. The labels $\{ {\it{0}},{\it{1/2}},{\it{1}} \} \in \text{su}(2)_2$ and $\widetilde{\it{0}},\widetilde{\it{1/2}} \in \text{su}(2)_1$ in the Kac table stem from the realization of $\mc{D}_\text{Ising}$ as the coset
\begin{equation}
\mc{D}_\text{Ising} = \frac{\text{su}(2)_1 \otimes \text{su}(2)_1}{\text{su}(2)_2},
\end{equation}
which is nothing but a means to construct the data of $\mc{D}_\text{Ising}$ from the known data of the $\text{su}(2)_k$ models. The partition function on a torus can be twisted with the topological defects corresponding to these primary states \cite{petkova2001generalised}:

\begin{align}
\begin{split}
Z_{\mathbf{1}}^\text{tor} &= |\chi_{0}(q)|^2 + |\chi_{1/2}(q)|^2 + |\chi_{1/16}(q)|^2 \\
Z_{\mathbf{\psi}}^\text{tor} &= |\chi_{1/16}(q)|^2 + \chi_{0}(q)\overline{\chi}_{1/2}(\overline{q}) + \chi_{1/2}(q)\overline{\chi}_{0}(\overline{q}) \\
Z_{\mathbf{\sigma}}^\text{tor} &= \chi_{0}(q)\overline{\chi}_{1/16}(\overline{q}) + \chi_{1/16}(q)\overline{\chi}_{0}(\overline{q}) + \chi_{1/2}(q)\overline{\chi}_{1/16}(\overline{q}) + \chi_{1/16}(q)\overline{\chi}_{1/2}(\overline{q})
\end{split}
\label{isingTorus}
\end{align}

with

\begin{align}
\begin{split}
\chi_{0}(q) &= q^{-c/24}(1 + q^2 + q^3 + 2q^4 + 2q^5 + 3q^6 + ...)\\
\chi_{1/2}(q) &= q^{1/2-c/24}(1 + q + q^2 + q^3 + 2q^4 + 2q^5 + 3q^6 + ...)\\
\chi_{1/16}(q) &= q^{1/16-c/24}(1 + q + q^2 + 2q^3 + 2q^4 + 3q^5 + 4q^6 + ...).
\end{split}
\label{IsingCharacters}
\end{align}

The $S$-matrix for the Ising model reads

\begin{align}
S_{\text{Ising}} = \frac{1}{2}
\begin{pmatrix}
1& \sqrt{2}& 1\\
\sqrt{2}&0&-\sqrt{2}\\
1&-\sqrt{2}&1
\end{pmatrix},
\end{align}

with the order of the primary states: $\bf{1}$, $\sigma$ and $\psi$. \\

Similarly, for the non-unitary case of the Yang-Lee CFT, with primary states $\mathbf{1}$ and $\tau$ and corresponding conformal weights $h_{\bf{1}} = 0$ and $h_{\tau} = -1/5$ we get the following Kac table:

\begin{center}
	\begin{tabular}{c|c c c c}
		\rule{0pt}{1.2em}%
		$\tilde{\it{0}}$ & $0$ & \cellcolor{blue!25}$-1/5$ & $-1/5$ & \cellcolor{blue!25}$0$ \\[0.1em]
		\hline
		\rule{0pt}{1.2em}%
		&$\it{0}$&$\it{1/2}$&$\it{1}$&$\it{3/2}$
	\end{tabular}
\end{center}
with corresponding $\mc{D}_\text{YL}$ that has two simple objects satisfying the non-trivial fusion rule
\begin{equation}
\tau \otimes \tau = 1 + \tau.
\end{equation}
We get two corresponding twisted partition functions:

\begin{align*}
Z_{\mathbf{1}}^\text{tor} &=  |\chi_{0}(q)|^2 + |\chi_{-1/5}(q)|^2\\
Z_{\mathbb{\tau}}^\text{tor} &=  |\chi_{-1/5}(q)|^2 + \chi_{0}(q)\overline{\chi}_{-1/5}(\overline{q}) + \chi_{-1/5}(q)\overline{\chi}_{0}(\overline{q})
\end{align*}

with

\begin{align}
\begin{split}
\chi_{0}(q) &= q^{-c/24}(1 + q^2 + q^3 + q^4 + q^5 + 2q^6 + ...)\\
\chi_{-1/5}(q) &= q^{-1/5-c/24}(1 + q + q^2 + q^3 + 2q^4 + 2q^5 + 3q^6 + ...).
\end{split}
\label{YLCharacters}
\end{align}

The $S$-matrix for the Yang-Lee model reads

\begin{align}
S_{\text{Yang-Lee}} = \frac{2}{\sqrt{5}}
\begin{pmatrix}
-\sin \frac{2\pi}{5}&\sin \frac{4\pi}{5}\\
\sin \frac{4\pi}{5}&\sin \frac{2\pi}{5}
\end{pmatrix}.
\end{align}

\section{Klein bottle partition functions with topological defects}
\label{Sec:TorusvsKlein}

\begin{figure}[h]
	\centering
	\begin{minipage}[b]{0.4\linewidth}
		\includegraphics[width=\textwidth]{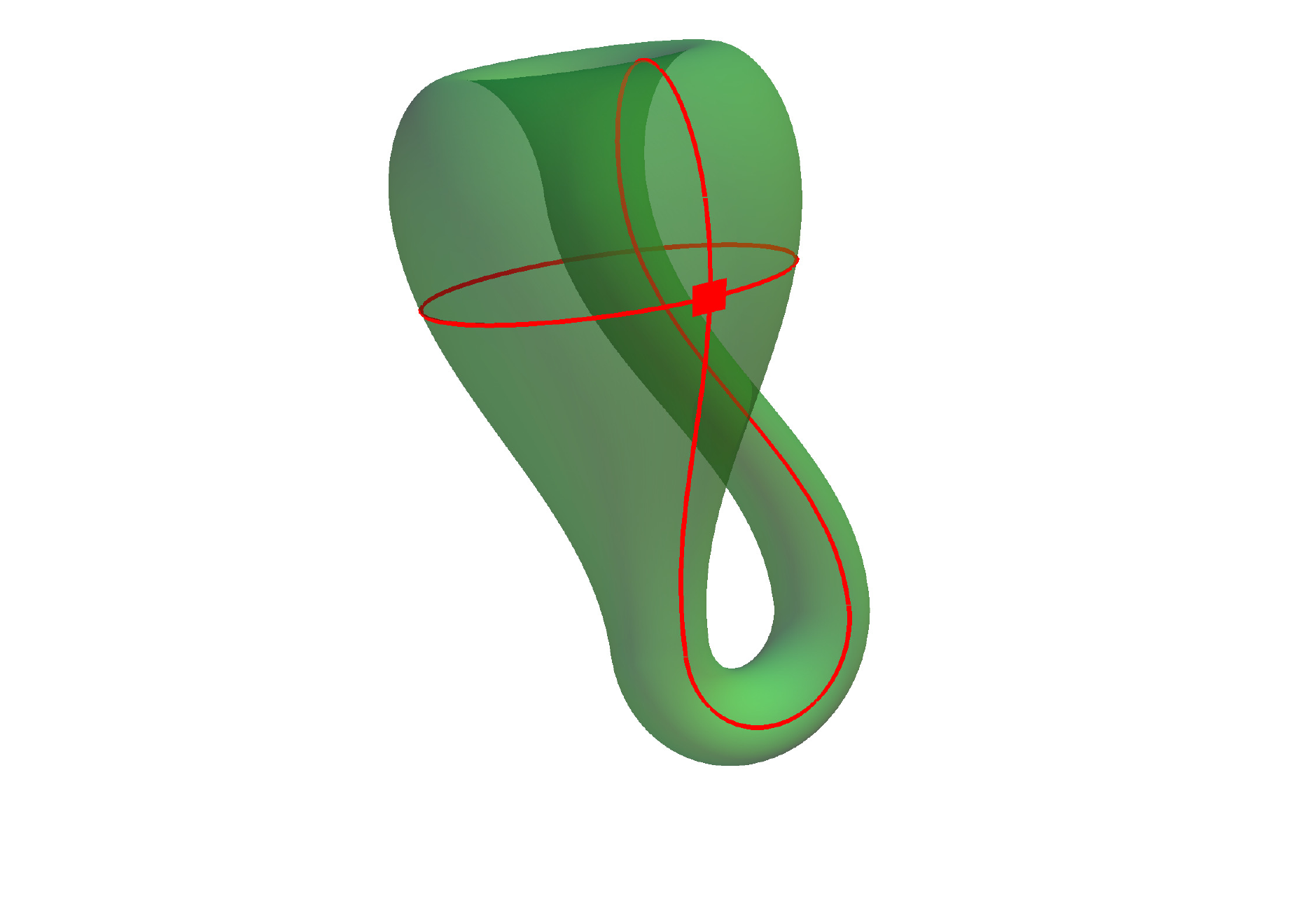}%
	\end{minipage}
	\begin{minipage}[b]{0.4\linewidth}
		\includegraphics[page=2]{figures/kbtorus}%
	\end{minipage}
	\caption{The Klein bottle partition function with the insertion of topological defects in both x- and y-directions. The reflection in the y-direction on the Klein bottle is presented by the different arrow directions. The crossing of the defects is marked with a red square, indicating the insertion of a simple idempotent.}
	\label{klein}
\end{figure}

The idea of defining CFT on a Klein bottle dates back to works on string theory (see \cite{fuchs2004tft} and references therein), and was recently extensively discussed in a series of papers \cite{tu2017universal, tang2017universal, tang2019klein} for (1+1) dimensional critical spin systems. The Klein bottle is shown in figure \ref{klein}. The partition function of the critical 2D model can be written in terms of its transfer matrix $T^{L_y}$ as \cite{blumenhagen2009introduction}:

\begin{align}
	Z^\text{KB} &= \tr(\mathcal{R}T^{L_y}), \quad \mathcal{R} = \includegraphics[valign=c,page=15]{figures/kbtorus},
\end{align}
which differs from the torus partition function by the spatial reflection operator $\mc{R}$, which implements the change of arrow direction in the right hand side of Figure \ref{klein} when identifying the top and bottom edges. The effect of the operator $\mathcal{R}$ on the spectrum is to interchange the left and right moving parts in the eigenstates: $\mathcal{R}\ket{\alpha,\overline{\beta}} = \ket{\beta, \overline{\alpha}}$. The left- and right-movers are swapped on the Klein bottle before the trace is taken, so that only left-right symmetric states $\ket{\alpha,\overline{\alpha}}$ survive. The partition function on the Klein bottle can therefore be written as a trace over just the left-right symmetric part of the Hilbert space in the following way:

\begin{align}
	Z^\text{KB} &=  \tr_{\mathcal{H}\otimes\overline{\mathcal{H}}}(\mathcal{R}q^{L_0-(c/24)}\overline{q}^{\overline{L}_0-(c/24)}) = \tr_{(\mathcal{H}\otimes\overline{\mathcal{H}})_{sym}}(q^{2L_0-(c/12)}).
\end{align}

Similarly as on the torus, the Klein bottle partition function is a sum of characters, but instead of the sesquilinear form we have a sum of single characters (due to the above formula):
\begin{align}
	Z^\text{KB} &=  \sum_{\alpha}M_{\alpha\bar{\alpha}}\chi_{\alpha}(q^2) =  q^{-\frac{c}{6}}\sum_{\alpha,n} M_{\alpha\bar{\alpha}} c_{\alpha}(n) q^{2h_{\alpha} + 2n}.
\end{align}
In both the case of a diagonal ($M = \mathbb{1}$) and non-diagonal partition function, the Klein bottle partition function will only contain single characters that were part of the diagonal part of the original torus partition function. \\

Topological defects can be inserted in both directions of the Klein bottle on the lattice, just like for the torus. However, in contrast to the torus, only certain defects on the Klein bottle yield a non-zero partition function. Inserting a topological defect labeled by $\gamma$ in the $y$-direction on the Klein bottle (just like on the torus) selects the diagonal part of the twisted partition function:

\begin{align}
	Z_\gamma^\text{KB} &=  \sum_{\alpha}\widetilde{M}^{\gamma}_{\alpha\bar{\alpha}}\chi_{\alpha}(q^2).
	\label{kleinSumChar}
\end{align}
It is clear that such twisted Klein bottle partition functions will only be non-zero if the original twisted torus partition function contains left-right symmetric parts. The simple idempotents $P_{c,i}^{\alpha\bar{\alpha}}$ of the tube algebra can then be used to project the Klein bottle partition function onto its single characters:
\begin{align}
	(Z_\gamma^{\alpha})^{\text{KB}} = \tr(\mathcal{R}P_{\gamma,i}^{\alpha\bar{\alpha}}\widetilde{T}^{L_y}_\gamma) = \chi_{\alpha}(q^2).
	\label{singlechar}
\end{align}
This requires that the simple idempotents commute with the reflection operator; we argue that this is the case in Appendix \ref{App:reflection}. \\

Numerically extracting a Klein bottle spectrum can be done as follows: when the transfer matrix and the simple idempotents commute with the reflection operator, all the zero and $\pi$-momentum eigenstates can consistently be labeled by the eigenvalues under reflection. Not all zero momentum states will contribute to the Klein bottle partition function however, since exactly degenerate eigenstates with reflection quantum numbers $+1$ and $-1$ cancel one another after the final trace is taken. The single characters corresponding to a primary state $\alpha$ emerge from such a cancellation. Starting from a twisted transfer matrix, we first project on a left-right symmetric single term (Formula \ref{torusProjected}):
\begin{align*}
P_\gamma^{\alpha\alpha}\tilde{T}_\gamma \rightarrow \chi_{\alpha}(q)\overline{\chi}_{\alpha}(\overline{q}) = \ q^{2h_{\alpha} - c/12} \sum_{n,m} c_{\alpha}(n)c_{\alpha}(m)q^n\overline{q}^m. \\
\end{align*}

Next, we project on momentum zero and $\pi$
\begin{align*}
\begin{split}
\stackrel{p=0,\pi}{\rightarrow} \ &q^{2h_\alpha - c/12}\sum_{n} c_{\alpha}(n)^2 |q|^{2n}
\end{split}
\end{align*}

and finally, the spectrum is labeled by the eigenvalues under reflection and exactly degenerate eigenvalues with different reflection quantum numbers ($+1$ or $-1$) are thrown out, since they will be canceled in the final trace of the Klein bottle partition function:

\begin{align*}
\begin{split}
\stackrel{\mathcal{R}^{\pm}}{\rightarrow} \  &q^{2h_\alpha - c/12}\sum_{n} c_\alpha(n) \ q^{2n}
= \ \chi_{\alpha}(q^2).
\end{split}
\end{align*}

The reflection operator leads to a cancellation of the crossterms at some level whenever $c_{\alpha}(n) > 1$, leading in the end to a single character. This is exactly how the single characters for the Klein bottle are numerically reproduced in Section \ref{Sec:results}. \\

For the Ising CFT, the Klein bottle partition function with a $\sigma$-twist is zero, because there are no diagonal character combinations in the $\sigma$-twisted torus partition function. For the trivial and the $\psi$-twist, the Klein bottle partition functions become:

\begin{align}
Z_{\mathbf{1}}^\text{KB} &= \chi_{\mathbf{1}}(q^2) + \chi_{\psi}(q^2) + \chi_{\sigma}(q^2)
\label{kleinIsing} \\
Z_{\psi}^\text{KB} &= \chi_{\sigma}(q^2)
\label{kleinIsingPsi}
\end{align}

In section \ref{Sec:results}, this will be illustrated numerically through exact diagonalization techniques, where the projectors that appear in \eqref{singlechar} will be explicitly constructed in terms of the MPOs, in order to further decompose \eqref{kleinIsing} into single characters.\\

For the Yang-Lee CFT, we get:

\begin{align}
Z_{\mathbf{1}}^\text{KB} &= \chi_{\mathbf{1}}(q^2) + \chi_{\tau}(q^2)
\label{kleinYangLee1} \\
Z_{\tau}^\text{KB} &= \chi_{\tau}(q^2).
\label{kleinYangLee2}
\end{align}

A similar projection will allow for the decomposition of \eqref{kleinYangLee1}.\\

\subsection{Klein bottle entropy}
\label{Sec:KBE}

\begin{figure}
	\centering
	\includegraphics[page=5,width=\textwidth]{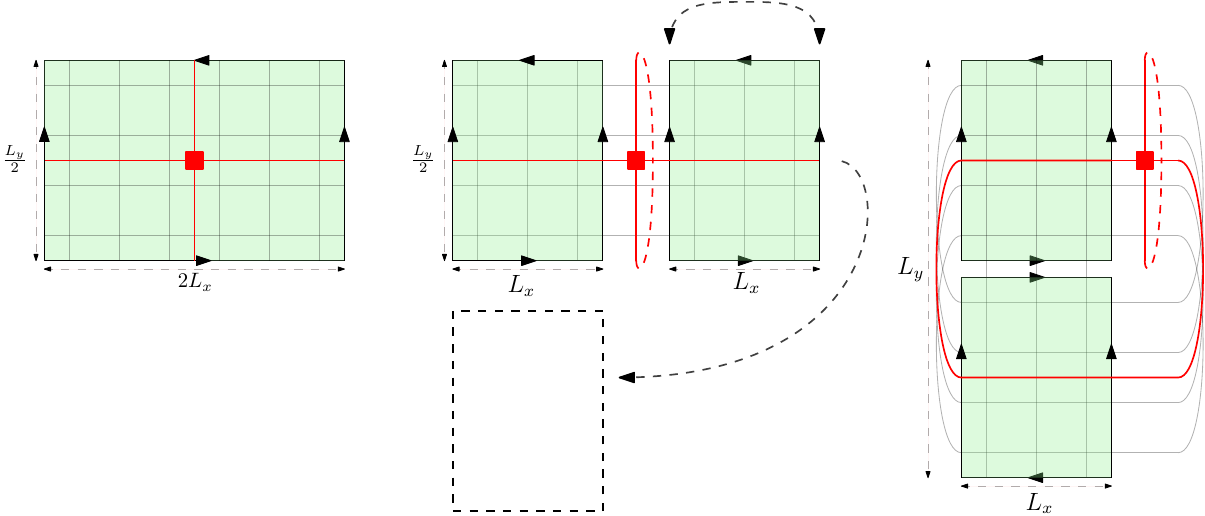}%
	\caption{The cutting-folding-sewing trick to map the Klein bottle of size $(2L_x,L_y/2)$ to a cylinder of size $(L_x,L_y)$ with non-local interactions at the endpoints in the $x$-directions, including a simple idempotent. The reflection is performed around the twist, such that the tube itself remains a ring of size $L_y/2$.}
	\label{foldingTrick}
\end{figure}

We now turn to the implication of formula \ref{singlechar} on the Klein bottle entropy, for rational (non-)unitary CFTs \cite{tu2017universal}. We therefore allow for at least one negative conformal weight and will denote the smallest one as $h_{\text{min}}$ and the corresponding character as $\chi_{\rho_{\text{min}}}$. We will need the $S$-transformation for the character \eqref{Strans} to rewrite the Klein bottle partition function this time with a topological defect $\gamma$:

\begin{align*}
	Z_\gamma^\text{KB} &=  \sum_{\alpha}\widetilde{M}^{\gamma}_{\alpha\bar{\alpha}}\chi_{\alpha}(q^2)\\
	&= \sum_{\alpha,\beta} \widetilde{M}^{\gamma}_{\alpha\bar{\alpha}} S_{\alpha,\beta}\chi_{\beta}(\widetilde{q}^{1/2}).
\end{align*}

In the limit $L_x \gg L_y$, the right hand side will be dominated by the contributions from the primary states. Furthermore, in the limit where $L_y \rightarrow \infty$ only the character corresponding to the primary state with the smallest (possibly negative) conformal weight ($\chi_{\rho_{\text{min}}}$) survives, since $\widetilde{q} \rightarrow 0$. In these limits $\chi_{\rho_{\text{min}}}(\widetilde{q}^{1/2}) \simeq (\widetilde{q}^{1/2})^{h_{\text{min}} - c/24} = (\widetilde{q}^{1/2})^{-c_{\text{eff}}/24}$, with $c_{\text{eff}} = c - 24h_{\text{min}}$. This means that we can write:
\begin{align*}
	Z_\gamma^\text{KB} &\simeq \sum_{\alpha} \widetilde{M}^{\gamma}_{\alpha\bar{\alpha}} S_{\alpha,\rho_{\text{min}}}\chi_{\rho_{\text{min}}}(\widetilde{q}^{1/2})\\
	&\simeq \left( \sum_{\alpha} \widetilde{M}^{\gamma}_{\alpha\bar{\alpha}} S_{\alpha,\rho_{\text{min}}} \right) e^{\frac{\pi c_{\text{eff}}}{24L_y}L_x}.
\end{align*}
Taking into account the non-universal terms due to the finite lengths \cite{tu2017universal}:
\begin{align*}
	Z_\gamma^\text{KB}(L_x,L_y) &\simeq \left( \sum_{\alpha} \widetilde{M}^{\gamma}_{\alpha\bar{\alpha}} S_{\alpha,\rho_{\text{min}}} \right) e^{-f_0L_xL_y + \frac{\pi c_{\text{eff}}}{24L_y}L_x},
\end{align*}
with $f_0$ the free-energy density. The analogue for the torus (without twists) was obtained in \cite{blote1986conformal,affleck1988universal}:
\begin{align*}
	Z^\text{tor}(L_x,L_y) &\simeq e^{-f_0L_xL_y + \frac{\pi c_{\text{eff}}}{6L_y}L_x}.
\end{align*}
This leads us to define the universal ratio for the Klein bottle entropy \cite{tu2017universal}, with the inclusion of topological defects
\begin{align}
	g_\gamma = \frac{Z_\gamma^\text{KB}(2L_x,\frac{L_y}{2})}{Z^\text{tor}(L_x,L_y)} \simeq \sum_{\alpha} \widetilde{M}^{\gamma}_{\alpha\bar{\alpha}} S_{\alpha,\rho_{\text{min}}}
	\label{KBE}
\end{align}
However, as \eqref{singlechar} implies, the projectors $P_{\gamma,i}^{\alpha\bar{\alpha}}$ can be applied to the Klein bottle, singling out the single character terms. The same can be done for the Klein bottle entropy

\begin{align}
	g_\gamma^{\alpha} = \frac{(Z_\gamma^{\alpha})^{\text{KB}}(2L_x,\frac{L_y}{2})}{Z^\text{tor}(L_x,L_y)} \simeq S_{\alpha,\rho_{\text{min}}},
	\label{gca}
\end{align}

allowing us to identify individual $S$-matrix elements.\\

In the case of unitary CFTs, the primary state with smallest conformal weight is the identity field with $h_{\bf{1}} = 0$ and $c_{\text{eff}} = c$, the $S$-matrix elements in the above sum are $S_{a,\bf{1}}$ and correspond to the quantum dimensions $d_a$ (up to the normalization $\mathcal{D} = \sqrt{\sum_{a}d_a^2}$). \\

In practice, it will be important to map the non-orientable Klein bottle partition function to an orientable manifold in order to calculate the Klein bottle entropies and single $S$-matrix elements \eqref{gca}. To this end, by first cutting the Klein bottle, then flipping it and finally sewing it back together, it was shown in \cite{tang2017universal} that the Klein bottle can be mapped to a cylinder with non-local interactions at the boundaries. We will use the same trick here, but we have to be careful in the presence of defects, as we will show now. \\

We will demonstrate the mapping on a Klein bottle with a simple idempotent ($P_{\gamma,i}^{\alpha\bar{\alpha}}$), where we immediately superimpose a lattice on the Klein bottle. The mapping is shown in figure \ref{foldingTrick} and it can be directly used to rewrite the numerator in the expression of $g_{\gamma}^{\alpha}$ in \eqref{gca} as a decomposition in terms of the tube elements:

\begin{multline}
		(Z_{\gamma}^{\alpha})^{\text{KB}}(2L_x,L_y/2) =
		 \vcenter{\hbox{\includegraphics[page=7,scale=0.7]{figures/kbtorusNew}}} = \vcenter{\hbox{\includegraphics[page=6,scale=0.7]{figures/kbtorusNew}}}.
	\label{KBentropy}
\end{multline}

In the second step the resolution of the identity has been inserted to obtain the overlap of a twisted transfer matrix $\widetilde{T}^{L_x}$ with a non-local boundary state on the left and on the right. In the limit $L_x \rightarrow \infty$, only the largest magnitude eigenvectors of the twisted transfer matrix $\ket{v_i}$ (possibly degenerate due to the presence of the twist) contribute to the Klein bottle entropy, normalized by the untwisted unique eigenvector $\ket{v^{\bf{1}}}$:

\begin{align}
g_\gamma^{\alpha} = \sum_{i}\vcenter{\hbox{\includegraphics[page=8,scale=0.7]{figures/kbtorusNew}}},
\label{KBentropyFixedPoint}.
\end{align}

We allow for a different left and right fixed point. This is important for the Yang-Lee model, where the transfer matrix is non-Hermitian.
This is how in practice the Klein bottle entropies $g_{\gamma}^{\alpha}$ can be calculated by only keeping these fixed-point terms. Finite-size effects will limit the precision of $g_{\gamma}^{\alpha}$, because of the finite size in the $y$-direction, but going to large cylinder sizes allows for accurate calculations. This will be shown in section \ref{Sec:results} through exact diagonalization of the twisted transfer matrix. \\

We again illustrate the above for the Ising and the Yang-Lee model. The Klein bottle entropies (\ref{KBE}) for the Ising model become:

\begin{align}
g_{\mathbf{1}} &= S_{\mathbf{1},\mathbf{1}} + S_{\psi,\mathbf{1}} + S_{\sigma,\mathbf{1}} = 1+\frac{\sqrt{2}}{2},\\
\label{KBEIsing1}
g_{\psi} &= S_{\sigma,\mathbf{1}} = \frac{1}{\sqrt{2}}.
\end{align}

Using the projectors, \eqref{KBEIsing1} can also be split up into its three terms by virtue of (\ref{gca}). \\

In the non-unitary Yang-Lee case, the primary state with the smallest conformal weight does not equal the identity and the Klein bottle entropy does not probe the first column of the $S$-matrix, but rather the column corresponding to the primary state with smallest conformal weight. From equations \eqref{kleinYangLee1} and \eqref{kleinYangLee2}, one can write:

\begin{align}
g_{\mathbf{1}} = &S_{\mathbf{1},\tau} + S_{\tau,\tau} = \sqrt{1+\frac{2}{\sqrt{5}}} \label{KBEYL1},\\
g_{\tau} = &S_{\tau,\tau}.
\end{align}

It is again \eqref{KBEYL1} that will be split up in its terms. \\

\section{Cylinder partition functions with topological defects}
\label{Sec:cylinder}

\begin{figure}[ht]
	\centering
	\begin{minipage}[c]{0.4\linewidth}
		\centering
		\includegraphics[width=\textwidth]{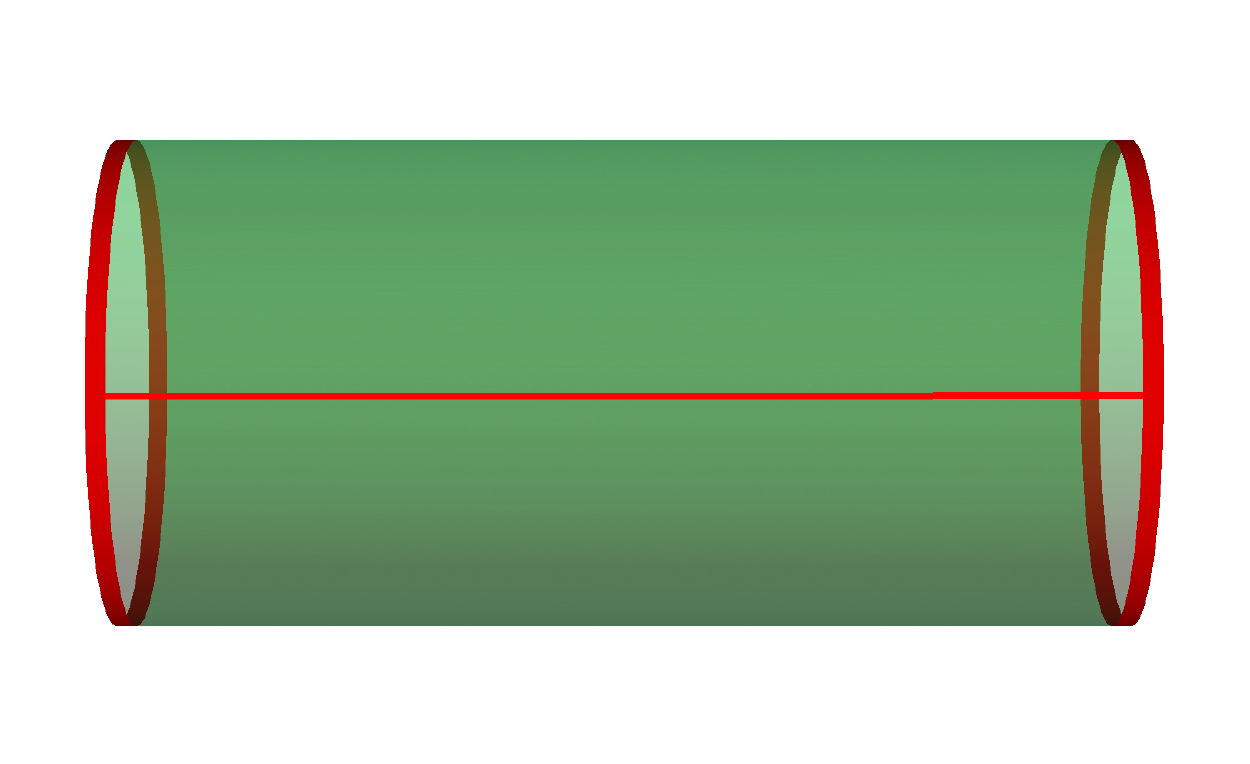}%
	\end{minipage}\quad
	\begin{minipage}[c]{0.4\linewidth}
		\includegraphics[width=\textwidth]{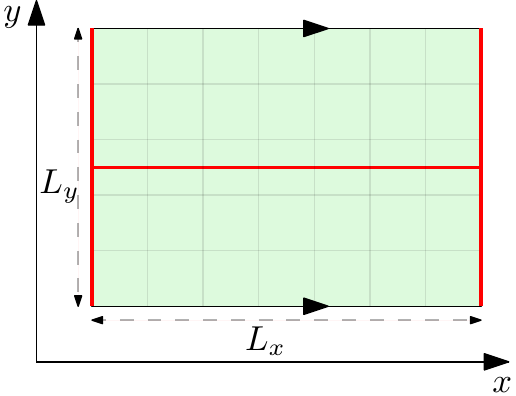}%
	\end{minipage}
	\caption{The cylinder partition function with the insertion of a topological defect in the $x$-direction and two unspecified boundaries.}
	\label{cylinder}
\end{figure}

The study of CFTs on manifolds with boundaries and its relation to the fusion rules of the CFT was initiated by Cardy in \cite{cardy1989boundary}. Much like the reflection operator on the Klein bottle, the boundary on the cylinder imposes non-trivial relations between the left- and right-movers by requiring no momentum flows across the boundary. For this reason, only single characters appear in CFT partition functions on the cylinder, just like in the case of the Klein bottle, instead of the sesquilinear combination on the torus. The cylinder partition function is shown in figure \ref{cylinder} with a horizontal defect and two unspecified boundary conditions.

\subsection{Ishibashi and Cardy states}
\label{Sec:IshibashiCardy}
On the boundary state $\ket{B}$, the requirement that the off-diagonal components of the stress-energy tensor of the CFT should vanish translate to $L_n\ket{B} = \overline{L}_{-n}\ket{B}$. The states that satisfy this constraint for diagonal theories are the so-called Ishibashi states \cite{ishibashi1989boundary}, and there is one for every primary state:

\begin{align*}
	\ket{I_\alpha} = \sum_{n} \sum_{m_n=1}^{c_{\alpha}(n)}\ket{h_\alpha+n;m_n, \overline{h}_\alpha + n;m_n},
\end{align*}

where $n$ indicates the level of the state and $m_n$ is the degeneracy label at level $n$. This Ishibashi state is an equal weight superposition of all left-right symmetric states in one particular tower. Not coincidentally, these left-right symmetric states are precisely the states from the conformal towers that survived on the Klein bottle, and similarly we can write for the cylinder partition function:

\begin{align}
	\begin{split}
		Z_{I_\alpha I_\beta}^{\text{cyl}} &= \bra{I_\alpha} \vcenter{\hbox{\includegraphics[page=1,scale=1]{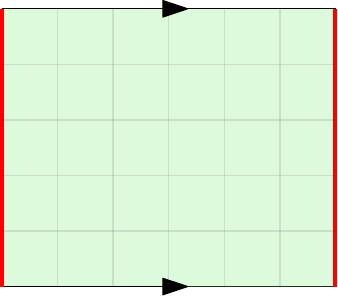}}} \ket{I_\beta}\\
		&= \delta_{\alpha\beta} \chi_{\alpha}(q^2),
	\end{split}
	\label{cylinderIshi}
\end{align}

since states in different towers are linearly independent and form a complete basis in the space of conformal blocks (at least in the diagonal case $M=\mathbb{1}$). If we define the reference state $\ket{\Sigma} = \sum_{\alpha} \ket{I_\alpha}$, it is clear that the projector $\ket{\Sigma}\bra{\Sigma}$ projects onto the states that are present in the Klein bottle partition function, so that we can formally identify the Klein bottle partition function with the cylinder partition function with boundary states $\ket{\Sigma}$: $Z^{\text{KB}} = \tr{(\ket{\Sigma}\bra{\Sigma} T^{L_y})} =  Z_{\Sigma\Sigma}^{\text{cyl}}$. \\

The role of the projectors $P_{\mathbf{1}}^{\alpha\bar{\alpha}}$, which project the untwisted Klein bottle partition function onto the single characters, is clear here: they project the reference state $\ket{\Sigma}$ onto the corresponding Ishibashi state. \\

One may wonder what happens when topological defects $O_{\alpha}$ are inserted in the cylinder, instead of the projectors $P_{\mathbf{1}}^{\alpha\bar{\alpha}}$, which are made from linear combinations of $O_{\alpha}$ (essentially inverting the relation between $P$ and $O$). To this end, Cardy showed that by identifying two different ways of interpreting the cylinder partition function, it is possible to relate the Ishibashi states to the so-called Cardy states. Where the previous cylinder partition functions could be interpreted as the propagation of a periodic transfer matrix from the left state to the right, making the relation to the Klein bottle obvious, a different interpretation is the trace of a transfer matrix, in the other direction, with fixed boundaries $\alpha$ at $y=0$ and $\beta$ at $y = L_y$:

\begin{align}
	Z_{\alpha\beta}^{\text{cyl}} = \tr{(T^{(\alpha\beta)L_x}_{L_y})} = \bra{\alpha} \vcenter{\hbox{\includegraphics[page=1,scale=1]{figures/cylinderSquare}}} \ket{\beta}.
	\label{secondPicture}
\end{align}

The transfer matrix with boundaries $\alpha$ and $\beta$, can be formally written (similarly as in equation \ref{transferHam} for the periodic case) as

\begin{align}
	T^{(\alpha\beta)}_{L_y} = e^{-H^{(\alpha\beta)}_{L_y}} = \includegraphics[valign=c,page=16]{figures/kbtorus}.
	\label{transferHamFixed}
\end{align}	

Just like on the torus (\ref{Ham}), one can isolate a universal CFT part from the Hamiltonian $H^{(\alpha\beta)}_{\text{CFT}}$ and write it in terms of a Virasoro generator:

\begin{align}
H^{(\alpha\beta)}_{\text{CFT}} = \frac{\pi}{L_y}\left(L_0 - \frac{c}{24}\right).
\label{CFThamCylinder}
\end{align}

Writing the partition function as a sum of single characters

\begin{align}
	Z_{\alpha\beta}^{\text{cyl}} = \sum_{\gamma} n_{\alpha\beta}^{\gamma} \ \chi_{\gamma}(\widetilde{q}^{\frac{1}{2}}),
	\label{cardyPartition}
\end{align}	

the dependence of the partition function on the boundary states (``Cardy states") $\alpha$ and $\beta$ is entirely contained in the (positive) integers $n_{\alpha\beta}^{\gamma}$. Just like on the torus - where the final trace together with the guiding principle of modular invariance dictated how the characters were combined through the matrix $M$ - the final trace on the cylinder selects the characters present in the partition function, through the data $n_{\alpha\beta}^{\gamma}$. Note the difference of the $q$-dependence in \eqref{cardyPartition} compared to \eqref{cylinderIshi}. This is because the roles of the $x$- and $y$-direction have been swapped going from the first picture to the second and the exponent $1/2$ in \eqref{cardyPartition} can be explained by the different factor of $2$ in \eqref{CFThamCylinder} compared to \eqref{transferHam}. \\

The different pictures of the cylinder, leading to \eqref{cylinderIshi} and \eqref{cardyPartition}, are related to one another by an $S$-transformation, so that the `guiding principle' for characterizing the data $n_{\alpha\beta}^{\gamma}$ should be that the states $\ket{\alpha}$ and $\ket{\beta}$ are expressible as linear combinations of the Ishibashi states, since we still require conformal invariance, and that $n_{\alpha\beta}^{\gamma}$ are positive integers. Cardy showed that the solution is

\begin{align}
	\ket{\alpha} = \sum_{\beta} \frac{S_{\alpha,\beta}}{\sqrt{S_{\bf{1},\beta}}}\ket{I_{\beta}}
	\label{cardyIshi}
\end{align}

and that by performing an $S$ transformation on \eqref{cardyPartition}

\begin{align*}
	Z_{\alpha\beta}^{\text{cyl}} = \sum_{\gamma} n_{\alpha\beta}^{\gamma} \ \chi_{\gamma}(\widetilde{q}^{\frac{1}{2}}) = \sum_{\gamma\delta} n_{\alpha\beta}^{\gamma} S^{-1}_{\gamma\delta} \chi_{\delta}(q^2)
\end{align*}

and comparing it to \eqref{cylinderIshi}, the Verlinde formula dictates that $n_{\alpha\beta}^{\gamma} = N_{\alpha\beta}^{\gamma}$, the fusion rules of the model. Note that we did not assume unitarity of the $S$-matrix in order to keep the results as general as possible and to include non-unitary CFTs.\\

For the case of diagonal models, we can think of the Cardy states $\ket{\alpha}$ as being created from a ``vacuum'' reference state by acting on it with a topological defect: $O_{\alpha} \ket{\bf{1}} = \ket{\alpha}$. Indeed, in this case we have

\begin{align*}
	Z_{\alpha\beta}^{\text{cyl}} &= \bra{\bf{1}}\vcenter{\hbox{\includegraphics[page=2,scale=1]{figures/cylinderSquare}}} \ket{\bf{1}} = \sum_{\gamma}N_{\alpha\beta}^{\gamma}  \bra{\bf{1}}\vcenter{\hbox{\includegraphics[page=3,scale=1]{figures/cylinderSquare}}}\ket{\bf{1}}.
\end{align*}

and in particular $Z_{\alpha\bf{1}}^{\text{cyl}} = Z_{\bf{1}\alpha}^{\text{cyl}} = \chi_{\alpha}(\widetilde{q}^{\frac{1}{2}})$. We will generalize this appropriately for non-diagonal models in section \ref{Sec:Orbifold}.\\

\subsection{Twisted Cardy states and the ladder algebra}
\label{Sec:twistedCardy}
The results discussed above are not new, but we have tried to emphasize the role of the topological defects. Similar to the torus and Klein bottle, we can consider twisted cylinder partition functions; a similar discussion can be found in \cite{aasen2020topological}. It is our aim however to set the scene in order to generalize the discussion to include non-diagonal orbifold models as well in section \ref{Sec:Orbifold}. Defining the partition function on the cylinder with a defect $\gamma$ in the horizontal direction:
\begin{align}
	\begin{split}
		Z_{I_{\alpha}^{\gamma}I_{\beta}^{\gamma}}^{\text{cyl}} &= \bra{I_{\alpha}^{\gamma}} \vcenter{\hbox{\includegraphics[page=4,scale=1]{figures/cylinderSquare}}} \ket{I_{\beta}^{\gamma}}\\
		&= \delta_{\alpha\beta} \chi_{\alpha}(q^2),
	\end{split}
\end{align}

where the Ishibashi states $\ket{I_{\alpha}}$ have been generalized to an equal weight superposition $\ket{I_{\alpha}^{\gamma}}$ of all left-right symmetric states in a tower $\alpha$, present in the $\gamma$-twisted sector. In exactly the same manner as in the case without a defect, one can define a $\gamma$-twisted reference state $\ket{\Sigma^{\gamma}}$, from which the twisted Ishibashi states can be obtained by using the simple idempotents. This case is similar as the Klein bottle with a twist, so that we can write

\begin{align*}
	Z_{I_{\alpha}^{\gamma}I_{\beta}^{\gamma}}^{\text{cyl}} = \delta_{\alpha\beta}(Z_{\gamma}^{\beta})^{\text{KB}},
\end{align*}

for diagonal partition functions. Which symmetric towers will be present in the reference state $\ket{\Sigma^{\gamma}}$ can be read off from the torus and Klein bottle partition functions.\\

The twisted cylinders can be understood in the second picture, reinterpreting the trace in \eqref{secondPicture} with a twist. We rely on the ladder algebra \cite{kitaev2012models, barter2019domain, clark2009fixing} (explained in appendix \ref{Sec:App:ladder}) to derive the effect of the twist before the trace is taken. We can rewrite equation \eqref{secondPicture} with the twist as

\begin{align}
Z_{\alpha^{\gamma}\beta^{\gamma}}^{\text{cyl}} = \tr{(\mc{L}_{\alpha\beta,\alpha\beta}^{\gamma} \ T^{(\alpha\beta)L_x}_{L_y})},
\label{secondPictureTwist}
\end{align}

with ladders defined as

\begin{align*}
\mc{L}_{\alpha\beta,\alpha\beta}^{\gamma} = \includegraphics[valign=c,page=17]{figures/kbtorus},
\end{align*}

where we have dropped the degeneracy labels $j$ and $k$ in \eqref{ladderDefinition}. \\

One can derive the structure factors of this algebra and again diagonalize it, to obtain so-called ladder idempotents. The role of these projectors can already be guessed: just like the idempotents of the tube algebra project on the product-of-characters terms on the torus, the idempotents of the ladder algebra will project onto the single-character terms in \eqref{cardyPartition} on the cylinder. We can rewrite \eqref{secondPictureTwist}:

\begin{align}
	\begin{split}
		Z_{\alpha^{\gamma} \beta^{\gamma}}^{\text{cyl}} &= \bra{\alpha^{\gamma}} \vcenter{\hbox{\includegraphics[page=4,scale=1]{figures/cylinderSquare}}} \ket{\beta^{\gamma}} = \ \bra{\bf{1}} \vcenter{\hbox{\includegraphics[page=5,scale=1]{figures/cylinderSquare}}} \ket{\bf{1}}\\[1em]
		&= \sum_{\delta} \tilde{n}_{\alpha^{\gamma}\beta^{\gamma}}^{\delta} \  \bra{\bf{1}} \vcenter{\hbox{\includegraphics[page=6,scale=1]{figures/cylinderSquare}}} \ket{\bf{1}} = \sum_{\delta} \tilde{n}_{\alpha^{\gamma}\beta^{\gamma}}^{\delta} \ \chi_{\delta}(\widetilde{q}^{\frac{1}{2}}).
	\end{split}
	\label{tubeCylinder}
\end{align}

The factors $\tilde{n}_{\alpha^{\gamma}\beta^{\gamma}}^{\delta}$ generalize the fusion coefficients $N_{\alpha\beta}^{\gamma}$ for the untwisted case to the $\gamma$-twisted case. This last argument is more in the spirit of the first picture of the cylinder: the twisted transfer matrix is propagated from the left tube to the right tube, with trivial boundaries at the edges. \\

Turning to the cylinder partition functions for the Ising model, we have three Ishibashi states ($\ket{I_{\mathbf{1}}}$, $\ket{I_{\psi}}$, $\ket{I_{\sigma}}$) and three Cardy states ($\ket{\mathbf{1}}$, $\ket{\psi}$, $\ket{\sigma}$) at our disposal in the untwisted sector. We can only apply a $\psi$-twist horizontally, since it is the only non-trivial defect with fusion compatibility with the twisted Cardy states $\ket{\sigma}$ at the boundaries ($N_{\sigma,\psi}^{\sigma}>0$). Looking at \eqref{kleinIsingPsi}, only the symmetric states belonging to the $\sigma$ tower are contained in the $\psi$-twisted sector, such that we can write $\ket{\Sigma^{\psi}} = \ket{I_{\sigma}^{\psi}}$. The cylinder, just like the Klein bottle, does not allow a $\sigma$-twist. \\

The two ladder algebra elements $\mc{L}_{\sigma\sigma,\sigma\sigma}^{\bf{1}}$ and $\mc{L}_{\sigma\sigma,\sigma\sigma}^{\psi}$ can be diagonalized to obtain the corresponding $\mathds{Z}_2$ irreps, projecting a cylinder with $\ket{\sigma}$ states at the two boundaries on the character $\chi_{\mathbf{1}}$ and $\chi_{\psi}$. \\

For the Yang-Lee model, we have two Ishibashi states ($\ket{I_{\mathbf{1}}}$, $\ket{I_{\tau}}$) and two Cardy states ($\ket{\mathbf{1}}$, $\ket{\tau}$) at our disposal. We can apply a horizontal $\tau$-twist and get only one Ishibashi state in the twisted sector, so we can write $\ket{\Sigma^{\tau}} = \ket{I_{\tau}^{\tau}}$. \\

The two ladder algebra elements $\mc{L}_{\tau\tau,\tau\tau}^{\bf{1}}$ and $\mc{L}_{\tau\tau,\tau\tau}^{\tau}$ can be diagonalized to obtain corresponding ladder idempotents, projecting a cylinder with $\ket{\tau}$ states at the two boundaries on the character $\chi_{\mathbf{1}}$ and $\chi_{\tau}$. This projection for the two models will be numerically shown in Section \ref{Sec:results}.

\section{Orbifolds and non-diagonal partition functions}
\label{Sec:Orbifold}

In the above discussion we have focused on diagonal CFT, i.e. models where the modular invariant torus partition function is a diagonal combination of the holomorphic and anti-holomorphic characters, which is sometimes referred to as the Cardy case. In the TFT construction of CFT, this case is obtained by choosing the right $\mc{D}$-module category $\mc{M}$ to be equal to the modular tensor category $\mc{D}$, in which case the topological defects are also labeled by $\mc{C} = \mc{D}$. To generalize this to non-diagonal modular invariants we need to relax this constraint and consider generic right $\mc{D}$-module categories $\mc{M}$. The topological defects of such a CFT are then given by $\mc{C} = \mc{D}_\mc{M}^*$, the ``dual'' of $\mc{D}$ with respect to $\mc{M}$.\\

Recently, it was shown that exactly the same mathematical structure is present in PEPS descriptions of string-net ground states \cite{lootens2020matrix}: to define the PEPS ground state tensors for some string-net model $\mc{D}$, one has to choose a right $\mc{D}$-module category $\mc{M}$. The MPO symmetries are then given by $\mc{C} = \mc{D}_\mc{M}^*$. These different PEPS representations are locally indistinguishable, but on the torus they might represent different ground states. This can be made very explicit through the existence of MPO intertwiners that act as the interface between two different PEPS representations, which like the MPO symmetries can be freely moved through the lattice. In particular, it allows us to map ground states in one representation to ground states in another representation. For details on these constructions, we refer to Appendix \ref{App:StringNets}.\\

The above suggests that the generalized PEPS representations of \cite{lootens2020matrix} are exactly the right language to understand non-diagonal partition functions of CFT. The fact that different representations correspond to different torus ground states corresponds to the different possible modular invariant partition functions on the torus after the strange correlator map. The topological defects are indeed described by the appropriate fusion category $\mc{C}=\mc{D}_\mc{M}^*$, and we can use the MPO intertwiners to map non-diagonal partition functions to twisted diagonal partition functions. In CFT, such a relation between two different modular invariants is known as orbifolding, which essentially corresponds to modding out a (set of) topological defects. In the case that the latter are group-like, this is known as simple current extension, but the description in terms of right $\mc{D}$-module categories $\mc{M}$ allows for more general orbifolds as well, such as needed for e.g. the exceptional minimal model modular invariants $E_6$, $E_7$ and $E_8$.\\

\subsection{Orbifold on the torus}
\label{Sec:OrbifoldTorus}

As the intertwiners map different modular invariant partition functions to one another, we will write down the transfer matrix for these models in the tensor network representation, just like we did on the torus:
\begin{align*}
T_{\mc{M}} = \vcenter{\hbox{\includegraphics[valign=c,page=17]{figures/kbtorusIntertwiner}}}, \quad T_{\mc{D}} =  \vcenter{\hbox{\includegraphics[valign=c,page=16]{figures/kbtorusIntertwiner}}}.
\end{align*}

The blue and green network are the tensor network representation of the two different modular invariant partition functions. As discussed above, the topological defects of the diagonal model are simply labeled by $\alpha,\beta,\gamma,... \in \mc{D}$. For the non-diagonal model, the topological defects are now labeled by objects $a,b,c,... \in \mc{C}$. The intertwiners, which we again represent as MPOs, are labeled by objects $A,B,C,... \in \mc{M}$ and can be freely pulled through the lattice, locally changing between the different partition functions:

\begin{equation}
	\includegraphics[valign=c,page=14]{figures/kbtorusIntertwiner}.
\end{equation}

Since the product of an MPO symmetry and an MPO intertwiner is again an MPO intertwiner, we get the following relations:
\begin{align}
\includegraphics[page=24,valign=c]{figures/kbtorusIntertwiner} &= \sum_{B} N_{a A}^B \, \includegraphics[page=25,valign=c]{figures/kbtorusIntertwiner},\\
\includegraphics[page=26,valign=c]{figures/kbtorusIntertwiner} &= \sum_{B} N_{A \alpha}^B \, \includegraphics[page=27,valign=c]{figures/kbtorusIntertwiner},
\end{align}
with $N_{aA}^B$ and $N_{A\alpha}^\beta$ again non-negative integers. These two relations imply the existence of fusion tensors satisfying
\begin{equation}
\includegraphics[page=28,valign=c]{figures/kbtorusIntertwiner}, \quad \includegraphics[page=29,valign=c]{figures/kbtorusIntertwiner}.
\end{equation}

Let us now take a torus partition function of the non-diagonal model$_\mc{M}$, including a simple idempotent (with defects labeled by the objects in $\mc{C}$). A closed intertwiner loop $I \in \mc{M}$ can be freely created without changing the partition function (up to a quantum dimension factor) on the torus and pulled through around both directions, switching from model$_\mc{M}$ to model$_\mc{D}$ everywhere, except between the defects of model$_\mc{M}$. Finally, the tube in model$_\mc{M}$ together with the intertwiner can be mapped to a linear combination of tubes in model$_\mc{D}$, for which the defects can be labeled by category $\mathcal{D}$:

\begin{multline}
\vcenter{\hbox{\includegraphics[valign=c,page=18, scale=0.8]{figures/kbtorusIntertwiner}}} \rightarrow \vcenter{\hbox{\includegraphics[valign=c,page=19, scale=0.8]{figures/kbtorusIntertwiner}}} \rightarrow
\vcenter{\hbox{\includegraphics[valign=c,page=20, scale=0.8]{figures/kbtorusIntertwiner}}}.
\label{torusIntertwiner}
\end{multline}

A crucial step here is the action of the intertwiners on the defects in $\mc{C}$. For this, we use the identity:
\begin{align}
	\vcenter{\hbox{\includegraphics[page=12,scale=0.5]{figures/kbtorusIntertwiner}}} = \sum_{A,\gamma} \vcenter{\hbox{\includegraphics[page=13,scale=0.5]{figures/kbtorusIntertwiner}}},
	\label{fusing}
\end{align}
which allows us to map a tube $\mc{T}_{aba}^c$ in $\mc{C}$ to a tube $\mc{T}_{\alpha\beta\alpha}^\gamma$ in $\mc{D}$; this is the \emph{tube map} $T_{\alpha\beta\gamma;I}^{abc}$ described in \eqref{tubeMap}, and for an invertible bimodule category, this provides an isomorphism between the distinct torus partition functions. Using this map, we can write down the torus partition function with a tube of model$_\mc{M}$ in terms of the characters of model$_\mc{D}$. A particular case of this is when we take the empty model$_\mc{M}$ partition function; this partition function is modular invariant since the action of the modular transformation only affects the topological defects. By mapping this empty model$_\mc{M}$ partition function to a twisted model$_\mc{D}$ partition function, we obtain a modular invariant twisted model$_\mc{D}$ partition function. This generalizes to surfaces of arbitrary genus, and provides a way of constructing partition functions on some closed surface that are invariant under the mapping class group of that surface.\\

\subsection{Orbifold on the Klein bottle}

We now want to perform a similar mapping on the Klein bottle. It only makes sense to start from a Klein bottle with a symmetric simple idempotent (any other projection will lead to a zero Klein bottle partition function). We grow an intertwiner loop $I$ on the Klein bottle, presenting model$_\mc{M}$, in the presence of a symmetric simple idempotent:

\begin{multline}
\vcenter{\hbox{\includegraphics[valign=c,page=21, scale=0.8]{figures/kbtorusIntertwiner}}} \rightarrow \vcenter{\hbox{\includegraphics[valign=c,page=22, scale=0.8]{figures/kbtorusIntertwiner}}} \rightarrow
\vcenter{\hbox{\includegraphics[valign=c,page=23, scale=0.8]{figures/kbtorusIntertwiner}}}.
\end{multline}

In the last step, the reflection operator is commuted through the intertwiner loops, turning the model into model$_\mc{D}$ everywhere, except between the defects of model$_\mc{M}$, labeled by simple objects in category $\mathcal{C}$, and the intertwiner $I$. This last step is only possible if the generalized intertwiner tube

\begin{align}
		\sum_{A} \quad \vcenter{\hbox{\includegraphics[page=11,scale=0.8]{figures/kbtorusIntertwiner}}}
	\label{generalizedTube}
\end{align}

is reflection invariant. The proof of this is completely analogous to the proof that the symmetric simple idempotents $P^{a\bar{a}}_{c,i}$ (themselves consisting of sums of tube elements) are reflection invariant. \\

The calculation of the Klein bottle entropy is analogous to the diagonal partition function case. The Klein bottle entropy with a twist $c$ can be written as

\begin{align}
g_c = \frac{Z_{c,{\mathcal{M}}}^\text{KB}(2L_x,\frac{L_y}{2})}{Z^{\text{tor}}_{\mathcal{M}}(L_x,L_y)} \simeq \sum_{\alpha} \widetilde{M}^{c}_{\alpha\bar{\alpha}} S_{\alpha,\rho_{\text{min}}},
\end{align}

generalizing equation \eqref{KBE}, where the matrix $\widetilde{M}^c$ can be obtained from the map in \eqref{torusIntertwiner} and the $S$-matrix is the one for the diagonal theory for model$_\mc{D}$.

\subsection{Orbifold on the cylinder}
\label{Sec:CylinderOrbifold}

For the cylinder, a general discussion can be held as in section \ref{Sec:Orbifold}: model$_\mc{M}$ as an orbifold of model$_\mc{D}$. The Ishibashi states of model$_\mc{D}$ can be constructed by the application of the projectors on the topological superselection sectors on the reference state $\ket{\Sigma_{M}}$. In the end we have both Ishibashi states and Cardy states for model$_\mc{M}$ and model$_\mc{D}$ at our disposal and the central question is: how can cylinders of one model be mapped to cylinders of the other model using the lattice orbifold procedure that was used for the torus and Klein bottle case? \\

Let us start with a cylinder partition function of model$_\mc{M}$ ($Z^{\text{cyl}}_{\mathcal{M}}$). An approach very much in the spirit of \cite{affleck1998boundary} is to grow an intertwiner and end up with twisted cylinders of model$_\mc{D}$:

\begin{align}
Z^{\text{cyl}}_{\mathcal{M}} \propto \vcenter{\hbox{\includegraphics[page=10]{figures/cylinderSquare}}} = \vcenter{\hbox{\includegraphics[page=11]{figures/cylinderSquare}}} = \sum_{\gamma} N_{II}^{\gamma} \,\vcenter{\hbox{\includegraphics[page=12]{figures/cylinderSquare}}}.
\label{AtoBCylinder}
\end{align}

This mapping signals that the Ishibashi states in model$_\mc{M}$ are made from both untwisted and twisted Ishibashi states from model$_\mc{D}$, exactly what was found for the Potts model in \cite{affleck1998boundary}. In both pictures the Ishibashi and Cardy states of the final model can be recovered by application of the intertwiners on the states of the first model \cite{pearce1993intertwiners}. \\

The different boundary states in model$_\mc{M}$ are labeled by all the possible intertwiners $I$ that we can apply on the cylinder (simple objects in the category $\mc{M}$). Clearly, the discussion is completely analogous to the twisted case for diagonal models, where the ladder algebra was introduced. Here, a generalization of the ladder algebra pops up, containing intertwiners in the vertical direction (see \eqref{ladderDefinition}):

\begin{align}
\mc{L}_{II,II}^{\gamma} &= \includegraphics[valign=c,page=15]{figures/kbtorusIntertwiner}.
\label{laddersCylinder}
\end{align}

Only those twists $\gamma \in \mc{D}$ that are compatible with fusion at the two boundaries can be applied horizontally.\\

Let us now do the reverse and start with the known cylinder partition functions of model$_\mc{D}$ ($Z^{\text{cyl}}_{\mathcal{D}}$) and work towards the cylinder partition functions of model$_\mc{M}$. This approach is natural since we have already done the cylinder analysis of diagonal partition functions and we have assumed in the orbifold procedure that model$_\mc{D}$ is diagonal. If one starts from an empty cylinder of model$_\mc{D}$ and grows an intertwiner around the cylinder, the intertwiners in the middle can be fused to defects in model$_\mc{M}$, mapping the partition function to model$_\mc{M}$, possibly with horizontal twists:

\begin{align}
	Z^{\text{cyl}}_{\mathcal{D}} \propto \vcenter{\hbox{\includegraphics[page=7,scale=1]{figures/cylinderSquare}}} = \vcenter{\hbox{\includegraphics[page=8,scale=1]{figures/cylinderSquare}}} = \sum_{c} N_{II}^c\vcenter{\hbox{\includegraphics[page=9,scale=1]{figures/cylinderSquare}}}.	
\label{BtoACylinder}
\end{align}

This trick relies on the invertibility of the bimodule, going from step two to step three in the above formula. If one started from a cylinder (model$_\mc{D}$) with Ishibashi states $I^{\alpha}$ at the boundaries (producing the single character $\chi_{\alpha}$), one ends up with a cylinder of model$_\mc{M}$ with Ishibashi states containing multiple twists, because the intertwiners at the boundaries can only map Ishibashi states of model$_\mc{D}$ to those of model$_\mc{M}$. The same applies if one started from Cardy states at the boundaries, since the intertwiners can only map Cardy states from one model to the other. This result is consistent, since for the orbifold model (model$_\mc{M}$) it is known that not all cylinders with untwisted Ishibashi states lead to single characters of model$_\mc{D}$, but rather to sums of characters.

\subsection{Example: the three-state Potts model}
Let us now turn to the specific example of the three-state Potts model, used in this work to illustrate the orbifolding procedure on the lattice. It is the non-diagonal modular invariant of the $c = 4/5$ minimal model CFT, and is obtained from the diagonal tetracritical Ising (TCI) CFT by an orbifold. The modular fusion category $\mc{D}_\text{TCI}$ is obtained as the coset
\begin{equation}
\mc{D}_\text{TCI} = \frac{\text{su}(2)_3 \otimes \text{su}(2)_1}{\text{su}(2)_4},
\end{equation}

which has 10 simple objects corresponding to the primary fields of the CFT. These fields can be conveniently represented in the Kac table:

\begin{center}
	\begin{tabular}{c|c c c c c}
		\rule{0pt}{1.2em}
		$\widetilde{\it{3/2}}$ & \cellcolor{blue!25}$3$ & $13/8$ & \cellcolor{blue!25}$2/3$ & $1/8$ &\cellcolor{blue!25} $0$\\[0.1em]
		\rule{0pt}{1.2em}%
		$\widetilde{\it{1}}$ & $7/5$ & \cellcolor{blue!25}$21/40$ & $1/15$ & \cellcolor{blue!25}$1/40$ & $2/5$\\[0.1em]
		\rule{0pt}{1.2em}%
		$\widetilde{\it{1/2}}$ & \cellcolor{blue!25}$2/5$ & $1/40$ & \cellcolor{blue!25}$1/15$ & $21/40$ & \cellcolor{blue!25}$7/5$\\[0.1em]
		\rule{0pt}{1.2em}%
		$\tilde{\it{0}}$ & $0$ & $\cellcolor{blue!25}1/8$ & $2/3$ & \cellcolor{blue!25}$13/8$ & $3$\\[0.1em]
		\hline
		\rule{0pt}{1.2em}%
		&$\it{0}$&$\it{1/2}$&$\it{1}$&$\it{3/2}$&$\it{2}$
	\end{tabular}
	\label{KacTetra}
\end{center}

The bottom row represents the simple objects of $\text{su}(2)_4$ and the left column those of $\text{su}(2)_3$, which has the same fusion rules as $\text{Fib} \otimes \mathds{Z}_2$. If one identifies the repeated fields (colored in blue) in the table with the same conformal weight, one obtains the 10 fields of the tetracritical Ising model, which have the same fusion rules as $\text{su}(2)_4 \otimes \text{Fib}$. In contrast however, the Frobenius-Schur indicators of $\mc{D}_\text{TCI}$ are all positive while this is not the case for $\text{su}(2)_4$; we will therefore (formally) write $\mc{D}_\text{TCI} = |\text{su}(2)_4| \otimes \text{Fib}$. The tetracritical Ising model, which is obtained by choosing $\mc{D}_\text{TCI}$ as a module category over itself, has a diagonal modular invariant torus partition function:
\begin{align}
	Z_{\text{TCI}}^\text{tor} =  &|\chi_{0}(q)|^2 + |\chi_{1/8}(q)|^2 + |\chi_{2/3}(q)|^2 + |\chi_{13/8}(q)|^2 + |\chi_{3}(q)|^2 + \\
	&|\chi_{2/5}(q)|^2 + |\chi_{1/40}(q)|^2 + |\chi_{1/15}(q)|^2 + |\chi_{21/40}(q)|^2 + |\chi_{7/5}(q)|^2.
\end{align}
Starting from the fusion category $\mc{D}_\text{TCI}$, there is a second modular invariant partition function one can write down, which is known as the three-state Potts model, obtained by choosing a different module category $\mc{M}_\text{Potts}$ of $\mc{D}_\text{TCI}$. Due to the tensor product structure of $\mc{D}_\text{TCI}$, we can write $\mc{M}_\text{Potts} = \mc{M}_\text{TY} \otimes \text{Fib}$, where the $\text{Fib}$ factor is obtained by taking $\text{Fib}$ as a module category over itself. The other factor $\mc{M}_\text{TY}$ is a module category over $|\text{su}(2)_4|$ and has 4 simple objects denoted as $\{A,\sigma,B,C\}$, where the action of $|\text{su}(2)_4|$ on $\mc{M}_\text{TY}$ is given by
\begin{equation}
\begin{tabular}{|c|c c c c c|}
\hline
$\ract$ & $\it{0}$ & $\it{1/2}$&$\it{1}$&$\it{3/2}$&$\it{2}$\\
\hline
$A$ & $A$ & $\sigma$&$B+C$&$\sigma$&$A$\\
$\sigma$ & $\sigma$ & $A+B+C$&$2 \sigma$&$A+B+C$&$\sigma$\\
$B$ & $B$ & $\sigma$&$A+C$&$\sigma$&$B$\\
$C$ & $C$ & $\sigma$&$A+B$&$\sigma$&$C$\\
\hline
\end{tabular}
\end{equation}

The fusion category describing the topological defects is now given by $\mc{C}_\text{Potts}$, which is obtained as the unique fusion category that turns $\mc{M}_\text{Potts}$ into an invertible $(\mc{C}_\text{Potts},\mc{D}_\text{TCI})$-bimodule category. It again factorizes as $\mc{C}_\text{Potts} = \mc{C}_{S_3} \otimes \text{Fib}$, where $\mc{C}_{S_3}$ is a fusion category with 8 simple objects $\{A^+,\sigma^+,B^+,C^+,A^-,\sigma^-,B^-,C^-\}$. The topological defects labeled by $\{A^+,B^+,C^+,A^-,B^-,C^-\}$ are group-like, where $\{A^+,B^+,C^+\}$ and $\{A^+,A^-\}$ form a $\mathds{Z}_3$ and $\mathds{Z}_2$ subgroup, respectively, the semi-direct product of which yields $S_3$. The simple objects $\{\sigma^+,\sigma^-\}$ are dualities, with fusion rules
\begin{equation}
\sigma^\pm \otimes X^\pm = \sigma^+, \quad \sigma^\mp \otimes X^\pm = \sigma^-, \quad \sigma^\pm \otimes \sigma^\pm = \sum_{X} X^+, \quad \sigma^\pm \otimes \sigma^\mp = \sum_{X} X^-,
\end{equation}
where $X \in \{A,B,C\}$. From this we see that the superscripts $+$ and $-$ act as a $\mathds{Z}_2$ grading in the fusion rules; the action of $\mc{C}_{S_3}$ on $\mc{M}_\text{TY}$ is given in Appendix \ref{App:data}. We note that $\{A^+,\sigma,B^+,C^+\}$ form a subcategory of $\mc{C}_{S_3}$; this fusion category is known as the $\mathds{Z}_3$ Tambara-Yamagami (TY) category \cite{tambara1998tensor}.\\

The orbifold procedure for the three-state Potts model was extensively discussed in \cite{affleck1998boundary}. The three-state Potts partition function is obtained from the tetracritical Ising partition function by applying the projector, projecting on the even subspace corresponding to the $\mathds{Z}_2$ symmetry generated by the $O_{(\it{2},\mathbf{1})}$ topological defect, with a projector $P_{+} = \frac{1}{2}(O_{(\it{0},\mathbf{1})} + O_{(\it{2},\mathbf{1})}$, in both the horizontal and vertical direction of the torus. This projector acts trivially in the $\text{Fib}$ component of $\mc{C}_\text{Potts}$, as denoted by $\mathbf{1}$ ($\{\mathbf{1},\tau \} \in \text{Fib}$), since the orbifold only involves the $\text{Fib}$ component of $\mc{C}_\text{TCI} = \mc{D}_\text{TCI}$. The three-state Potts partition function therefore consists of two parts: the $\mathds{Z}_2$ even terms of the horizontally untwisted partition function and the $\mathds{Z}_2$ even terms of the $O_{(\it{2},\mathbf{1})}$-twisted partition function:

\begin{align}
	Z_{\text{Potts}}^\text{tor} = P_{+}Z_{\text{TCI},(\it{0},\mathbf{1})}^\text{tor} + \tilde{P}_{+}Z_{\text{TCI},(\it{2},\mathbf{1})}^\text{tor}.
	\label{PottsOrbif}
\end{align}

Note the tilde on the twisted part indicating that the projector also has a nontrivial $(\it{2},\mathbf{1})$-twist. This construction can indeed be recovered by starting with the empty Potts partition function and growing an intertwiner $I=(A,\mathbf{1})$:

\begin{align}
		Z_{\text{Potts}}^\text{tor} &=  \vcenter{\hbox{\includegraphics[page=2,scale=1]{figures/kbtorusIntertwiner}}} = \vcenter{\hbox{\includegraphics[page=3,scale=1]{figures/kbtorusIntertwiner}}}
	\label{torusOrbifoldProjection}.
\end{align}

The resulting partition function \ref{PottsOrbif} can be read off from the tetracritical idempotent tables \ref{TetraTable1} and \ref{TetraTable4} in appendix \ref{pottsData}), keeping only those linear combination of tube elements that are even under the $\mathds{Z}_2$ subgroup generated by $O_{(\it{2},\mathbf{1})}$:

\begin{align}
	P_{+}Z_{\text{TCI},\it{0}} &= |\chi_{0}(q)|^2+|\chi_{3}(q)|^2+|\chi_{2/5}(q)|^2+|\chi_{7/5}(q)|^2+|\chi_{2/3}(q)|^2 + |\chi_{1/15}(q)|^2, \\[1em]
	\tilde{P}_{+}Z_{\text{TCI},\it{2}} &= \chi_{0}(q)\overline{\chi}_{3}(\overline{q})+\chi_{3}(q)\overline{\chi}_{0}(\overline{q})+\chi_{2/5}(q)\overline{\chi}_{7/5}(\overline{q})+\chi_{7/5}(q)\overline{\chi}_{2/5}(\overline{q})+|\chi_{2/3}(q)|^2\nonumber\\
	&+|\chi_{1/15}(q)|^2.
\end{align}

At this point, we are ready to write down the twisted partition functions of the Potts model on the torus and Klein bottle, like we did for the Ising model and Yang-Lee model. In light of the specific lattice model we will study later, we will restrict ourselves to topological defects in $\mc{C}_\text{Potts}$ that are trivial in the $\text{Fib}$ component, but this is merely for the sake of brevity. The specific coefficients $T_{\alpha\beta\gamma;A}^{abc}$ in \ref{tubeMap} for the case $\alpha,\beta,\gamma \in |\text{su}(2)_4|$, $A \in \mc{M}_{TY}$ and $a,b,c \in \mc{C}_{S_3}$ are given in the attached file in the supplementary material.
 The coefficients $(P^{-1})^{\delta\overline{\kappa}}_{\alpha\beta\gamma\beta}$ can be found by splitting the rows of the $|\text{su}(2)_4|$ component of the tetracritical Ising idempotents (Tables \ref{TetraTable1}-\ref{TetraTable4} in appendix \ref{pottsData}) into its non-central components and inverting the resulting square matrix.\\

In this way the following eight twisted partition functions on the torus can be calculated:

\begin{align}
	\begin{split}
		Z_{(A^{+},\mathbf{1})}^\text{tor} &= Z_{\text{Potts}}^\text{tor},\\
		Z_{(B^{+},\mathbf{1})}^\text{tor} &= |\chi_{1/15}(q)|^2 + |\chi_{2/3}(q)|^2 + \chi_{2/3}(q)\overline{\chi}_{0}(\overline{q}) +
		\chi_{0}(q)\overline{\chi}_{2/3}(\overline{q}) + \chi_{7/5}(q)\overline{\chi}_{1/15}(\overline{q}) + \\ &\chi_{1/15}(q)\overline{\chi}_{7/5}(\overline{q}) + \chi_{2/3}(q)\overline{\chi}_{3}(\overline{q}) +
		\chi_{3}(q)\overline{\chi}_{2/3}(\overline{q}) + \chi_{1/15}(q)\overline{\chi}_{2/5}(\overline{q}) +
		\chi_{2/5}(q)\overline{\chi}_{1/15}(\overline{q}), \\
		Z_{(C^{+},\mathbf{1})}^\text{tor} &= Z_{(B^{+},\mathbf{1})}^\text{tor}, \\
		Z_{(\mathbb{\sigma^{+}},\mathbf{1})}^\text{tor} &= \chi_{0}(q)\overline{\chi}_{1/8}(\overline{q}) + \chi_{7/5}(q)\overline{\chi}_{21/40}(\overline{q}) +
		\chi_{3}(q)\overline{\chi}_{13/8}(\overline{q}) + \chi_{2/5}(q)\overline{\chi}_{1/40}(\overline{q}) + \\
		&\chi_{3}(q)\overline{\chi}_{1/8}(\overline{q}) + \chi_{2/5}(q)\overline{\chi}_{21/40}(\overline{q}) +
		\chi_{0}(q)\overline{\chi}_{13/8}(\overline{q}) + \chi_{7/5}(q)\overline{\chi}_{1/40}(\overline{q}) + \\
		& 2\chi_{2/3}(q)\overline{\chi}_{1/8}(\overline{q}) + 2\chi_{1/15}(q)\overline{\chi}_{21/40}(\overline{q}) +
		2\chi_{2/3}(q)\overline{\chi}_{13/8}(\overline{q}) + 2\chi_{1/15}(q)\overline{\chi}_{1/40}(\overline{q}), \\
		Z_{(\mathbb{\sigma^{-}},\mathbf{1})}^\text{tor} &= \overline{Z_{(\mathbb{\sigma^{+}},\mathbf{1})}^\text{tor}}, \\
		Z_{(A^{-},\mathbf{1})}^\text{tor} &= |\chi_{1/40}(q)|^2 + |\chi_{21/40}(q)|^2 +
		|\chi_{1/8}(q)|^2 + |\chi_{13/8}(q)|^2 +
		\chi_{1/40}(q)\overline{\chi}_{21/40}(\overline{q}) + \\ & \chi_{21/40}(q)\overline{\chi}_{1/40}(\overline{q}) + \chi_{1/8}(q)\overline{\chi}_{13/8}(\overline{q}) + \chi_{13/8}(q)\overline{\chi}_{1/8}(\overline{q}),\\
		Z_{(A^{-},\mathbf{1})}^\text{tor} &= Z_{(B^{-},\mathbf{1})}^\text{tor} = Z_{(C^{-},\mathbf{1})}^\text{tor}
	\end{split}
	\label{PottsTorus}
\end{align}

with
\begin{align}
	\begin{split}
		\chi_{0}(q) = &q^{-c/24}(1 + q^2 + q^3 + 2q^4 + 2q^5 + 4q^6  + ...)\\
		\chi_{2/5}(q) = &q^{2/5-c/24}(1 + q + q^2 + 2q^3 + 3q^4 + 4q^5 + 6q^6 + ...)\\
		\chi_{7/5}(q) = &q^{7/5-c/24}(1 + q + 2q^2 + 2q^3 + 4q^4 + 5q^5 + 8q^6 + ...)\\
		\chi_{2/3}(q) = &q^{2/3-c/24}(1 + q + 2q^2 + 2q^3 + 4q^4 + 5q^5 + 8q^6 + ...)\\
		\chi_{3}(q) = &q^{3-c/24}(1 + q + 2q^2 + 3q^3 + 4q^4 + 5q^5 + 8q^6 + ...)\\
		\chi_{1/15}(q) = &q^{1/15-c/24}(1 + q + 2q^2 + 3q^3 + 5q^4 + 7q^5 + 10q^6 + ...) \\
		\chi_{1/8}(q) = &q^{1/8-c/24}(1 + q + q^2 + 2q^3 + 3q^4 + 4q^5 + 6q^6 + ...) \\
		\chi_{13/8}(q) = &q^{13/8-c/24}(1 + q + 2q^2 + 3q^3 + 5q^4 + 6q^5 + 9q^6 + ...) \\
		\chi_{1/40}(q) = &q^{1/40-c/24}(1 + q + 2q^2 + 3q^3 + 4q^4 + 6q^5 + 9q^6 + ...) \\
		\chi_{21/40}(q) = &q^{21/40-c/24}(1 + q + 2q^2 + 3q^3 + 5q^4 + 7q^5 + 10q^6 + ...).
	\end{split}
	\label{TetraChars}
\end{align}

The above twisted partition functions are not new and can be found in \cite{petkova2001generalised} - where non-trivial Fib twists are also included, yielding a total of 16 twisted partition functions - but the bimodule category framework we have used here allows us to translate these results directly to the lattice. The strength of this approach is that it can deal with lattice partition functions with intersecting defects and using the relationship between the tubes of both representations \eqref{tubeMap}, one can write down general twisted partition functions as a sum of characters of the original model before the orbifold procedure (the tetracritical Ising model in this case).\\

The twisted Klein bottle partition functions are simply given by selecting the left-right symmetric terms in the twisted torus partition functions, just like in the case of the other models:

\begin{align}
	\begin{split}
		Z_{(A^{+},\mathbf{1})}^\text{KB} =  &\chi_{0}(q^2) + \chi_{3}(q^2) + \chi_{2/5}(q^2) + \chi_{7/5}(q^2) + 2\chi_{1/15}(q^2) + 2\chi_{2/3}(q^2) \\
		Z_{(B^{+},\mathbf{1})}^\text{KB} = &Z_{(C^{+},\mathbf{1})}^\text{KB} =  \chi_{1/15}(q^2) + \chi_{2/3}(q^2) \\
		Z_{(A^{-},\mathbf{1})}^\text{KB} = &Z_{(B^{-},\mathbf{1})}^\text{KB} =  Z_{(C^{-},\mathbf{1})}^\text{KB} = \chi_{1/40}(q^2) + \chi_{21/40}(q^2) +
		 \chi_{1/8}(q^2) + \chi_{13/8}(q^2).
	\end{split}
	\label{PottsKlein}
\end{align}

Note that the Klein bottles with duality twists ($(\sigma^{+},\mathbf{1})$ and $(\sigma^{-},\mathbf{1})$) are zero, just like in the Ising model, because they contain no left-right symmetric terms. \\

Let us now turn to the Klein bottle entropies for the Potts model. The $S$-matrix for the tetracritical Ising CFT is given by the tensor product of the Fibonacci $S$-matrix and the $|\text{su}(2)_4|$ $S$-matrix: $S_\text{TCI} = S_{|\text{su}(2)_4|} \otimes S_\text{Fib}$, with (see for example \cite{bonderson2007non}):

\begin{align}
	S_{\text{Fib}} = \frac{1}{D_\text{Fib}}
	\begin{pmatrix}
		1&\phi\\
		\phi&-1
	\end{pmatrix},
\end{align}

where $D_{\text{Fib}} = \sqrt{1+\phi^2}$ and $\phi = \frac{1}{2}(1+\sqrt{5})$, and

\begin{align}
	S_{|\text{su}(2)_4|} = \frac{1}{\sqrt{12}}
	\begin{pmatrix}
		1&\sqrt{3}&2&\sqrt{3}&1\\
		\sqrt{3}&\sqrt{3}&0&-\sqrt{3}&-\sqrt{3}\\
		2&0&-2&0&-2\\
		\sqrt{3}&-\sqrt{3}&0&\sqrt{3}&-\sqrt{3}\\
		1&-\sqrt{3}&2&-\sqrt{3}&1
	\end{pmatrix}.
\end{align}

The Klein bottle entropy for the Potts model with no idempotent projection and no non-trivial twist was given and numerically calculated in \cite{tang2017universal}:

\begin{align}
\begin{split}
g = \frac{Z^{\text{KB}}_{\text{Potts}}(2L_x,\frac{L_y}{2})}{Z^{\text{tor}}_{\text{Potts}}(L_x,L_y)} &=
g^{(\it{0},\bf{1})} + g^{(\it{0},\tau)} + g^{(\it{2},\bf{1})} + g^{(\it{2},\tau)}\\  &+ 2g^{(\it{1},\bf{1})} + 2g^{(\it{1},\tau)} \\
& = \sqrt{3+\frac{6}{\sqrt{5}}}.
\end{split}
\end{align}

The projected Klein bottle entropies with the projectors according to the Potts idempotents can be immediately read off from \eqref{PottsIdem}. One simply selects, in the above sum, the fields from \eqref{PottsIdem} that belong to a left-right symmetric term.

\subsubsection{The Potts model on the cylinder}

Finally, we discuss the cylinder partition functions for the Potts model. We have 10 untwisted tetracritical Ising Ishibashi and Cardy states at our disposal. Let's look at the mapping from an empty tetracritical Ising cylinder to a twisted Potts cylinder \eqref{BtoACylinder}:

\begin{align}
Z^{\text{cyl}}_{\text{TCI}} \propto \vcenter{\hbox{\includegraphics[page=18,scale=1]{figures/cylinderSquare}}} = \vcenter{\hbox{\includegraphics[page=19,scale=1]{figures/cylinderSquare}}} = \sum_{c} N_{II}^c\vcenter{\hbox{\includegraphics[page=20,scale=1]{figures/cylinderSquare}}}.
\label{cylinderMapPottsTCI}
\end{align}

We can choose $I=A$ (the identity intertwiner). The allowed horizontal twists in the resulting Potts cylinder are $c=A^+$ and $c=A^-$, since they are the only allowed fusion channels in $I \otimes \overline{I}$. Different horizontal twists are allowed if different intertwiners are chosen. For example, if one chooses $I=\sigma$, all six $S_3$ ladder elements ($\mc{\tilde{L}}^{A^+}$, $\mc{\tilde{L}}^{B^+}$, $\mc{\tilde{L}}^{C^+}$, $\mc{\tilde{L}}^{A^-}$, $\mc{\tilde{L}}^{B^-}$, $\mc{\tilde{L}}^{C^-}$, in the simplified notation $\mc{\tilde{L}}^{a} = \mc{L}_{\sigma\sigma,\sigma\sigma}^{a}$) are allowed in \eqref{laddersCylinder}. This ladder algebra can be diagonalized to form four simple idempotents (consistent with the irreps of $S_3$). \\

One may wonder what tetracritical Ising cylinder one should start from to obtain an empty (untwisted) Potts cylinder. It is exactly the tetracritical Ising cylinder, projected on the orbifold subspace in both directions, just like on the torus (\ref{torusOrbifoldProjection}).\\

\begin{figure}[ht]
	\centering
	\includegraphics[page=13]{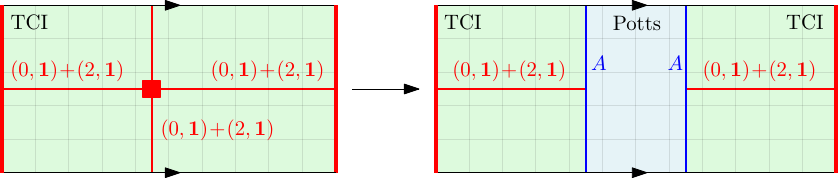}%
	\caption{Projecting the tetracritical Ising partition function to the orbifold subspace in both directions, one ends up with an empty Potts cylinder. }
	\label{ishibashiTetra}
\end{figure}

Going to the second mapping \eqref{AtoBCylinder}, the horizontal twists in the resulting tetracritical Ising cylinder (having started from an empty Potts cylinder) are $\gamma = (\it{0},\mathbf{1})$ and $\gamma = (\it{2},\mathbf{1})$. This is consistent with \cite{affleck1998boundary}, where it was shown that the Potts Ishibashi states are made from both the untwisted sector and the $(\it{2},\mathbf{1})$-twisted sector in the tetracritical Ising model. Looking at the last column of Table \ref{TetraTable1} (appendix \ref{pottsData}), one can see that there is only one symmetric idempotent in a $(\it{2},\mathbf{1})$-twisted tube, namely $P^{\it{1},\overline{\it{1}}}$, which contains the two twisted Ishibashi states $\ket{I_{1/15}^{(\it{2},\mathbf{1})}}$ and $\ket{I_{2/3}^{(\it{2},\mathbf{1})}}$ (this can be read off from the Kac table). Therefore, in order to construct the Potts Ishibashi states one can use these two twisted states on top of the six untwisted Ishibashi states ($\ket{I_{0}}$, $\ket{I_{3}}$, $\ket{I_{2/5}}$, $\ket{I_{7/5}}$, $\ket{I_{1/15}}$, $\ket{I_{2/3}}$), making a total of eight Potts Ishibashi states. Stated alternatively: in order to get an untwisted Potts partition function one needs both untwisted and $(\it{2},\mathbf{1})$-twisted tetracritical Ising Ishibashi states at the boundaries (see figure \ref{ishibashiTetra}).\\

We now turn towards the Potts cylinders as in figure \ref{ishibashiTetra} and evaluate the action of all the four possible intertwiners ($A$,$B$,$C$ and $\sigma$) at the boundaries. The eight distinct Potts Cardy states are known and related to each other by fusion according to the tensor product of the Tambara-Yamagami (TY) category and the Fibonacci algebra. The choice of the intertwiner determines the Potts state that will be recovered at the boundary.\\

Seven states are labeled by $\ket{A,\bf{1}}$, $\ket{B,\bf{1}}$, $\ket{C,\bf{1}}$, $\ket{\sigma,\bf{1}}$, $\ket{A,\tau}$, $\ket{B,\tau}$, $\ket{C,\tau}$ and were found by Cardy in \cite{cardy1989boundary}. The last state $\ket{\sigma,\tau}$ was found in \cite{affleck1998boundary} and coined the ``New'' state. It is special in the sense that its lattice representation in the 3-state Potts model is no longer a product state, as was the case for all Cardy states of the previous models, but a bond dimension 2 MPS (see section \ref{Sec:results} and for example \cite{verniernot}).\\

Just like in the case of the Potts Ishibashi states, the Potts Cardy states are constructed not only from the untwisted tetracritical Ising states, but also from the $(\it{2},\mathbf{1})$-twisted states. The mapping can be derived from \eqref{AtoBCylinder} (the superscript indicates the horizontal twist):

\begin{align}
	\begin{split}
		&\ket{A,\bf{1}} \rightarrow \ket{(\it{0},\bf{1})^{(\it{0},\bf{1})}} + \ket{(\it{2},\bf{1})^{(\it{0},\bf{1})}} \\
		&\ket{B,\bf{1}} \rightarrow \ket{(\it{1},\bf{1})^{(\it{0},\bf{1})}} - \ket{(\it{1},\bf{1})^{(\it{2},\bf{1})}}	\\
		&\ket{C,\bf{1}} \rightarrow \ket{(\it{1},\bf{1})^{(\it{0},\bf{1})}} + \ket{(\it{1},\bf{1})^{(\it{2},\bf{1})}}	\\
		&\ket{\sigma,\bf{1}} \rightarrow \ket{(\it{1/2},\bf{1})^{(\it{0},\bf{1})}} + \ket{(\it{3/2},\bf{1})^{(\it{0},\bf{1})}}	\\
		&\ket{A,\tau} \rightarrow \ket{(\it{0},\tau)^{(\it{0},\bf{1})}} + \ket{(\it{2},\tau)^{(\it{0},\bf{1})}}\\
		&\ket{B,\tau} \rightarrow \ket{(\it{1},\tau)^{(\it{0},\bf{1})}} - \ket{(\it{1},\tau)^{(\it{2},\bf{1})}}	\\
		&\ket{C,\tau} \rightarrow \ket{(\it{1},\tau)^{(\it{0},\bf{1})}} + \ket{(\it{1},\tau)^{(\it{2},\bf{1})}}	\\
		&\ket{\sigma,\tau} \rightarrow \ket{(\it{1/2},\tau)^{(\it{0},\bf{1})}} + \ket{(\it{3/2},\tau)^{(\it{0},\bf{1})}}.	
	\end{split}
	\label{boundarymap}
\end{align}

The states $\ket{B,\bf{1}}$ and $\ket{C,\bf{1}}$, just like $\ket{B,\tau}$ and $\ket{C,\tau}$, are distinguished by the sign of the term with a horizontal $(\it{2},\mathbf{1})$-twist.
Using this map and the known partition functions of the tetracritical Ising cylinder with a trivial horizontal defect and a $(\it{2},\mathbf{1})$-defect, we can evaluate 12 distinct Potts cylinder partition functions in terms of the tetracritical Ising characters:

\begin{align}
	Z_{(A,\textbf{1}),(A,\bf{1})}^{\text{cyl}} &= \chi_{0}(q) + \chi_{3}(q) \nonumber\\
	Z_{(A,\textbf{1}),(B,\bf{1})}^{\text{cyl}} &= \chi_{2/3}(q) \nonumber\\
	Z_{(A,\textbf{1}),(C,\tau)}^{\text{cyl}} &= \chi_{1/15}(q) \nonumber\\
	Z_{(A,\textbf{1}),(A,\tau)}^{\text{cyl}} &= \chi_{2/5}(q) + \chi_{7/5}(q) \nonumber\\
	Z_{(A,\bf{1}),(\sigma,\bf{1})}^{\text{cyl}} &= \chi_{13/8}(q) + \chi_{1/8}(q) \nonumber\\
	Z_{(A,\textbf{1}),(A,\tau)}^{\text{cyl}} &= \chi_{1/40}(q) + \chi_{21/40}(q) \nonumber\\
	Z_{(C,\tau),(C,\tau)}^{\text{cyl}} &= \chi_{0}(q) + \chi_{2/5}(q) +  \chi_{7/5}(q) + \chi_{3}(q) \label{PottsCylinderCardy} \\
	Z_{(C,\tau),(B,\tau)}^{\text{cyl}} &= \chi_{1/15}(q) + \chi_{2/3}(q) \nonumber\\
	Z_{(C,\tau),(\sigma,\tau)}^{\text{cyl}} &= \chi_{1/40}(q) + \chi_{21/40}(q) +  \chi_{13/8}(q) + \chi_{1/8}(q) \nonumber\\
	Z_{(\sigma,\bf{1}),(\sigma,\bf{1})}^{\text{cyl}} &= \chi_{0}(q) + \chi_{3}(q) +  2\chi_{2/3}(q) \nonumber\\
	Z_{(\sigma,\bf{1}),(\sigma,\tau)}^{\text{cyl}} &= \chi_{2/5}(q) + \chi_{7/5}(q) +  2\chi_{1/15}(q) \nonumber\\
	Z_{(\sigma,\tau),(\sigma,\tau)}^{\text{cyl}} &= \chi_{0}(q) + \chi_{2/5}(q) +  \chi_{7/5}(q) + \chi_{3}(q) + 2\chi_{1/15}(q) + 2\chi_{2/3}(q),\nonumber
\end{align}

recovering the known partition functions found in \cite{behrend1998integrable}. The unlisted partition functions can similarly be obtained from (\ref{boundarymap}). Alternatively, one can easily show for example that $Z_{(A,\textbf{1}),(B,\bf{1})}^{\text{cyl}} = Z_{(A,\textbf{1}),(C,\bf{1})}^{\text{cyl}}$, by applying a vertical twist $A^{-} \in \mc{C}_{S_3}$ on the cylinder (this can be done freely since the left boundary ($\bra{A,\textbf{1}}$) is invariant under this action) and fusing it with the right boundary ($\ket{B,\bf{1}}$), changing it to $\ket{C,\bf{1}}$ (see Table \ref{Table:N_CMM}). Similar equalities between the other partition functions can be shown in the same way.

\section{Strange correlators and the minimal models}
\label{Sec:results}

In this section we present examples of three classical two-dimensional statistical mechanics models to illustrate the torus, Klein bottle and cylinder partition functions discussed in the previous sections. These partition functions are obtained as a \emph{strange correlator}, i.e., as the overlap of a PEPS description of a string-net ground state $\ket{\psi_\text{SN}}$ given in \eqref{peps} with a suitable product state $\bra{\Omega}$:
\begin{equation}
\ket{\psi_\text{SN}} = \includegraphics[valign=c,page=16]{figures/transferMatrices},\quad \braket{\Omega|\psi_\text{SN}} =  \includegraphics[valign=c,page=17]{figures/transferMatrices}.
\end{equation}
In this way, the MPO symmetries and intertwiners of the classical statistical mechanics model are inherited from the PEPS description of string-net models \cite{lootens2020matrix}. The specific choice of $\bra{\Omega}$ then determines the local details of the statistical mechanics model; paraphrasing \cite{fendley2021integrability}, it ``adds geometry to topology". The choice of $\bra{\Omega}$, both in the bulk and in the presence of boundaries has been determined for integrable lattice models by solving the Yang-Baxter equation, and we will use these results here. After the strange correlator map, the following color convention for the PEPS tensors, MPO symmetries, fusion tensors (\ref{fusion}) and MPO intertwiners will be adopted in the remainder of the text:

\begin{align*}
	\includegraphics[scale=0.7,valign=c,page=12]{figures/transferMatrices}\enspace, \quad \includegraphics[scale=0.7,valign=c,page=13]{figures/transferMatrices}\enspace, \quad \includegraphics[scale=0.7,valign=c,page=14]{figures/transferMatrices}\enspace, \quad \includegraphics[scale=0.7,valign=c,page=15]{figures/transferMatrices}\enspace,
\end{align*}
where again we use green/blue for the transfer matrix tensors to indicate a diagonal/non-diagonal model.\\

Exact diagonalization is performed on the transfer matrices of these models to obtain the torus, Klein bottle and cylinder partition functions with their character decomposition as well as the Klein bottle entropies. The 2d critical classical Ising model is given as the simplest example of a unitary CFT. The second example is the non-unitary Yang-Lee model (as a non-unitary version of the critical hard hexagon model on a honeycomb lattice \cite{baxter1980hard, lootens2020galois}). This model is presented as an illustration of the non-unitary single character projection. These first two cases are straightforward, in the sense that their CFT partition functions are diagonal in the characters. \\

The third example is the three-state Potts model and it is more subtle since the CFT partition function on the torus is non-diagonal and as made clear in section \ref{Sec:Orbifold}, it requires the full orbifolding procedure on the lattice. The three-state Potts model also illustrates the fact that on the lattice, not necessarily all topological defects of the continuum CFT are present at finite size \cite{belletete2020topological}, and owing to this we can not fully decompose the partition functions into single terms and end up with sums of characters instead.\\

\subsection{The Ising model}

The strange correlator construction for the Ising model is explained in \cite{vanhove2018mapping}, together with exact diagonalization for the torus partition function with different topological defects and with a consistent topological superselection sector (idempotent) labeling. The relevant fusion category $\mc{D}$ here is the Ising category, with three simple objects satisfying
\begin{equation}
\sigma \otimes \sigma = 1 + \psi, \quad \psi \otimes \sigma = \sigma, \quad \psi \otimes \psi = 1.
\end{equation}
This fusion category has a $\mathds{Z}_2$-grading on the objects $\{\bf{1},\psi\}\oplus\{\sigma\}$ and nontrivial $F$-symbols

\begin{align*}
	[F^{\sigma\sigma\sigma}_{\sigma}]_{ij}=\frac{1}{\sqrt{2}}\begin{pmatrix}
		1 & 1\\
		1 & -1
	\end{pmatrix},~ [F^{\sigma\psi\sigma}_{\psi}]_{\sigma}^{\sigma}=[F^{\psi\sigma\psi}_{\sigma}]_{\sigma}^{\sigma}=-1.
\end{align*}

The lattice on which the model is defined is the -colloquially called- `bathroom' lattice. The model is effectively the Ising RSOS model, which has been extensively discussed \cite{bazhanov1989critical, klumper1992conformal}, however using the bathroom lattice makes the interpretation of the Ising RSOS model as a doubled version of the normal Ising partition function ($Z=\sum_{\braket{ij}} \exp \left(-\beta \sigma_{i}\sigma_{j}\right)$) containing both the primal and dual lattices geometrically clear. The transfer matrix can be interpreted as a direct sum of a primal matrix and a dual matrix, shifted over ``half" a lattice site. The choice of strange correlator can be diagrammatically shown (constructed from the PEPS tensors \eqref{peps} with $\mc{C}= \mc{M} = \mc{D}=\text{Ising}$):

\begin{align}
	\vcenter{\hbox{\includegraphics[page = 1]{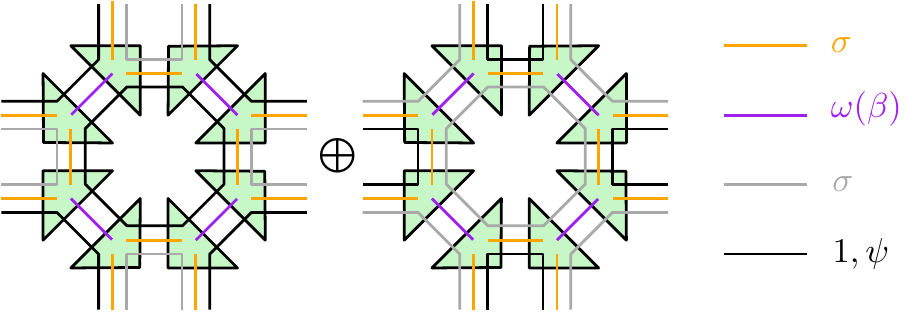}}}
	\label{eq:isingdirectsum}
\end{align}

where the purple diagonal physical legs are fixed on the product state $\bra{\omega(\beta)}= \sqrt{2}(\cosh(\beta)\bra{1} + \sinh(\beta)\bra{\psi})$, with $\beta = \beta_c=\log(1+\sqrt{2})/2$, such that the Kramers-Wannier duality (or $\sigma$-MPO) converts the primal lattice into the dual lattice and vice versa, consistent with the high-low temperature duality of the critical Ising model. A detailed discussion of this construction can further be found in \cite{bal2018real}.

\subsubsection{Torus results}

\begin{figure}[t]
	\centering
	\includegraphics[height=17em]{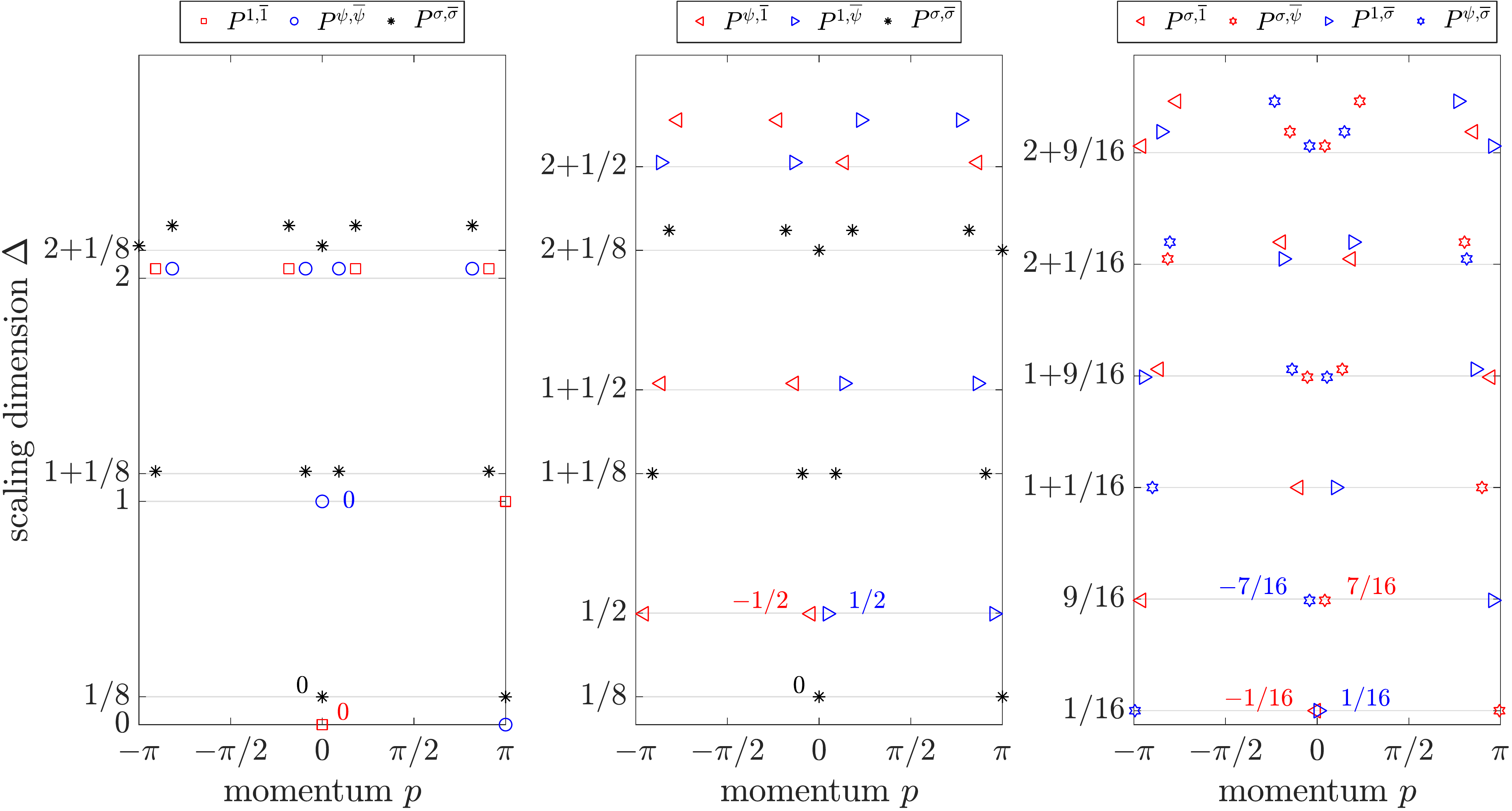}
	\caption{Topological sector labeling of the Ising model of finite-size CFT spectra (scaling dimension $\Delta$ versus momentum) on a cylinder with a circumference of 20 sites with no defect line on the left, a $\psi$ defect in the middle and a $\sigma$ defect on the right. The exact topological correction to the conformal spin is denoted next to the first appearance of the respective idempotents. This figure captures the finite size spectra consistent with equations \ref{isingTorus}.}
	\label{fig:isingspectra}
\end{figure}

The periodic transfer matrix spectra were obtained by a momentum labeling corresponding to the eigenvalues of the ``half" shift matrix, switching between primal and dual transfer matrices. The resulting torus spectra for the three possible twists are shown in figure \ref{fig:isingspectra}. The indicated topological superselection sectors (tube algebra idempotents) can be found in Table \ref{IsingTable} in appendix \ref{App:data}.\\

Two towers emerge in this way at momentum $0$ and $\pi$, both containing the single CFT towers $\bf{1}$, $\psi$ and $\sigma$, which can be identified by the idempotent labeling, consistent with the Kac table of the Ising RSOS model:

\begin{center}
	\begin{tabular}{c|c c c}
		\rule{0pt}{1.2em}%
		$\widetilde{\it{1/2}}$ & \cellcolor{blue!25}$1/2$ & $1/16$ & \cellcolor{blue!25}$0$ \\[0.1em]
		\rule{0pt}{1.2em}%
		$\tilde{\it{0}}$ & $0$ & \cellcolor{blue!25}$1/16$ & $1/2$ \\[0.1em]
		\hline
		\rule{0pt}{1.2em}%
		&$\it{0}$&$\it{1/2}$&$\it{1}$
	\end{tabular}
\end{center}

In the continuum, the repeated fields in this table are identified, but this is not the case at finite size. Instead, there is an additional $\mathds{Z}_2$ operator corresponding to the $(\it{1},\widetilde{\it{1/2}})$ field in this table. On the lattice, this operator is implemented as a half-shift, switching between primal and dual lattices, and is responsible for the fact that the characters in the spectrum are doubled. This half-shift operator can be implemented as an MPO symmetry, but it does not square to the identity and the size of the associated MPO algebra depends on the system size. This means we can not find its associated idempotents, but we can use the half-shift operator to separate the different characters based on their momentum, as shown in Figure \ref{fig:isingspectra}.\\

\begin{figure}[t]
	\centering
	\includegraphics[page=3]{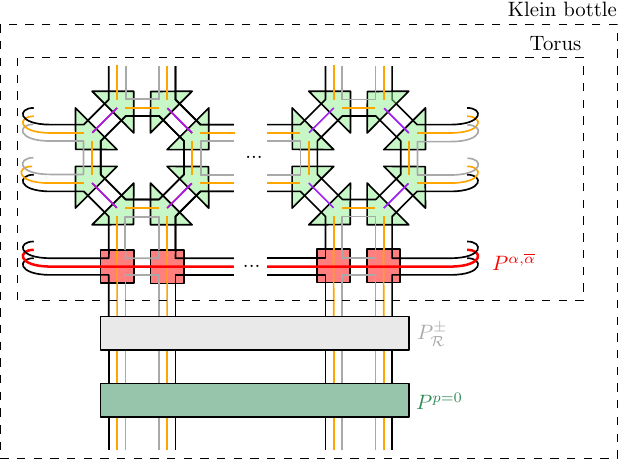}
	\caption{The diagonalization procedure on the torus and Klein bottle for the primal lattice of the Ising model. The transfer matrix on the cylinder is projected on a topological sector (idempotent) $P^{\alpha,\overline{\alpha}}$. On the Klein bottle, the spectrum is projected on the momentum $0$ subspace ($P^{p=0}$) and labeled by the reflection quantum numbers ($P^{\pm}_{\mathcal{R}}$).}
	\label{fig:torusKleinIsing}
\end{figure}

\subsubsection{Klein bottle results}

\begin{figure}[h]
	\centering
	\includegraphics[height = 17em,keepaspectratio=true]{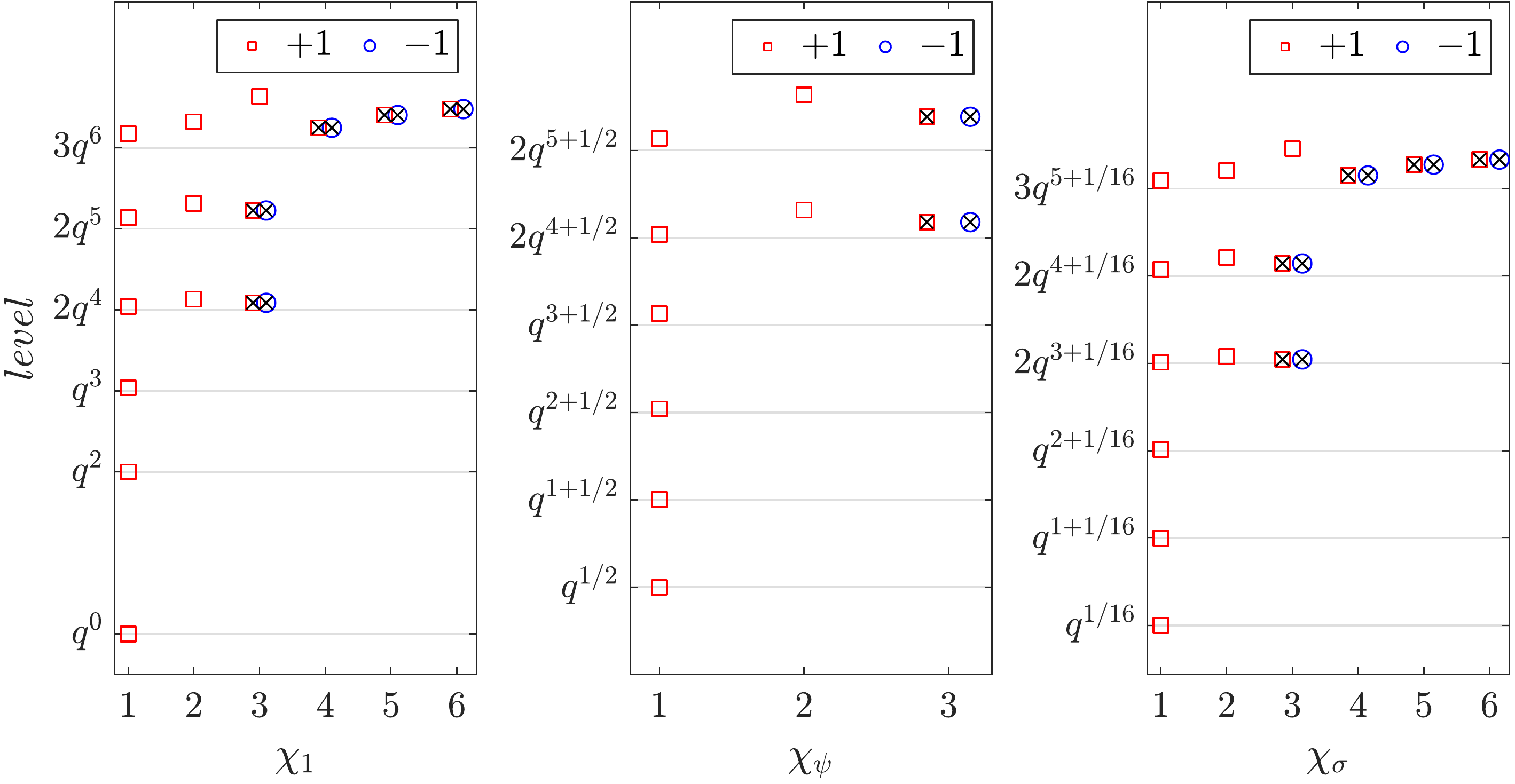}
	\caption{The three characters for the Ising model obtained numerically on the Klein bottle with length $L=22$. The explicit cancellation is shown at every level between exactly degenerate eigenvalues with positive and negative quantum number under reflection.}
	\label{fig:isingspectraKlein}
\end{figure}

The diagonalization procedure on the Klein bottle requires an extra projection on the different sectors under reflection and a projection on momentum $0$. The difference between the torus and Klein bottle diagonalization is schematically shown in figure \ref{fig:torusKleinIsing}. The result for the Ising model on the Klein bottle, as explained above, is shown in figure \ref{fig:isingspectraKlein}. Up to finite-size effects, the degeneracies of the three Ising characters (\ref{IsingCharacters}) are recovered. \\

Next, we look at the result for the Klein bottle entropy for the Ising model (see \eqref{KBentropy} and \eqref{KBentropyFixedPoint}). As is clear from the previous discussion and figure \ref{fig:isingspectra}, we need momentum information in order to separate the towers belonging to the superselection sectors $P^{\bf{1}\overline{\bf{1}}}$ and $P^{\psi\overline{\psi}}$. Without this momentum information, the rows of the Kac table are indistinguishable. On the Klein bottle, the rows could be separated by the projection onto momentum 0, for the Klein bottle entropy calculations however, this is no longer possible, as it is impossible to represent momentum projectors as a finite bond dimension MPO in the thermodynamic limit. Therefore, as we project onto the trivial idempotent $P^{\bf{1}\overline{\bf{1}}}$, we will obtain the sum of the first two elements of the first row of the $S$-matrix. The three symmetric idempotents in Table \ref{IsingTable} in appendix \ref{App:data} reduce to the two simple $\mathds{Z}_2$ irreducible representation under the spin flip operator. The result of the calculation is shown in Table \ref{tab:KleinBottleEntropyIsing}. \\

\begin{center}
	\begin{table}[ht]
		\centering
		\resizebox{0.5\textwidth}{!}{
			\begin{tabular}{ |c|c|c| }
				\hline
				$L_y$&$S(\bf{1},\bf{1})$ + $S(\bf{1},\psi)$&$S(\bf{1},\sigma)$\\
				\hline
				$10$&$1.0000000000$&$0.707001643$\\
				$12$&$1.0000000000$&$0.707088743$\\
				$14$&$1.0000000000$&$0.707103686$\\
				$16$&$1.0000000000$&$0.707106250$\\
				$18$&$1.0000000000$&$0.707106690$\\
				$20$&$1.0000000000$&$0.707106765$\\
				$22$&$1.0000000000$&$0.707106778$\\
				$24$&$1.0000000000$&$0.707106780$\\
				\hline
				Exact&$1$&$\frac{\sqrt{2}}{2}=0.707106781$\\
				\hline
		\end{tabular}}
		\caption{The Ising Klein bottle entropy $g$ in function of system size $L_y$ with the projection onto the $\mathds{Z}_2$ spin flip sectors: $P^{\bf{1}\overline{\bf{1}}} + P^{\psi\overline{\psi}}$ (second column) and $P^{\sigma\overline{\sigma}}$ (third column), calculated according to \eqref{KBentropyFixedPoint}.}
		\label{tab:KleinBottleEntropyIsing}
	\end{table}
\end{center}

\subsubsection{Cylinder results}

Let us turn to the conformal boundaries on the cylinder for the Ising model. The three (untwisted) Cardy states on the lattice were first given by Cardy \cite{cardy1984conformal}. On a lattice of $N$ sites, these are defined as: $\ket{\bf{1}} = \ket{\uparrow}^{\otimes N}$, $\ket{\psi} = \ket{\downarrow}^{\otimes N}$ and $\ket{\sigma} = \ket{+}^{\otimes N}$ called the up, down and free boundary conditions, respectively. In the RSOS (strange correlator) model, the $\sigma$-fixing of the physical indices leads to a staggering at the boundaries, where not all loops can be fixed on the vacuum. The vacuum state is instead given by fixing every other loop to $\mathbf{1}$ alternated with loops fixed to $\sigma$. Because the RSOS model is the direct sum of a primal $(p)$ and dual $(d)$ model, we end up with the following six Cardy states:

\begin{align}
\begin{split}
\ket{\mathbf{1}_p}\!\rangle &= \ket{\mathbf{1} \sigma \mathbf{1} \sigma \mathbf{1} \sigma ...},\\
\ket{\mathbf{1}_d}\!\rangle &= \ket{\sigma \mathbf{1} \sigma \mathbf{1} \sigma \mathbf{1} ...},\\
\ket{\psi_p}\!\rangle &= O_{\psi}\ket{\mathbf{1}_p}\!\rangle = \ket{\psi \sigma \psi \sigma \psi \sigma ...},\\
\ket{\psi_d}\!\rangle &= O_{\psi}\ket{\mathbf{1}_d}\!\rangle = \ket{\sigma \psi \sigma \psi \sigma \psi ...},\\
\ket{\sigma_p}\!\rangle &= O_{\sigma}\ket{\mathbf{1}_d}\!\rangle = \ket{(\mathbf{1}\!+\!\psi) \sigma (\mathbf{1}\!+\!\psi) \sigma (\mathbf{1}\!+\!\psi) \sigma ...},\\
\ket{\sigma_d}\!\rangle &= O_{\sigma}\ket{\mathbf{1}_p}\!\rangle = \ket{\sigma (\mathbf{1}\!+\!\psi) \sigma (\mathbf{1}\!+\!\psi) \sigma (\mathbf{1}\!+\!\psi) ...}.
\end{split}
\end{align}

One can map between the two vacuum states $\ket{\mathbf{1}_p}\!\rangle$ and $\ket{\mathbf{1}_d}\!\rangle$ by acting with the half-site shift operator (see also \cite{behrend1998integrable}); as mentioned above however, this is not a topological defect. \\

For the $\psi$-twisted case, we have to insert a fusion tensor in the Cardy states $\ket{\sigma_p}$ and $\ket{\sigma_d}$ to obtain the twisted Cardy states:

\begin{align}
\ket{\sigma^{\psi}_p} \rightarrow \vcenter{\hbox{\includegraphics[page=1,scale=1]{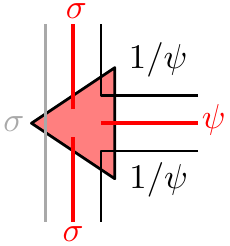}}} \quad \ket{\sigma^{\psi}_d} \rightarrow \vcenter{\hbox{\includegraphics[page=2,scale=1]{figures/cardyStates}}}
\label{fusionBoundary}
\end{align}

\begin{figure}[t]
	\centering
	\includegraphics[page=4,scale = 1]{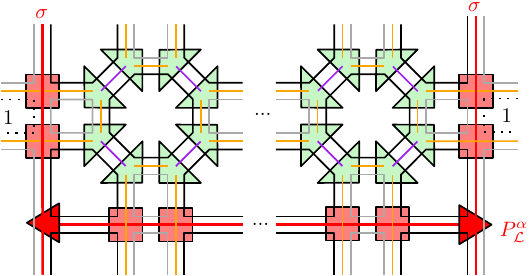}
	\caption{The diagonalization procedure on the cylinder with the MPO$_{\sigma}$ applied at both boundaries on the vacuum state. The transfer matrix on the cylinder is projected on the idempotents of the ladder algebra ($P^{\alpha}_{\mathcal{L}}$, $\alpha=\bf{1},\psi$), which in this case is just $\mathds{Z}_2$.}
	\label{fig:cylinderIsing}
\end{figure}

\begin{figure}[t]
	\centering
	\includegraphics[height=17em,keepaspectratio=true]{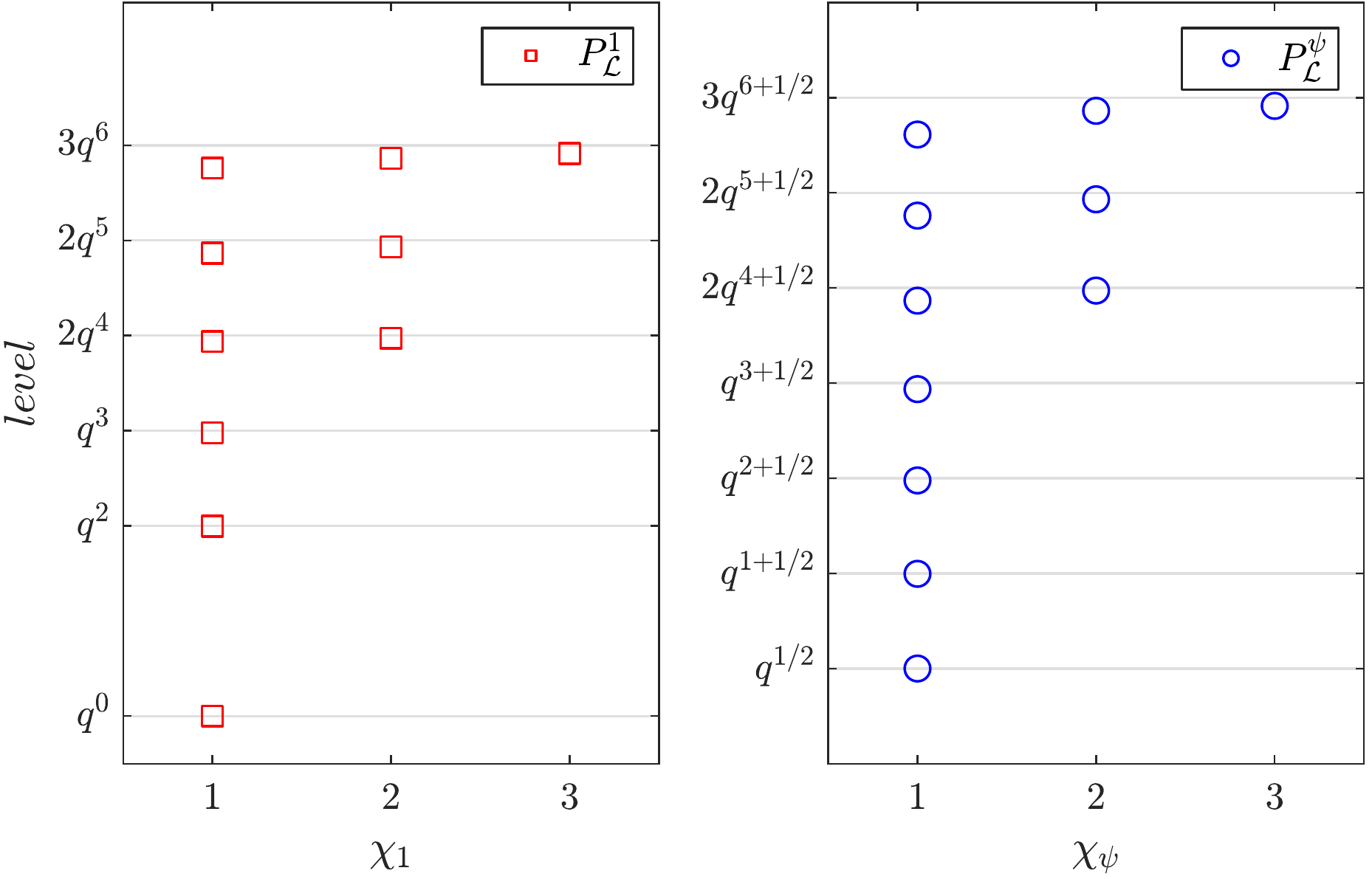}
	\caption{The two characters $\chi_{\bf{1}}$ and $\chi_{\psi}$ for the Ising model obtained numerically on the cylinder with length $L=22$ and the free boundary condition $\ket{\sigma_d}$ at both ends (figure \ref{fig:cylinderIsing}). The projection on the single characters is obtained through the idempotents of the ladder algebra with elements ($\mc{L}_{\sigma\sigma,\sigma\sigma}^{\bf{1}}$ and $\mc{L}_{\sigma\sigma,\sigma\sigma}^{\psi}$).}
	\label{fig:isingCylinderCharacters}
\end{figure}

Numerics were performed on the cylinder with boundary states $\ket{\sigma_p}$ at both ends. The diagonalization procedure is schematically shown in figure \ref{fig:cylinderIsing}. The results are presented in figure \ref{fig:isingCylinderCharacters}. The spectrum consists of the characters $\chi_{\bf{1}}$ and $\chi_{\psi}$. The ladder algebra, consisting of $\mc{L}_{\sigma\sigma,\sigma\sigma}^{\bf{1}}$ and $\mc{L}_{\sigma\sigma,\sigma\sigma}^{\psi}$ in this case, is isomorphic to $\mathds{Z}_2$ with the corresponding irreps projecting on the two characters in the spectrum.

\subsection{The non-unitary Yang-Lee model}

\subsubsection{Torus results}

The non-unitary Yang-Lee model is, besides its non-unitarity, quite simple. It contains only two primary states $\mathbf{1}$ and $\tau$ with conformal weights $0$ and $-1/5$. The lattice model we will consider is defined on a hexagonal lattice and can be considered the non-unitary version of the hard hexagon model. It is obtained from a strange correlator of the non-unitary Fibonacci string-net \cite{lootens2020galois}, which has two simple objects $\{\bf{1},\tau\}$, nontrivial fusion rule ${\tau \times \tau = \bf{1} + \tau}$ and nontrivial $F$-symbols

\begin{align}
[F^{\tau\tau\tau}_{\tau}]_{ij}=\frac{1}{\phi}\begin{pmatrix}
1 & \sqrt{\phi}\\
\sqrt{\phi} & -1
\end{pmatrix}
\end{align}

with $\phi = \frac{1}{2}(1-\sqrt{5})$ the negative solution of the golden ratio equation $\phi^2=1+\phi$.
The non-unitary hard-hexagon model is obtained by projecting all physical degrees of freedom onto the $\tau$-label, yielding a partition function constructed from the tensors
\begin{align}
\vcenter{\hbox{\includegraphics[page=5,scale=0.7]{figures/transferMatrices.pdf}}} = \phi^{1/2} \quad \text{and} \quad
\vcenter{\hbox{\includegraphics[page=6,scale=0.7]{figures/transferMatrices.pdf}}} = -(\phi)^{-1/2},\label{eq:hardhex}
\end{align}

and all rotations of the first tensor. The topological properties from a strange correlator of the non-unitary Fibonacci string-net were discussed in \cite{lootens2020galois}, together with the idempotent labeling of the torus spectrum.\\

\begin{figure}
	\center
	\includegraphics[height=17em]{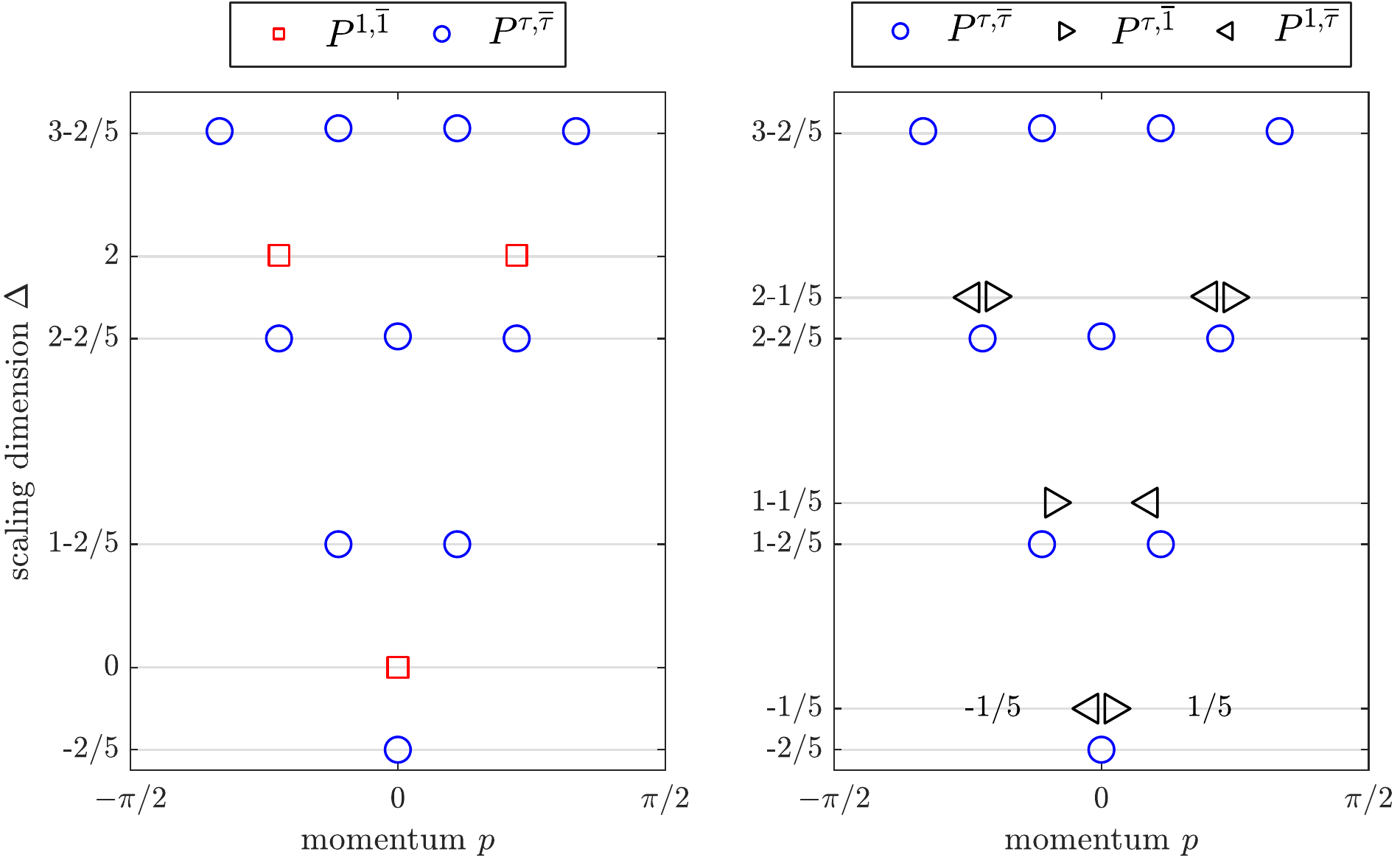}
	\caption{Topological sector labeling of the Yang-Lee model of finite-size CFT spectra (scaling dimension $\Delta$ versus momentum) on a cylinder with a circumference of 18 sites with no defect line on the left and a $\tau$ defect line on the right. The exact topological correction to the conformal spin is denoted next to the first appearance of the respective idempotents.}
	\label{spectra_YangLee}
\end{figure}

The results for the torus are given in figure \ref{spectra_YangLee}. The tube algebra is not closed under Hermitian conjugation, but nevertheless one can construct idempotents just as in the case of the unitary Fibonacci string-net \cite{lootens2020galois}. \\

\subsubsection{Klein bottle results}

\begin{figure}[t]
	\center
	\includegraphics[page=7]{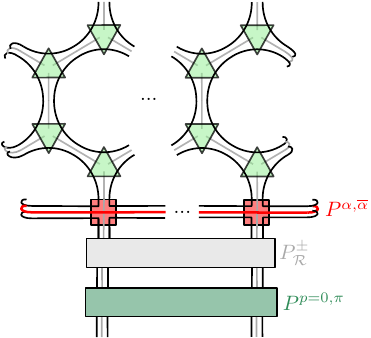}
	\caption{The diagonalization procedure on the Klein bottle for the Ising model on a hexagonal lattice. The transfer matrix on the cylinder is projected on a topological sector (idempotent) $P^{\alpha,\overline{\alpha}}$ and labeled by the reflection quantum numbers ($P^{\pm}_{\mathcal{R}}$).}
	\label{fig:torusKleinYL}
\end{figure}

\begin{figure}[t]
	\centering
	\includegraphics[height=17em]{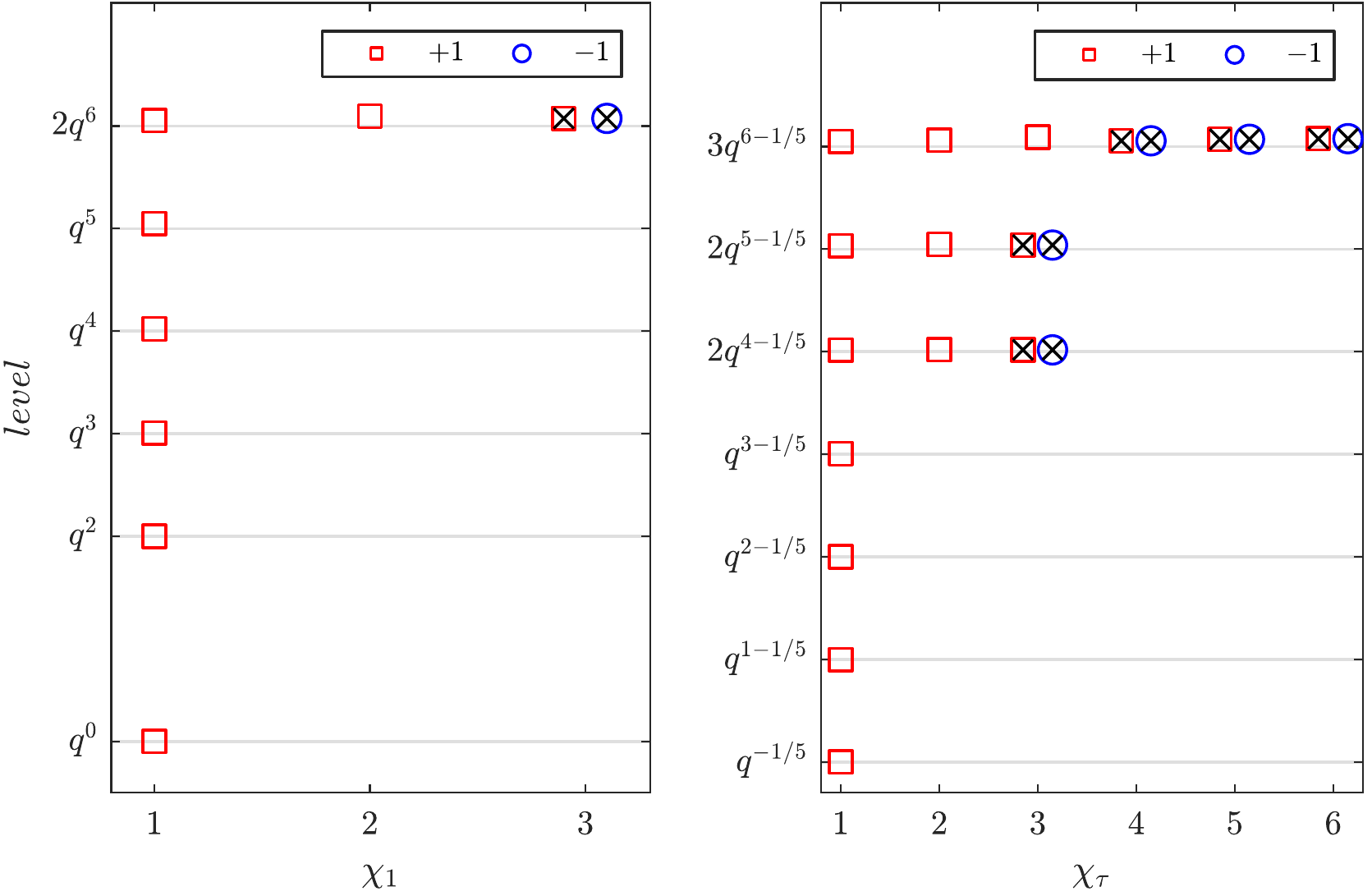}
	\caption{The two characters for the Yang-Lee model obtained numerically on the Klein bottle with length $L=30$. The explicit cancellation is shown at every level between exactly degenerate eigenvalues with positive and negative quantum number under reflection.}
	\label{fig:YangLeespectraKlein}
\end{figure}

The result on the Klein bottle is shown in figure \ref{fig:YangLeespectraKlein} for a trivial twist. An important subtlety in the non-unitary case is that the left handed MPO should not be defined by taking the complex conjugate of the $F$-symbol. Therefore, the conjugation in equations \eqref{mpo_reflect}-\eqref{idem_reflect} drops out. The diagonalization scheme is schematically shown in figure \ref{fig:torusKleinYL}. One can also apply a $\tau$-twist, but since the spectrum does not contain any new characters ($Z_{\tau}^\mathcal{K} = \chi_{\tau}(q^2)$) it is not shown. Up to finite-size effects, the degeneracies of the two Yang-Lee characters (\ref{YLCharacters}) are recovered. \\

The result for the Klein bottle entropy is shown in Table \ref{tab:KleinBottleEntropyYL}. Special care has to be taken for the Yang-Lee model, since its transfer matrix is non-Hermitian and therefore the right- and left leading eigenvectors of the transfer matrix are different in equation \eqref{KBentropyFixedPoint}.\\

\begin{center}
	\begin{table}[ht]
		\centering
		\resizebox{0.7\textwidth}{!}{
			\begin{tabular}{ |c|c|c| }
				\hline
				$L_y$&$S(\tau,\bf{1})$&$S(\tau,\tau)$\\
				\hline
				$8$&$0.5252487$&$0.8498703$\\
				$10$&$0.5254144$&$0.8501385$\\
				$12$&$0.5255105$&$0.8502939$\\
				$14$&$0.5255690$&$0.8503885$\\
				$16$&$0.5256070$&$0.8504499$\\
				$18$&$0.5256330$&$0.8504921$\\
				$20$&$0.5256516$&$0.8505222$\\
				$22$&$0.5256655$&$0.8505444$\\
				\hline
				Exact&$\frac{2}{\sqrt{5}}\sin(4\pi/5) = 0.5257311$&$\frac{2}{\sqrt{5}}\sin(2\pi/5) = 0.8506508$\\
				\hline
		\end{tabular}}
		\caption{The Yang-Lee Klein bottle entropy $g$ in function of system size $\beta$ with the projection onto the topological sectors: $P^{\bf{1}\overline{\bf{1}}}$ (second column) and $P^{\tau\overline{\tau}}$ (third column). The second row $S$-matrix elements are recovered.}
		\label{tab:KleinBottleEntropyYL}
	\end{table}
\end{center}

\subsubsection{Cylinder results}

The vacuum state $\ket{\mathbf{1}}\!\rangle$ is given by alternating $\tau$ and $1$ loops at the boundary. The non-trivial Cardy state $\ket{\mathbf{\tau}}\!\rangle$ can again be obtained by application of the MPO $O_{\tau}$:

\begin{align}
\begin{split}
\ket{\mathbf{1}}\!\rangle &= \ket{ \tau \mathbf{1} \tau \mathbf{1} \tau \mathbf{1} ...}\\
\ket{\tau}\!\rangle &= O_{\tau}\ket{\bf{1}} = \ket{(\mathbf{1}\!+\!\tau) \tau (\mathbf{1}\!+\!\tau) \tau (\mathbf{1}\!+\!\tau) \tau ...}.
\end{split}
\end{align}
For the $\tau$-twisted case, we have to insert fusion tensors in the $\ket{\tau}$ Cardy state:

\begin{align}
\ket{\tau^{\tau}} \rightarrow \vcenter{\hbox{\includegraphics[page=3,scale=0.8]{figures/cardyStates}}} \quad ; \quad \vcenter{\hbox{\includegraphics[page=4,scale=0.8]{figures/cardyStates}}}.
\end{align}

\begin{figure}[t]
	\centering
	\includegraphics[page=8,scale = 1]{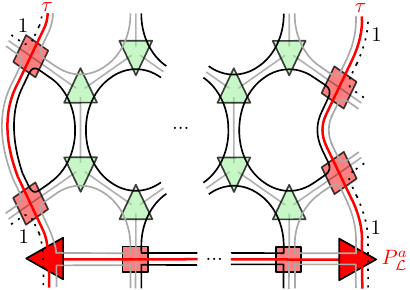}
	\caption{The diagonalization procedure on the cylinder with the MPO$_{\tau}$ applied at both boundaries on the vacuum state. The transfer matrix on the cylinder is projected on the idempotents of the ladder algebra ($P^{a}_{\mathcal{L}}$, $a=\bf{1},\tau$).}
	\label{fig:cylinderYL}
\end{figure}

\begin{figure}[t]
	\centering
	\includegraphics[height=17em]{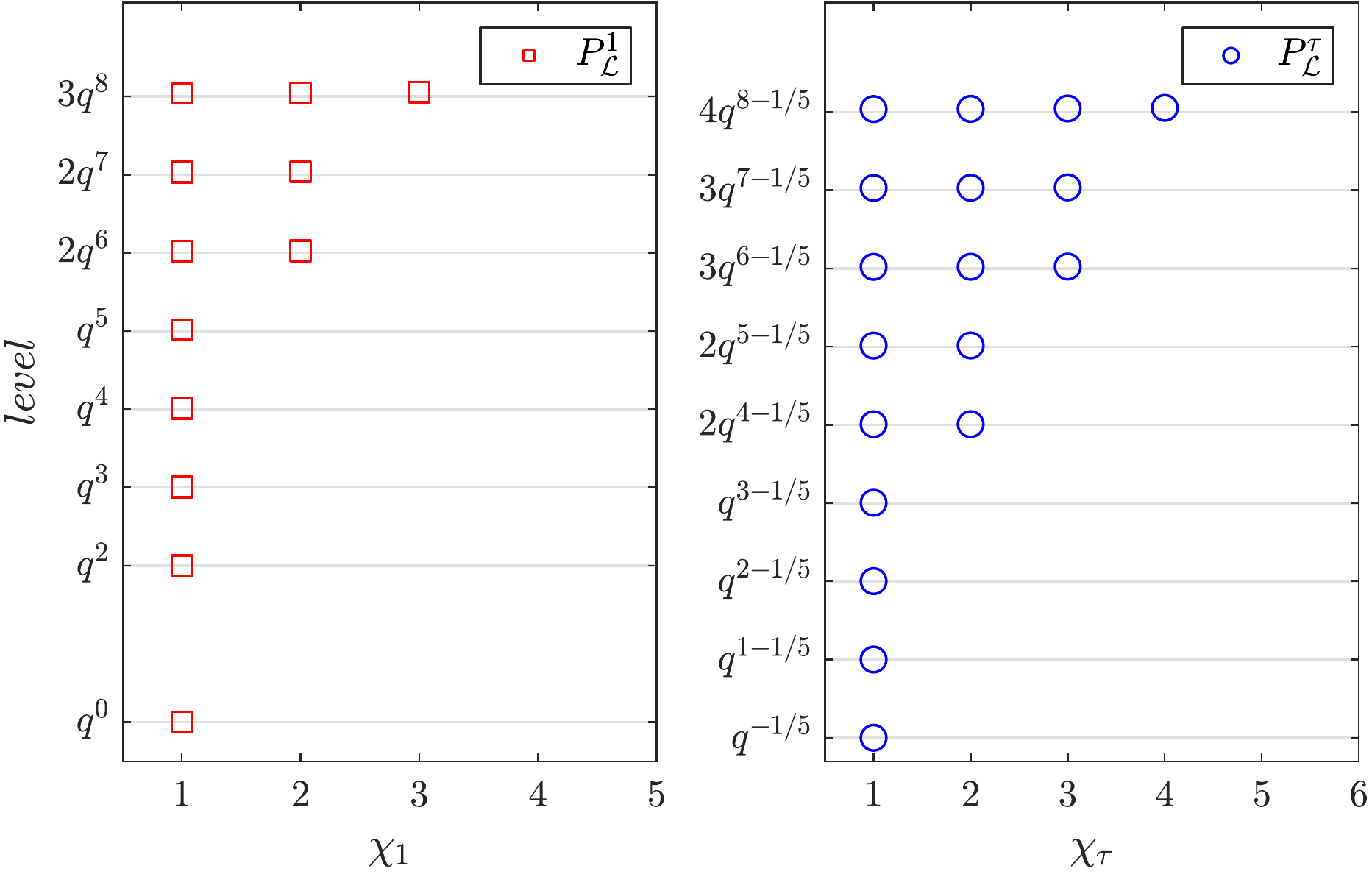}
	\caption{The two characters $\chi_{\bf{1}}$ and $\chi_{\tau}$ for the Yang-Lee model obtained numerically on the Cylinder with length $L=22$ and the $\ket{\tau}$ boundary at both ends (figure \ref{fig:cylinderYL}). The projection on the single characters is obtained through the idempotents of the ladder algebra with elements $\mc{L}_{\tau\tau,\tau\tau}^{\bf{1}}$ and $\mc{L}_{\tau\tau,\tau\tau}^{\tau}$.}
	\label{fig:YangLeeCylinder}
\end{figure}

Numerics were performed on the cylinder with boundary states $\ket{\tau}$ at both ends of the cylinder. The diagonalization procedure is schematically shown in figure \ref{fig:cylinderYL}. The results are presented in figure \ref{fig:YangLeeCylinder}. The spectrum consists of the characters $\chi_{\bf{1}}$ and $\chi_{\tau}$. The characters are singled out by using the idempotents of the ladder algebra, $\mc{L}_{\tau\tau,\tau\tau}^{\bf{1}}$ and $\mc{L}_{\tau\tau,\tau\tau}^{\tau}$ in this case.

\subsection{The three-state Potts model}

Finally, we turn to the three-state Potts model, which requires special attention, as explained in the orbifold section \ref{Sec:Orbifold}. It is the non-diagonal modular invariant of the $c=4/5$ minimal model CFT, and can be realized on the lattice as a generalized Ising model with three spins instead of two. As an RSOS model, it is based on the $D_4$ Dynkin diagram:
\begin{align*}
D_4 = \enspace \vcenter{\hbox{\includegraphics[page=1,scale=0.5]{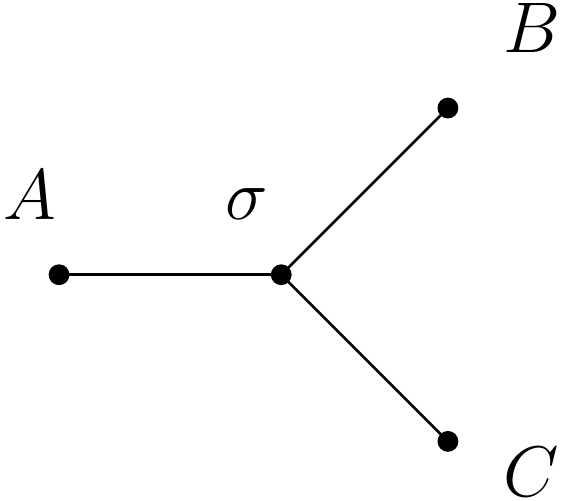}}}.
\end{align*}
The relevant input MFC $\mc{D}_\text{TCI}$ here is the same as for the tetracritical Ising model, and as discussed above can be written as $\mc{D}_\text{TCI} = |\text{su}(2)_4| \otimes \text{Fib}$. For the strange correlator construction of the $D_4$ RSOS model, we will not be concerned with the $\text{Fib}$ factor of $\mc{D}_\text{TCI}$. As a module category of $|\text{su}(2)_4|$, we take $\mc{M}_{TY}$ with simple objects $\{A,\sigma,B,C\}$. The strange correlator is entirely analogous to the Ising case, with the choice of the bathroom tiling to make the duality on the square lattice apparent. The transfer matrix is again a direct sum of a primal and dual checkering (now with blue tensors, indicating that we are dealing with a non-diagonal model with degrees of freedom in $\mc{M} \neq \mc{D}$):
\begin{align*}
	\vcenter{\hbox{\includegraphics[page = 2, width=0.75\linewidth]{figures/transferMatrices}}}
\end{align*}
In this construction, the PEPS physical legs are labeled by $|\text{su}(2)_4|$ simple objects. The middle legs (purple) are fixed to the state $(1+\sqrt{3})\ket{\it{0}} + \ket{\it{1}}$, while the outer physical legs (fixed on $\sigma$ in the Ising model) are fixed to $\it{1/2}$ (orange). The (virtual) loop degrees of freedom are labeled by the objects in $\mc{M}_{TY}$ and they represent the 3-state Potts degrees of freedom. The same strange correlator, but with the choice of $|su(2)_4|$ as a module category over itself, yields the $A_5$ RSOS model based on the corresponding Dynkin diagram
\begin{align*}
A_5 = \enspace \vcenter{\hbox{\includegraphics[page=2,scale=0.5]{figures/Dynken}}}.
\end{align*}
These two lattice models can be transformed into one another by MPO intertwiners, which are a topological version of the intertwiners discussed in \cite{pearce1993intertwiners}.\\

The topological defects of the $D_4$ RSOS model are described by the fusion category $\mc{C}_{S_3}$. The corresponding MPO symmetries associated to the $S_3$ subgroup $\{A^+, B^+, C^+, A^-, B^- C^-\}$ are tensor products of local operators that implement the $S_3$ permutations on the $\{A,B,C\}$ degrees of freedom. As we discussed above, the continuum CFT also has topological defects corresponding to the $\text{Fib}$ factor in $\mc{C}_\text{Potts}$. For the $D_4$ RSOS model however, it is believed and argued in \cite{belletete2020topological} that these defects are not topological at finite size, but rather become topological only in the continuum limit. As such, they will not be part of the discussion here, and their absence means that we will not be able to fully resolve the characters in the partition functions.

\subsubsection{Torus results}

\begin{figure}[t]
	\centering
	\includegraphics[width=\textwidth]{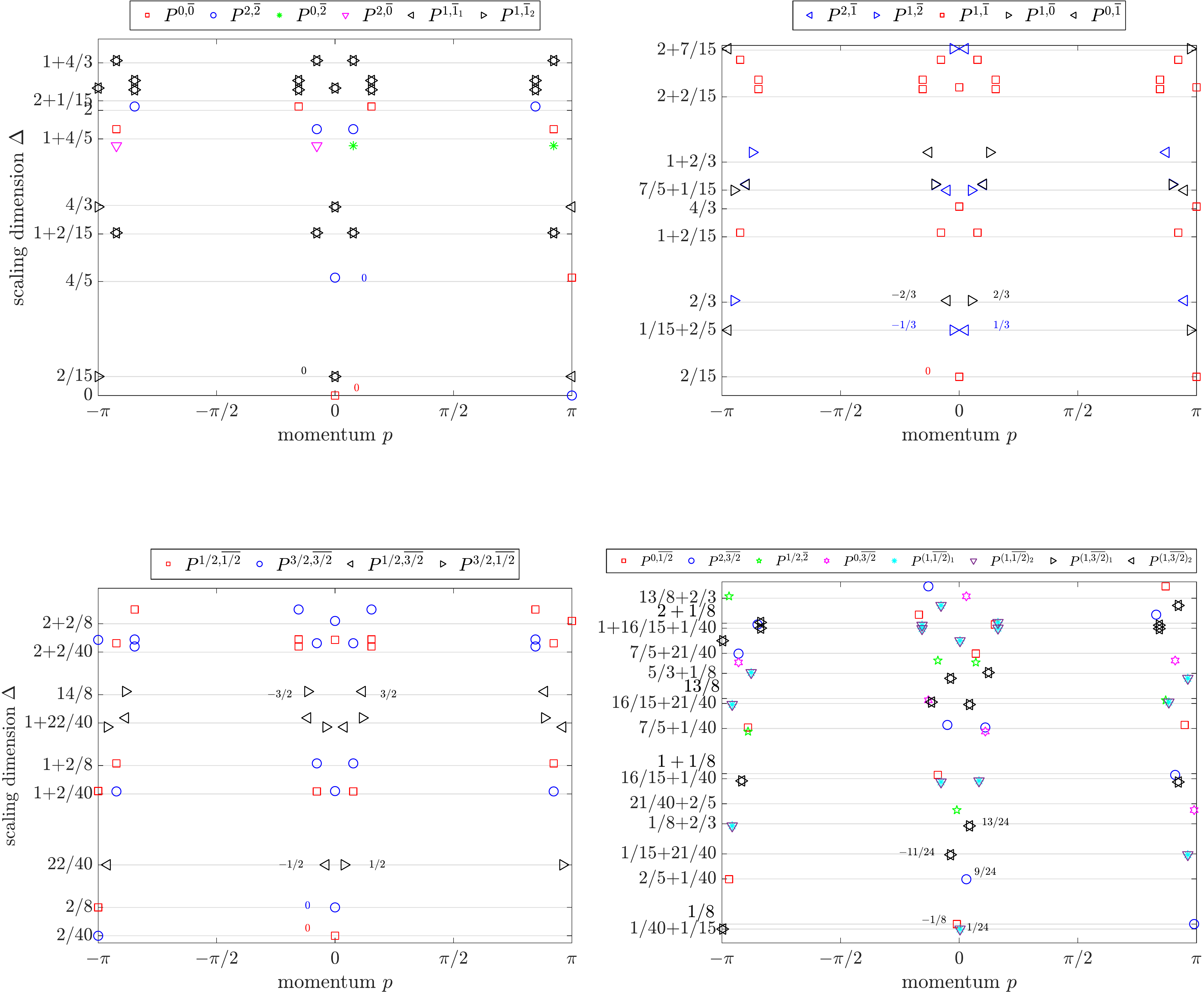}
	\caption{Exact diagonalization of the Potts transfer matrix with a trivial (upper left), $B^+$ (upper right), $A^-$ (lower left) and $\sigma^+$-twist (lower right) on $L=12$ sites. The $C^+$, $B^-$ and $C^-$-twists are not shown, since they do not lead to different spectra. The $\sigma^-$-twist is simply the complex conjugated version of the $\sigma^+$-twist. The spectrum is labeled by the idempotents of the tube algebra, according to tables \ref{PottsTableBlock1}-\ref{PottsTableBlock5} in appendix \ref{App:data}. The topological corrections to the conformal spins are shown for the lowest eigenvalues. The spectra are consistent with \ref{PottsTorus}.}
	\label{fig:PottsTorusAB}
\end{figure}

On the torus, the same trick as in the Ising case can be used and by decomposing the spectrum into the primal and dual momentum branches, a consistent labeling of the twisted CFT spectra with the full idempotent table can be performed. This is shown in figure \ref{fig:PottsTorusAB} for a trivial ($A^+$), $B^+$, $A^-$ and $\sigma^+$ twist. These spectra are consistent with \eqref{PottsTorus} and \cite{petkova2001generalised}, with finite size effects considerably larger than in the previous two examples. Furthermore, every first distinct eigenvalue is labeled by its corresponding topological spin, consistent with the conformal spin of the resulting term in the CFT partition function. \\

Since we have the decomposition of all Potts tube elements, we can also identify the projectors (idempotents) of $C_{S_3}$ in terms of the characters of the tetracritical Ising model. Let us illustrate this with the partition function with a trivial twist $Z_{\text{Potts}} = Z_{A^{+}}$. The idempotent table in appendix \ref{pottsData} (table \ref{PottsTableBlock1}) contains five idempotents, labeled by $Z(\mc{C}_{S_3}) = |\text{su}(2)_4| \boxtimes \overline{|\text{su}(2)_4|}$. The last idempotent $P^{\it{1},\overline{\it{1}}}$ is two-dimensional and can be further split into

\begin{align}
\begin{split}
P^{(\it{1},\overline{\it{1}})_{1}} = \frac{1}{3}(O_{A^{+}} + \omega \ O_{B^{+}} + \overline{\omega} \ O_{C^{+}}) \\
P^{(\it{1},\overline{\it{1}})_{2}} = \frac{1}{3}(O_{A^{+}} + \overline{\omega} \ O_{B^{+}} + \omega \ O_{C^{+}}),
\end{split}
\label{simpleIdems}
\end{align}

with $\omega = e^{2\pi i/3}$. The full projection onto the six simple idempotents results in

\begin{align}
\begin{split}
&P^{\it{0},\overline{\it{0}}} \rightarrow |\chi_{0}(q)|^2+
|\chi_{7/5}(q)|^2 \\
&P^{\it{2},\overline{\it{2}}} \rightarrow |\chi_{2/5}(q)|^2+
|\chi_{3}(q)|^2 \\
&P^{\it{0},\overline{\it{2}}} \rightarrow  \chi_{7/5}(q)\overline{\chi}_{2/5}(\overline{q}) + \chi_{0}(q)\overline{\chi}_{3}(\overline{q}) \\
&P^{\it{2},\overline{\it{0}}} \rightarrow  \chi_{2/5}(q)\overline{\chi}_{7/5}(\overline{q}) +
\chi_{3}(q)\overline{\chi}_{0}(\overline{q}) \\
&P^{(\it{1},\overline{\it{1}})_1} \rightarrow |\chi_{2/3}(q)|^2 + |\chi_{1/15}(q)|^2 \\
&P^{(\it{1},\overline{\it{1}})_2} \rightarrow |\chi_{2/3}(q)|^2 + |\chi_{1/15}(q)|^2.
\end{split}
\label{PottsIdem}
\end{align}

\subsubsection{Klein bottle results}

The group-like MPO symmetries of the Potts model commute with the reflection operator. The duality MPOs ($O_{\sigma^+}$ and $O_{\sigma^-}$) are not real and therefore they are not reflection invariant, however they transform into one another when the reflection operator is commuted through them. An important consequence of this is that not all idempotents commute with the reflection operator, but only those (sums of) idempotents that are symmetric under the exchange of the two dualities. Looking at the idempotent table \ref{PottsTableBlock1} in Appendix \ref{App:data}, one sees that only the projectors $P^{\it{0}\overline{\it{0}}}$, $P^{\it{2}\overline{\it{2}}}$ and $P^{\it{1}\overline{\it{1}}}$ satisfy this. Remembering from \eqref{simpleIdems} that the projector $P^{\it{1}\overline{\it{1}}}$ can be further split into two simple idempotents $P^{(\it{1},\overline{\it{1}})_1}$ and $P^{(\it{1},\overline{\it{1}})_2}$, we end up with four Klein bottle sectors that can be singled out. These are exactly those sectors in \eqref{PottsIdem} that contain left-right symmetric terms. The fact that the projectors $P^{\it{0}\overline{\it{2}}}$ and $P^{\it{2}\overline{\it{0}}}$ are not reflection invariant is not a surprise, since they lead to terms in the partition function that are not left-right symmetric.\\

\begin{figure}[t]
	\centering
	\includegraphics[height=17em]{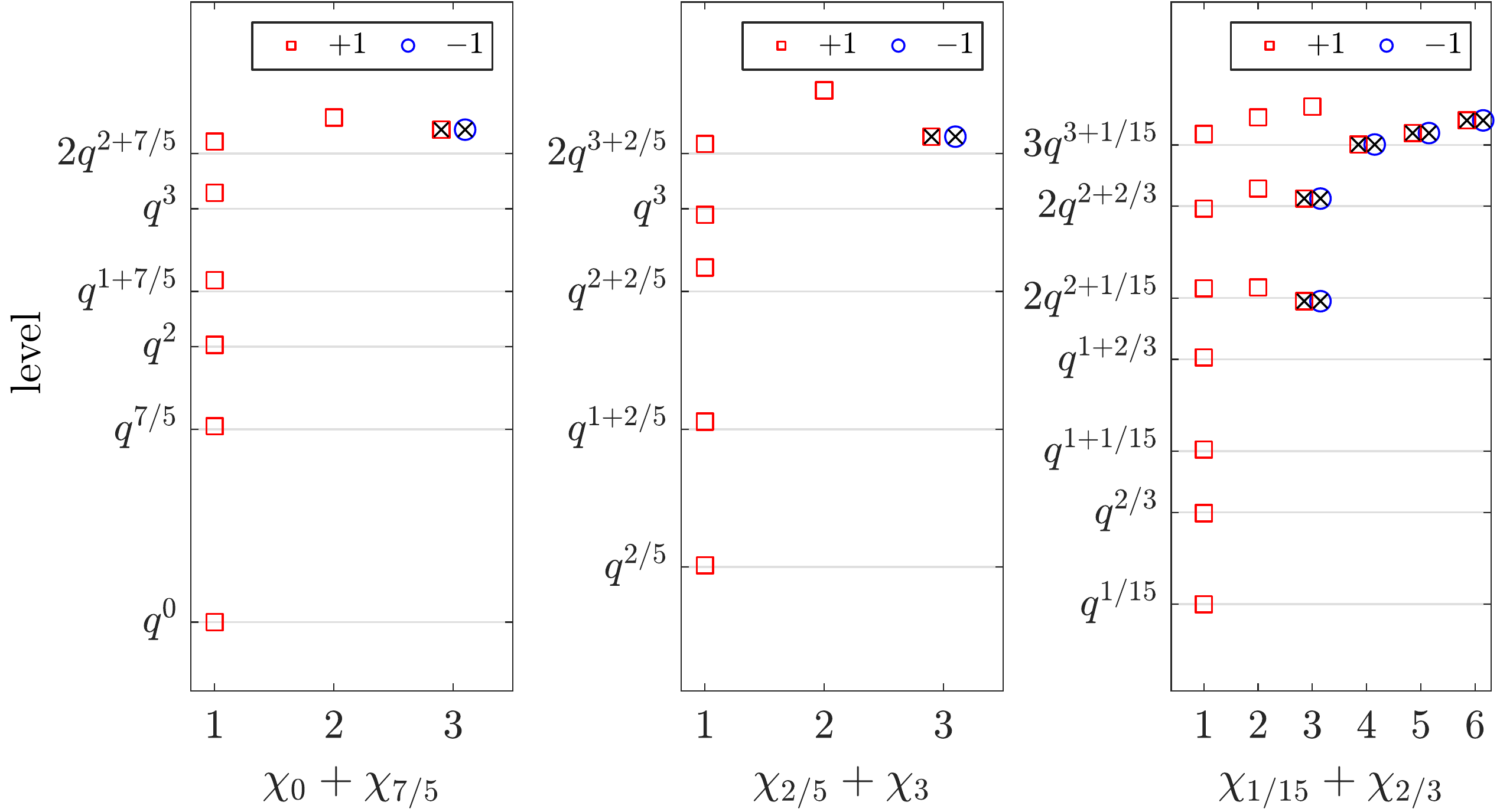}
	\caption{The different tetracritical Ising character decomposition (\ref{TetraChars}) of the Potts model obtained numerically on the Klein bottle with length $L=14$. From left to right the projections $P^{\it{0}\overline{\it{0}}}$, $P^{\it{2}\overline{\it{2}}}$, $P^{(\it{1},\overline{\it{1}})_1}$ are shown. The characters appearing in the $P^{(\it{1},\overline{\it{1}})_2}$ projection are the same as $P^{(\it{1},\overline{\it{1}})_1}$ and are not shown. The explicit cancellation is shown at every level between exactly degenerate eigenvalues with positive and negative quantum number under reflection.}
	\label{fig:PottsKlein}
\end{figure}

The diagonalization procedure is the same as for the Ising model, shown in Figure \ref{fig:torusKleinIsing}. Since a $\sigma^+$- or $\sigma^-$-twisted Klein bottle partition function is zero and the other twists are merely local operators, we illustrate the Klein bottle projection for the Potts model with a trivial horizontal twist together with the projection on $P^{\it{0}\overline{\it{0}}}$, $P^{\it{2}\overline{\it{2}}}$, $P^{(\it{1},\overline{\it{1}})_1}$ and $P^{(\it{1},\overline{\it{1}})_2}$. The result is shown in figure \ref{fig:PottsKlein}. The projected spectrum is consistent with what is expected for the diagonal selection of the corresponding trivially twisted torus partition function, with the relevant tetracritical Ising characters given in \eqref{TetraChars}. \\

Next, we turn to the Klein bottle entropy, considering again the four idempotents that commute with the reflection operator. The story is very similar as in the Ising case. We need momentum information to separate the rows of the tetracritical Kac table (\ref{KacTetra}), but in the Klein bottle entropy calculations we do not have a finite bond dimension operator that separates these rows. Therefore, we cannot separate the momentum $0$ and momentum $\pi$ contributions to $P^{\it{0}\overline{\it{0}}}$ and $P^{\it{2}\overline{\it{2}}}$. This effectively reduces the usable projectors to the projectors on the $\mathds{Z}_3$ irreps: $P^{\it{0}\overline{\it{0}}}+P^{\it{2}\overline{\it{2}}}$, $P^{(\it{1},\overline{\it{1}})_1}$ and $P^{(\it{1},\overline{\it{1}})_2}$.\\

Looking at the Klein bottle entropy terms present in \eqref{KBentropyFixedPoint} with a trivial twist:

\begin{align*}
\vcenter{\hbox{\includegraphics[page=9,scale=0.7]{figures/kbtorusNew}}},
\end{align*}

it is clear that both for the term $a=B^+$ and $a=C^+$, the resulting single twist in the transfer matrix is respectively $b=C^+$ and $b=B^-$. In the limit $L_x \rightarrow \infty$, both terms will be exponentially suppressed, since the largest eigenvalues $\lambda_{max}$ of the $B^+$- and $C^+$-twisted transfer matrix are strictly larger than the largest eigenvalue of the trivially twisted transfer matrix. This can be seen in the spectra in figure \ref{fig:PottsTorusAB}): the ground-state energy ($-\log{|\lambda_\text{max}|}$) of the trivially twisted partition function is smaller than the one of the $B^+$- and $C^+$ twisted partition functions. Note that in the Yang-Lee model, this was not the case, as a trivially twisted term ($b=1$) appears in the Klein bottle entropy when $a=\tau$. For the remaining terms with $a=A^-,B^-,C^-$ in the projector $P^{\it{0}\overline{\it{0}}}+P^{\it{2}\overline{\it{2}}}$, the only non-vanishing twist is $b=A^+$. The Klein bottle entropy is therefore split into the three equal terms $g^{\it{0}}+g^{\it{2}}$, $g^{\it{1}_1}$ and $g^{\it{1}_2}$.
In table \eqref{tab:PottsKBE} the Klein bottle entropy (without any projection) is shown  for several sizes. The finite-size effects are again noticeably bigger than in the previous two models, seeing a slower convergence towards the expected value of $\sqrt{3+\frac{6}{\sqrt{5}}}$ with increasing $\beta$.

\begin{center}
	\begin{table}[ht]
		\centering
		\resizebox{0.4\textwidth}{!}{
			\begin{tabular}{ |c|c| }
				\hline
				$L_y$&$g = g^{A}+g^{B}+g^{C}+g^{\sigma}$\\
				\hline
				$4$&$2.3694$\\
				$6$&$2.3720$\\
				$8$&$2.3738$\\
				$10$&$2.3751$\\
				$12$&$2.3761$\\
				$14$&$2.3768$\\
				$16$&$2.3774$\\
				\hline
				Exact&$\sqrt{3+\frac{6}{\sqrt{5}}} \approx 2.384$\\
				\hline
		\end{tabular}}
		\caption{The Potts Klein bottle entropy diagonalization for increasing size $L_y$, calculated according to \ref{KBentropyFixedPoint}. The large finite size effects for the Potts model make for a much slower convergence compared to the previous two models.}
		\label{tab:PottsKBE}
	\end{table}
\end{center}

\subsubsection{Cylinder results}

Lastly, we show some nontrivial results for the Potts model on the cylinder. The eight Cardy states are known on the lattice and four of them can be reproduced from the application of the four intertwiners ($O_A$, $O_B$, $O_C$ and $O_{\sigma}$) to the tetracritical Ising vacuum states. These vacuum state can be labeled by the Kac table elements as $\ket{(\it{0},\mathbf{1})_p}\rangle$ and is given by alternating the loops on $\it{0}$ and $\it{1/2}$ for the primal vacuum state and vice versa for the dual vacuum state $\ket{(\it{0},\mathbf{1})_d}\rangle$:

\begin{align}
\ket{(\it{0},\mathbf{1})_p}\rangle &= \ket{\it{0}\it{\frac{1}{2}}\it{0}\it{\frac{1}{2}}\it{0}\it{\frac{1}{2}}...},\\
\ket{(\it{0},\mathbf{1})_d}\rangle &= \ket{\it{\frac{1}{2}}\it{0}\it{\frac{1}{2}}\it{0}\it{\frac{1}{2}}\it{0}...}.
\end{align}
Using these, we obtain the following six distinct ``fixed'' Potts Cardy states:
\begin{align*}
\ket{(A,\mathbf{1})_p}\rangle = O_{A}\ket{(\it{0},\mathbf{1})_p}\rangle &= \ket{A \sigma A \sigma A \sigma ...},\\
\ket{(A,\mathbf{1})_d}\rangle = O_{A}\ket{(\it{0},\mathbf{1})_d}\rangle &= \ket{\sigma A \sigma A \sigma A ...},\\
\ket{(B,\mathbf{1})_p}\rangle = O_{B}\ket{(\it{0},\mathbf{1})_p}\rangle &= \ket{B \sigma B \sigma B \sigma ...},\\
\ket{(B,\mathbf{1})_d}\rangle = O_{B}\ket{(\it{0},\mathbf{1})_d}\rangle &= \ket{\sigma B \sigma B \sigma B ...},\\
\ket{(C,\mathbf{1})_p}\rangle = O_{C}\ket{(\it{0},\mathbf{1})_p}\rangle &= \ket{C \sigma C \sigma C \sigma ...},\\
\ket{(C,\mathbf{1})_d}\rangle = O_{C}\ket{(\it{0},\mathbf{1})_d}\rangle &= \ket{\sigma C \sigma C \sigma C ...},
\end{align*}
and two ``free'' \cite{cardy1989boundary} Cardy states
\begin{align*}
\ket{(\sigma,\mathbf{1})_d}\rangle = O_{\sigma}\ket{(\it{0},\mathbf{1})_p}\rangle &= \ket{\sigma (A+B+C) \sigma (A+B+C) \sigma (A+B+C) ...},\\
\ket{(\sigma,\mathbf{1})_p}\rangle = O_{\sigma}\ket{(\it{0},\mathbf{1})_d}\rangle &= \ket{(A+B+C) \sigma (A+B+C) \sigma (A+B+C) \sigma ...}.
\end{align*}

Since we do not have an MPO description for the Fibonacci topological defect, we cannot obtain the other four distinct Cardy states from the vacua $\ket{(\it{0},\mathbf{1})_p}\rangle$ and $\ket{(\it{0},\mathbf{1})_d}\rangle$ on the lattice. We can however obtain them by applying the intertwiners to the tetracritical Ising Cardy state $\ket{(\it{0},\tau)_p}\rangle$ and $\ket{(\it{0},\tau)_p}\rangle$, defined as
\begin{align}
\ket{(\it{0},\tau)_p}\rangle &= \ket{\it{1}\it{\frac{1}{2}}\it{1}\it{\frac{1}{2}}\it{1}\it{\frac{1}{2}}...},\\
\ket{(\it{0},\tau)_d}\rangle &= \ket{\it{\frac{1}{2}}\it{1}\it{\frac{1}{2}}\it{1}\it{\frac{1}{2}}\it{1}...}.
\end{align}
These boundary states are not related to the previously defined vacua by a topological defect, and have to be obtained by invoking the integrability of the model \cite{behrend1998integrable}. Applying the intertwiners, we now obtain the following six Cardy states:
\begin{align}
\ket{(A,\tau)_p}\rangle = O_{A}\ket{(\it{0},\tau)_p}\rangle &= \ket{(B+C) \sigma (B+C) \sigma (B+C) \sigma ...},\\
\ket{(A,\tau)_d}\rangle = O_{A}\ket{(\it{0},\tau)_p}\rangle &= \ket{\sigma (B+C) \sigma (B+C) \sigma (B+C) ...},\\
\ket{(B,\tau)_p}\rangle = O_{B}\ket{(\it{0},\tau)_p}\rangle &= \ket{(A+C) \sigma (A+C) \sigma (A+C) \sigma ...},\\
\ket{(B,\tau)_d}\rangle = O_{B}\ket{(\it{0},\tau)_p}\rangle &= \ket{\sigma (A+C) \sigma (A+C) \sigma (A+C) ...},\\
\ket{(C,\tau)_p}\rangle = O_{C}\ket{(\it{0},\tau)_p}\rangle &= \ket{(A+B) \sigma (A+B) \sigma (A+B) \sigma ...},\\
\ket{(C,\tau)_d}\rangle = O_{C}\ket{(\it{0},\tau)_p}\rangle &= \ket{\sigma (A+B) \sigma (A+B) \sigma (A+B) ...},
\end{align}
which are sometimes called the ``not-$A$'', ``not-$B$'' and ``not-$C$'' boundary conditions and were known to Cardy in his original paper. The remaining boundary condition was later obtained by Affleck, Oshikawa and Saleur \cite{affleck1998boundary} and was called the ``new'' boundary condition $\ket{N}$. We obtain it by acting with the $O_{\sigma}$ intertwiner on $\ket{(\it{0},\tau)_p}\rangle$ and $\ket{(\it{0},\tau)_p}\rangle$:
\begin{align}
\ket{(\sigma,\tau)_p}\rangle = O_{\sigma}\ket{(\it{0},\tau)_d}\rangle &= \ket{N_p},\\
\ket{(\sigma,\tau)_d}\rangle = O_{\sigma}\ket{(\it{0},\tau)_p}\rangle &= \ket{N_d},
\label{PottsNew}
\end{align}
completing the Potts Cardy states list. This final boundary condition is not a product state anymore, but rather a bond dimension two MPS (see for example \cite{verniernot}). This can be seen by drawing \eqref{PottsNew} in the MPO language, including the loop degrees of freedom:
\begin{align*}
	\centering
 \vcenter{\hbox{\includegraphics[page=9,scale=1]{figures/transferMatrices}}} \quad = \quad \vcenter{\hbox{\includegraphics[page=10,scale=1]{figures/transferMatrices}}}.
\end{align*}
Now, the multiplicity labels $j$ and $k$ in definition \eqref{intertwiner} have to be taken into account, since $N_{\sigma\sigma}^{\it{1}} = 2$. These multiplicities are the origin of the nontrivial bond dimension of the new boundary state. Explicitly, the MPS matrices are \cite{verniernot}:

\begin{equation}
N^{A} =	
\begin{pmatrix}
1&1\\
1&1\\
\end{pmatrix},
N^{B} =	
\begin{pmatrix}
1&\omega\\
\overline{\omega}&1\\
\end{pmatrix},
N^{C} =
\begin{pmatrix}
1&\overline{\omega}\\
\omega&1\\
\end{pmatrix},
\end{equation}
with $\omega = e^\frac{2\pi i}{3}$. The new boundary condition is invariant under the $S_3$ symmetry of the three-state Potts model, since we have $g \lact \sigma = \sigma$ for $g$ a group-like label in $\mc{C}_{S_3}$ (see Appendix \ref{App:data} for the table of fusion rules). \\

Using this, we want to illustrate the case of a nontrivial horizontal twist on the cylinder, stressing the strength of the MPO approach. Starting from the $C_{Potts}$ category, all the $S_3$ twists are allowed on the cylinder (\ref{Sec:CylinderOrbifold}), while the $\sigma^+$ or $\sigma^-$ twists are not. The Cardy states at the boundary must carry a $\sigma$ label and must therefore have been created by a $O_\sigma$ intertwiner from either the tetracritical vacuum sate or the $\ket{(\it{0},\tau)}$ state. The reason is again that all the six $S_3$ twists are allowed in the fusion $a \lact \sigma = \sigma, a \in \mc{C}_{S_3}$ (see Appendix \ref{App:data} for the table of fusion rules). \\

We will perform numerical simulations to illustrate the above on the three cylinder partition functions:

\begin{align}
\label{Z1}
&Z_{free,free}^{\mathcal{C}} \\
\label{Z2}
&Z_{free,New}^{\mathcal{C}} \\
\label{Z3}
&Z_{New,New}^{\mathcal{C}}.
\end{align}

The transfer matrix with such boundary conditions can be projected onto the four different $S_3$ irreps, consistent with the idempotents of the ladder algebra elements $\mc{\tilde{L}}^{A^+}$, $\mc{\tilde{L}}^{B^+}$, $\mc{\tilde{L}}^{C^+}$, $\mc{\tilde{L}}^{A^-}$, $\mc{\tilde{L}}^{B^-}$ and $\mc{\tilde{L}}^{C^-}$  (in the simplified notation $\mc{\tilde{L}}^{a} = \mc{L}_{\sigma\sigma,\sigma\sigma}^{a}$):

\begin{align}
\begin{split}
P_{\mc{\tilde{L}}}^{1} &= \frac{1}{6}(\mc{\tilde{L}}^{A^+} + \mc{\tilde{L}}^{B^+} + \mc{\tilde{L}}^{C^+} + \mc{\tilde{L}}^{A^-} + \mc{\tilde{L}}^{B^-} + \mc{\tilde{L}}^{C^-})\\
P_{\mc{\tilde{L}}}^{2} &= \frac{1}{6}(\mc{\tilde{L}}^{A^+} + \mc{\tilde{L}}^{B^+} + \mc{\tilde{L}}^{C^+} - \mc{\tilde{L}}^{A^-} - \mc{\tilde{L}}^{B^-} - \mc{\tilde{L}}^{C^-}) \\
P_{\mc{\tilde{L}}}^{3} &= \frac{1}{3}(\mc{\tilde{L}}^{A^+} + \omega \mc{\tilde{L}}^{B^+} + \overline{\omega}\mc{\tilde{L}}^{C^+})\\
P_{\mc{\tilde{L}}}^{4} &= \frac{1}{3}(\mc{\tilde{L}}^{A^+} + \overline{\omega} \mc{\tilde{L}}^{B^+} + \omega\mc{\tilde{L}}^{C^+}).
\end{split}
\label{ladderIdempotentsS3}
\end{align}

\begin{figure}[h]
	\begin{minipage}[b]{0.37\linewidth}
		\includegraphics[height=31em,page=11]{figures/transferMatrices.pdf}
	\end{minipage}%
	\begin{minipage}[b]{0.5\linewidth}
		\includegraphics[height=31em,]{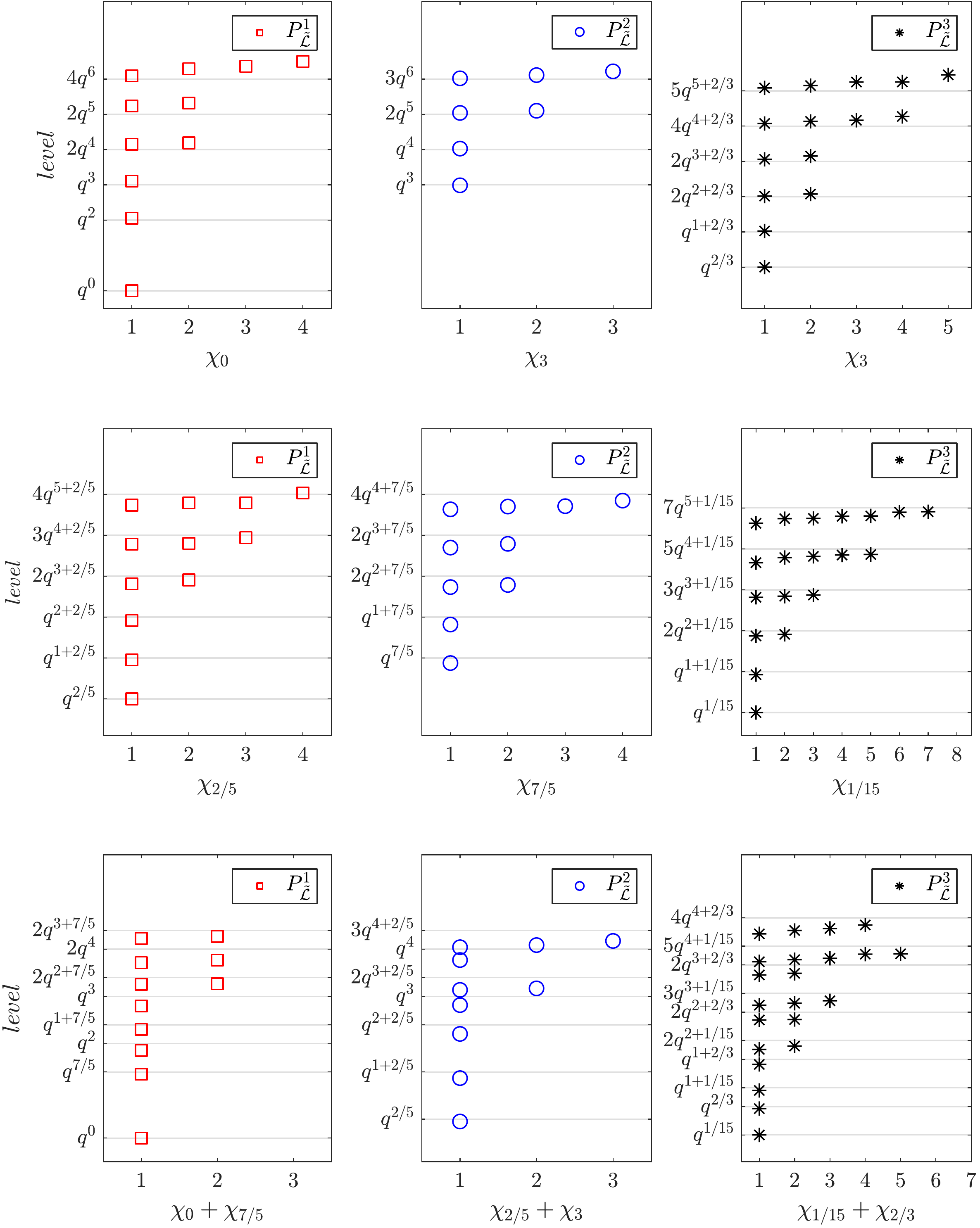}
	\end{minipage}
	\caption{Potts cylinder diagonalizations for the three partition functions \ref{Z1} (top), \ref{Z2} (middle) and \ref{Z3} (bottom). The cylinder lengths are respectively $13$, $12$ and $12$. Each cylinder is projected on the $S_3$ ladder idempotents (\ref{ladderIdempotentsS3}). The projection $P_{\tilde{\mathcal{L}}}^4$ is not shown, because it leads to the same spectrum as $P_{\tilde{\mathcal{L}}}^3$.}
	\label{fig:PottsCylinders}
\end{figure}

The diagonalization of the three distinct partition functions, together with the projections, is diagrammatically summarized in figure \ref{fig:PottsCylinders}, as well as the numerical results of the corresponding diagonalizations. Note that the desired sums of the tetracritical Ising characters for the distinct Potts cylinders with the right Cardy states (\ref{PottsCylinderCardy}) were obtained numerically using DMRG in \cite{chepiga2017excitation,chepiga2021critical}.

\section{Conclusion and outlook}

We have shown that the application of the strange correlator to generalized PEPS descriptions of string-net models yields a systematic characterization of the topological defects of the three-state Potts model in terms of MPO symmetries. We started by studying diagonal theories such as the Ising and the Yang-Lee model on the torus, Klein bottle and cylinder in the presence of topological defects that are allowed to intersect. Using these to construct transfer matrix projectors on the different superselection sectors of the original topological order, the separate CFT characters in the partition function of the torus and Klein bottle can be isolated. We have shown this explicitly through exact diagonalizaion of the transfer matrix of the different models. Additionally, this method has allowed us to further resolve the recently introduced Klein bottle entropies into their constituent $S$-matrix elements. Besides illustrating the general formalism, the simple examples of the Yang-Lee and Ising model reveal an important relation between the finite size effects in a critical lattice model and its MPO symmetries. Indeed, as argued in \cite{Buican2017}, the presence of topological defects restricts the allowed irrelevant perturbations away from the exact RG fixed point, which lie at the origin of finite size effects. In the case where all topological defects of the continuum CFT are present on the lattice, as is the case in the Yang-Lee model, the finite size effects are heavily suppressed, as can be observed from the excellent agreement between the numerical spectra and the CFT prediction. The fact that for a critical lattice model not all topological defects have to be present in finite size can be seen using the same reasoning. One can imagine perturbing a critical lattice model which does possess a certain topological defect with an irrelevant perturbation that breaks said topological defect; the lattice model remains critical, but no longer possesses that particular topological defect in finite size.\\

The non-diagonal case of the three-state Potts model required the use of a strange correlator based on a generalized PEPS description of string-net ground states in terms of bimodule categories. In this way, we have obtained a lattice version of the FRS construction of 2d rational CFT, whereby a local full CFT is determined by a modular tensor category $\mc{D}$ and a corresponding module category $\mc{M}$. This allowed us to obtain a full characterization of all MPO symmetries and provides a framework in which the well-known orbifold can be understood as the existence of MPO intertwiners, acting as the interface between the non-diagonal three-state Potts model and the diagonal tetracritical Ising model. These MPO intertwiners have allowed us to understand the non-diagonal model by mapping it to the diagonal model, recovering the relation between the torus and Klein bottle partition functions as well as the conformal boundary conditions of the two models directly on the lattice; we have explicitly recoverd the MPS description of the ``New'' boundary condition in the Potts model, starting from the MPO intertwiners, and applied it to cylinder exact diagionalization calculations. \\

Although we have mainly focused on the three-state Potts model, we have tried to keep the discussion on MPO symmetries and intertwiners in the context of critical lattice models as general as possible. The main challenge for performing an equally explicit study for general RSOS models is finding solutions to the relevant bimodule pentagon equations in these cases. For the $D$ series this problem was partly solved in \cite{runkel2000structure} where the $\F{3}$ symbols were obtained, but even given this, obtaining the remaining $F$ symbols requires some more work still and we have only done this explicitly for the $D_4$ or three-state Potts model.\\

There are many remaining aspects of the FRS construction that still require a lattice interpretation. Among these, the most prominent is the construction of bulk/boundary operators and their correlation functions, as well as their operator product expansions. Preliminary results in this direction have been described in \cite{lootens2019cardy}, and we will report on this in more detail in future work. The understanding of lattice versions of CFT operators has seen much progress in recent years, and explicit results have been obtained for a large class of models \cite{mong2014parafermionic,milsted2017extraction}. One particularly interesting development \cite{fendley2021integrability} is that by requiring that these operators satisfy a discrete version of the Cauchy-Riemann equations, a condition known as discrete holomorphicity, one can heavily constrain the wealth of lattice models that can be obtained from a strange correlator to a subclass of (critical) lattice models that satisfy the Yang-Baxter equations, and are therefore integrable. Paraphrasing \cite{fendley2021integrability}, integrability appears to be the combination of topology and geometry, and we expect our work to contribute in further understanding the precise nature of these connections.

\section*{Acknowledgements}
We are grateful to Paul Fendley, Tobias Osborne, J\"urgen Fuchs, Christoph Schweigert, Dominic Williamson and Nick Bultinck for various discussions and helpful comments. This work has received funding from the European Research Council (ERC) under  the  European  Union’s  Horizon  2020  research  and  innovation  programme  (grant agreement No 647905 (QUTE)). HHT is supported by the Deutsche Forschungsgemeinschaft (DFG) through project A06 of SFB 1143 (Project No. 247310070). RV and LL are supported by a PhD fellowship from the Research Foundation Flanders (FWO).

\newpage
\appendix
\section{Category theory}
\label{App:Category}

In this section we aim to review the basic notions of tensor categories used throughout this work; for more details, we refer to \cite{lootens2020matrix,etingof2016tensor}.
\subsection{Fusion categories}

A \textit{fusion category} $\mc{C}$ is a finite semi-simple rigid monoidal category with a functor $\otimes\colon \mc{C} \,{\times}\, \mc{C} \,{\rightarrow}\, \mc{C}$, a natural isomorphism $\F{0}\colon (a \,{\otimes}\, b) \,{\otimes}\, c \,{\xrightarrow{\,\cong\,}}\, a \,{\otimes}\, (b \,{\otimes}\, c)$ for $a,b,c \,{\in}\, \mc{C}$, called the \emph{associator} or associativity constraint, and with a distinguished \textit{unit object} $\mathbf{1} \,{\in}\, \mc{C}$. In terms of the simple objects, the functor $\otimes\colon \mc{C} \,{\times}\, \mc{C} \,{\rightarrow}\, \mc{C}$ can be expressed as
\begin{equation}
a \otimes b \cong \bigoplus_{c \in \mc{C}} N_{ab}^{c}\, c \,,
\end{equation}
where the multiplicities $N_{ab}^c$ are called the fusion rules of $\mc C$. The associator $\F{0}$ has an inverse which we denote as $\Fi{0}$, and can be expressed in terms of the simple objects in $\mc{C}$ as follows (using the diagrammatic language to be read from top to bottom):
\begin{equation}
\includegraphics[valign=c,page=1]{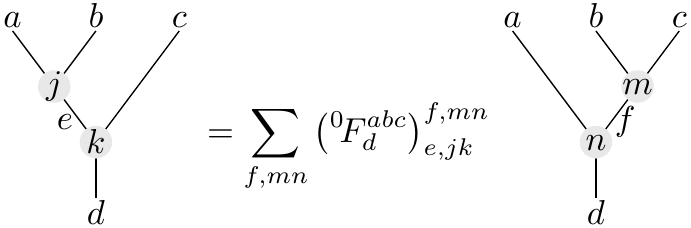}.
\end{equation}
Here, $m,n$ and $j,k$ are multiplicity labels. The associators $\F{0}$ and $\Fi{0}$ are required to satisfy the pentagon equations. In our context, we have to deal with two different fusion categories. We denote the second one by $\mc{D}$, its simple objects by $\alpha,\beta,...\,$, and the fusion rules of $\mc{D}$ by
\begin{equation}
\alpha \otimes \beta \cong \bigoplus_{\gamma\in\mc{D}} N_{\alpha\beta}^{\gamma}\, \gamma.
\end{equation}
We denote the associator of $\mc{D}$ by $\F{4}$ and its inverse by $\Fi{4}$; it satisfies
\begin{equation}
\includegraphics[valign=c,page=5]{figures/MTC}
\end{equation}
and the pentagon equations. We furthermore assume all fusion categories to possess a spherical pivotal structure, allowing us to define a unique dual $\bar{a}$ for every $a \in \mc{C}$ with dimension $d_a = d_{\bar{a}}$; the total dimension $D$ of a fusion category is then defined as $D^2 = \sum_i d_i^2$.

A \textit{braided} fusion category is a fusion category equipped with a braiding $R$, a natural isomorphism $R_{a,b}\colon a \,{\otimes}\, b \,{\xrightarrow{\cong}}\, b \,{\otimes}\, a$ that is compatible with the associators through the hexagon equations. In terms of simple objects, the braiding $R$ can graphically be written as
\begin{equation}
\includegraphics[valign=c,page=6]{figures/MTC}.
\end{equation}
Using the braiding, the topological spin is defined as
\begin{equation}
\theta_a = \theta_{\bar{a}} := \frac{1}{d_a} \includegraphics[valign=c,page=7]{figures/MTC} = \sum_{c}\frac{d_c}{d_a}\tr\left(R_{a,b}^c\right),
\end{equation}
which satisfies the ribbon property
\begin{equation}
R_{a,b}^c R_{b,a}^c = \frac{\theta_c}{\theta_a \theta_b} \mathds{1}.
\end{equation}
At this point, we are ready to define the topological $S$-matrix:
\begin{equation}
S_{ab} := \frac{1}{D^2} \, \includegraphics[valign=c,page=8]{figures/MTC},
\end{equation}
which can be reduced to
\begin{equation}
S_{ab} = \frac{1}{D^2}\sum N_{a\bar{b}}^c \frac{\theta_c}{\theta_a \theta_b} d_c.
\end{equation}
Through some straightforward diagrammatic manipulations \cite{kitaev2006anyons}, one can show
\begin{equation}
\frac{S_{ax} S_{bx}}{S_{1x}} = \sum_c N_{ab}^c S_{cx},
\label{Srep}
\end{equation}
which means that
\begin{equation}
\lambda_x(a) := \frac{S_{ax}}{S_{1x}}
\end{equation}
form representations of the fusion rules for all $x$. A braided fusion category is called \textit{modular} if its $S$-matrix is invertible, in which case Eq. \eqref{Srep} can be recast into the following form:
\begin{equation}
N_{ab}^c = \sum_x \frac{S_{ax}S_{bx}(S^{-1})_{xc}}{S_{1x}},
\end{equation}
which is known as the Verlinde formula.

\subsection{Module categories}

A left \textit{module category} $\mc{M}$ over a fusion category $\mc{C}$ is a (linear, semisimple) category $\mc{M}$ with a functor $\lact\colon \mc{C} \,{\times}\, \mc{M} \,{\rightarrow}\, \mc{M}$ (called the \textit{action} of $\mc{C}$ on $\mc{M}$) and a natural isomorphism $\F{1}\colon (a \,{\otimes}\, b) \lact A \,{\xrightarrow{\,\cong\,}}\, a \lact (b \lact A)$ with $a,b \,{\in}\, \mc{C}$ and $A \,{\in}\, \mc{M}$.

In terms of simple objects of $\mc{M}$ we write
\begin{equation}
a \lact A \cong \bigoplus_{B\in\mc{M}} N_{aA}^B\, B.
\end{equation}
The isomorphism $\F{1}$ has an inverse denoted as $\Fi{1}$ can be expressed as follows:
\begin{equation}
\includegraphics[valign=c,page=2]{figures/MTC}
\end{equation}
and it satisfies a mixed pentagon equation that enforces compatibility with $\F{0}$.

Analogously, a \textit{right} module category $\mc{M}$ over the fusion category $\mc{D}$ is a category $\mc{M}$ with a right action functor $\ract\colon \mc{M} \,{\times}\, \mc{D} \,{\rightarrow}\, \mc{M}$ and a natural isomorphism $\F{3}\colon (A \ract \alpha) \ract \beta \,{\xrightarrow{\,\cong\,}}\, A \ract (\alpha \,{\otimes} \beta)$ with $A \,{\in}\, \mc{M}$ and $\alpha,\beta \,{\in}\, \mc{D}$. In terms of simple objects, the right action $\ract$ is expressed as
\begin{equation}
A \ract \alpha \cong \bigoplus_{B\in\mc{M}} N_{A\alpha}^B\, B,
\end{equation}
while the isomorphism $\F{3}$ with inverse $\Fi{3}$ is described as
\begin{equation}
\includegraphics[valign=c,page=4]{figures/MTC},
\end{equation}
satisfying a pentagon equation expressing compatibility with $\F{4}$. In contrast to fusion categories, module categories do not come with an intrinsic duality, and therefore one can not a priori define dimensions for its objects. We can however take the duals $\bar{A}$ of $A \in \mc{M}$ to be objects in the opposite category $\mc{M}^\text{op}$, and define a generalized spherical structure on the combination of $\mc{M}$ and $\mc{M}^\text{op}$ that allows us to also define dimensions $d_A$ for the objects in $\mc{M}$.

\subsection{Bimodule categories}

A $(\mc{C},\mc{D})$-bimodule category $\mc{M}$ over a pair of fusion categories $\mc C$ and $\mc D$ is a category $\mc{M}$ with additional structure such that $\mc M$ is a left $\mc{C}$-module category a right $\mc{D}$-module category and such that there is a natural isomorphism $\F{2}\colon (a \lact A) \ract \alpha \,{\xrightarrow{\,\cong\,}}\, a \lact (A \ract \alpha)$ for $a \,{\in}\, \mc C$, $A \,{\in}\, \mc M$ and $\alpha \,{\in}\, \mc D$. In terms of simple objects, this imposes the compatibility condition
\begin{equation}
\sum_C N_{aA}^{C} N_{C\alpha}^B = \sum_D N_{aD}^B N_{A\alpha}^D
\end{equation}
on the left and right action functors. The isomorphism $\F{2}$ with inverse $\Fi{2}$ gives
\begin{equation}
\includegraphics[valign=c,page=3]{figures/MTC}
\end{equation}
and it satisfies two mixed pentagon equations for compatibility with $\F{1}$ and $\F{3}$.


\section{String-net models and tensor networks}
\label{App:StringNets}

The critical lattice models we consider in the main text are obtained as the overlap of a product state and a state with topological order. String-net models, as introduced by Levin and Wen, are believed to be a microscopic realization of all non-chiral topological orders in 2+1D. The input of these models is a fusion category $\mc{D}$, and tensor network descriptions for their ground states have been constructed in \cite{buerschaper2009explicit, gu2009tensor} using the F-symbols of this fusion category. Recently, these constructions were generalized using bimodule categories; we give a brief overview here, and refer to \cite{lootens2020matrix} for details.

\subsection{String-net models}
The Hilbert space of the string-net models is most easily defined as degrees of freedom $\alpha,\beta,... \in \mc{D}$ living on the edges of a hexagonal lattice. The Hamiltonian
\begin{equation}
H = - \sum_v A_v - \sum_p B_p
\end{equation}
is a sum of commuting terms, consisting of vertex operators $A_v$ and plaquette operators $B_p$. The vertex operators $A_v$ enforce the fusion rules at the vertices by penalizing configurations $\alpha,\beta\,\gamma$ for which $N_{\alpha\beta}^\gamma = 0$, while the plaquette operator is diagonalized by taking superpositions of all closed loop types around a plaquette weighed by their quantum dimensions. Diagonalizing these two operators simultaneously gives a ground state that is a superposition of all closed loop configurations that satisfy the fusion rules at every vertex. These ground states are RG fixed points and satisfy various properties expressing their scale invariance, the most prominent one being that one can recouple a given configuration using an F-move:
\begin{equation}
\includegraphics[valign=c,page=7]{figures/TN},
\label{RG}
\end{equation}
where $\F{4}$ is the associator of the fusion category $\mc{D}$ that serves as the input of the string-net construction.
\subsection{PEPS ground state and MPO symmetries}
In \cite{lootens2020matrix}, it was argued that a complete tensor network description for a string-net model $\mc{D}$ is determined by all possible choices of $(\mc{C},\mc{D})$ bimodule categories. A PEPS description for the string-net ground state can be obtained by using the following tensors:
\begin{equation}
\includegraphics[page=5, valign = c]{figures/TN} := \left(\frac{d_\alpha d_\beta}{d_\gamma}\right)^{\frac{1}{4}}\frac{ \left(\F{3}^{A\alpha\beta}_B\right)^{\gamma,km}_{C,jn}}{\sqrt{d_C}}, \quad
\includegraphics[page=6, valign = c]{figures/TN} := \left(\frac{d_\alpha d_\beta}{d_\gamma}\right)^{\frac{1}{4}}\frac{ \left(\Fi{3}^{A\alpha\beta}_B\right)^{\gamma,km}_{C,jn}}{\sqrt{d_C}},
\label{peps}
\end{equation}
where we have used the $\F{3}$ symbols of a right $\mc{D}$ module category $\mc{M}$. The fact that these PEPS tensors describe some topologically ordered state can be characterized by the existence of virtual MPO symmetries satisfying a local pulling-through condition:
\begin{equation}
\includegraphics[valign=c,page=8]{figures/TN},
\end{equation}
where the label $a$ is a simple object of a different fusion category $\mc{C}$. The pulling-through condition is solved by the following left and right handed MPO tensors:
\begin{equation}
		\includegraphics[page=3, valign = c]{figures/TN} := \frac{\left(\F{2}^{aC\alpha}_B\right)^{D,nk}_{A,jm}}{\sqrt{d_Ad_D}}, \qquad
		\includegraphics[page=4, valign = c]{figures/TN} := \frac{\left(\Fi{2}^{aC\alpha}_B\right)^{D,nk}_{A,jm}}{\sqrt{d_Ad_D}},
		\label{mpo}
\end{equation}
which depend on the $\F{2}$ symbols of the $(\mc{C},\mc{D})$-bimodule category $\mc{M}$. These MPO tensors can be used to construct closed loops of MPO symmetries which we denote $O_a^L$ and $O_a^R$ for the left and right handed case, respectively. The product of two such MPO symmetries is again an MPO symmetry:
\begin{equation}
O_a^L O_b^L = \sum_c N_{ab}^c O_c^L, \quad O_a^R O_b^R = \sum_c N_{ab}^c O_c^R,
\end{equation}
which can be locally expressed as the existence of a fusion tensor satisfying the zipper condition:
\begin{equation}
\includegraphics[valign=c,page=9]{figures/TN}.
\end{equation}
Explicit fusion tensors are obtained using the following definitions:
\begin{equation}
\includegraphics[page=1, valign = c]{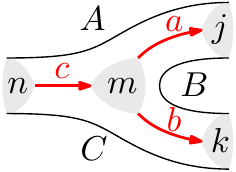} := \left(\frac{d_a d_b}{d_c}\right)^{\frac{1}{4}}\frac{\left(\F{1}^{abC}_A\right)^{B,kj}_{c,mn}}{\sqrt{d_B}}, \quad
\includegraphics[page=2, valign = c]{figures/TN} := \left(\frac{d_a d_b}{d_c}\right)^{\frac{1}{4}}\frac{\left(\Fi{1}^{abC}_A\right)^{B,kj}_{c,mn}}{\sqrt{d_B}}, \label{fusion}
\end{equation}
which use the $\F{1}$-symbol of the left $\mc{C}$-module category. Finally, this fusion process is associative, meaning that the fusion tensors satisfy the recoupling identity
\begin{equation}
\includegraphics[valign=c,page=10]{figures/TN},
\end{equation}
where $\F{0}$ is the associator of the fusion category $\mc{C}$. Using the following states to terminate or create an trivial MPO line
\begin{equation}
\includegraphics[valign=c,page=11]{figures/TN} \enspace = \enspace \includegraphics[valign=c,page=12]{figures/TN} \enspace = \delta_{n,1}\delta_{B,A} \sqrt{d_A},
\label{terminate}
\end{equation}
we can make the addition or removal of a trivial MPO to some other MPO symmetry trivial:
\begin{equation}
\begin{gathered}
\includegraphics[valign=c,page=15]{figures/TN} \quad = \quad \includegraphics[valign=c,page=13]{figures/TN} \quad = \delta_{k,n}\delta_{b,c} \enspace \includegraphics[valign=c,page=17]{figures/TN},\\[1em]
\includegraphics[valign=c,page=16]{figures/TN} \quad = \quad \includegraphics[valign=c,page=14]{figures/TN} \quad = \delta_{j,n}\delta_{a,c} \enspace \includegraphics[valign=c,page=17]{figures/TN}.
\end{gathered}
\label{tubeCap}
\end{equation}
Using the particular states in Eq. \eqref{terminate} we can define the following cup and cap operators
\begin{align}
	\includegraphics[valign=c,page=25]{figures/TN}  &= \varkappa_a \enspace \includegraphics[valign=c,page=18]{figures/TN} := \enspace \includegraphics[valign=c,page=24]{figures/TN},\\[1em]
	\enspace \includegraphics[valign=c,page=19]{figures/TN}  &= \varkappa_a^* \enspace \includegraphics[valign=c,page=22]{figures/TN}  := \enspace \includegraphics[valign=c,page=21]{figures/TN}.
\end{align}
where $\varkappa_a$ is a phase known as the \textit{Frobenius-Schur} indicator, defined by
\begin{equation}
\left(\F{0}^{a\bar{a}a}_a\right)^{\mathbf{1},11}_{\mathbf{1},11} = \frac{\varkappa_a}{d_a}.
\end{equation}
If the fusion category $\mc{C}$ associated to the MPO symmetries is unitary, closed loops of MPO symmetries $O_a^R$ satisfy $(O_a^R)^\dag = O_{\bar{a}}^R = O_a^L$. At the level of the MPO tensors, the cup and cap operators realize this by providing a relation between the left and right handed MPO tensors by acting as a gauge transformation on the internal MPO legs:
\begin{equation}
\includegraphics[valign=c,page=27]{figures/TN} \enspace = \enspace \includegraphics[valign=c,page=4]{figures/TN}.
\end{equation}
The above relation can be generalized to the nonunitary case, but the MPO symmetries are no longer closed under Hermitian conjugation so the dagger must be changed for a different operation; in \cite{lootens2020galois}, this was done by using the transpose instead.\\
In a completely analogous way, cup and cap tensors can also be defined by terminating one of the legs of the PEPS tensor, as opposed to the fusion tensor. This allows us to again relate the two types of MPO tensors, now by acting with a gauge transformation on the external MPO legs:
\begin{equation}
\includegraphics[valign=c,page=28]{figures/TN} \enspace = \enspace \includegraphics[valign=c,page=29]{figures/TN}
\label{virtual_transpose}
\end{equation}

\subsection{Torus ground states and tube algebra} \label{Sec:App:TorusOcneanu}
One of the most prominent features of a system exhibiting topological order is a ground state degeneracy that is independent of the system size but instead depends solely on the topology of the underlying manifold. This property is naturally explained in the tensor network formalism by the fact that the PEPS tensors for such models have nontrivial MPO symmetries. To see this, consider defining the PEPS wave function on a torus and adding a closed loop MPO symmetries along the non-contractible cycles of the torus:
\begin{equation}
\ket{\psi} = \enspace \includegraphics[valign=c,page=30,scale=0.6]{figures/TN}
\end{equation}
The state $\ket{\psi}$ has the same energy regardless of which closed MPO symmetries are present on the virtual level. Indeed, the Hamiltonian is a sum of local terms, and since the MPO symmetries can be moved freely through the lattice, we can evaluate all terms of the Hamiltonian such that they never see the presence of these MPOs by moving them appropriately. Of course, this reasoning requires that the tensor at the intersection of the two MPO symmetries along the two different cycles of the torus can be moved freely through the lattice as well. Using the fusion tensors, we can define such a tensor as
\begin{equation}
\includegraphics[valign=c,page=31,scale=1]{figures/TN},
\end{equation}
which is in fact the only possibility. Being more explicit, we can define \textit{tubes} as
\begin{equation}
\mc{T}_{t_1} := \mc{T}_{abc}^{d,jk} = \includegraphics[valign=c,page=32,scale=1]{figures/TN},
\end{equation}
where it is understood that the red lines labeled by $d$ are connected allowing for an arbitrary number of MPO tensor insertions along the connecting line and $t_1$ is shorthand for the labels in the tube. When interpreted as matrices from left to right, these tubes can be multiplied and form a closed algebra called the \textit{tube algebra}:
\begin{equation}
\mc{T}_{t_1} \mc{T}_{t_2} = \sum_{t_3} \Omega_{t_1,t_2}^{t_3} \mc{T}_{t_3},
\end{equation}
with a structure factor $\Omega$ that can be worked out using $\F{0}$ moves. On the torus, only tubes with $a=c$ will contribute, and one might wonder why we allow for tubes with $a \neq c$ at all. It is important however to note that not all of these tubes will give independent torus ground states. A correct characterization of the ground states is instead given by the central idempotents $P_i$ of the tube algebra, satisfying
\begin{gather}
	P_i = \sum_{abd,jk} c_i^{abd,jk} \mathcal{T}_{aba}^{d,jk}, \label{eq:idempotents}\\
	\quad P_i P_j = \delta_{ij} P_i, \quad [ \mathcal{T}_{abc}^{d,jk},P_i ] = 0.
\end{gather}
The tubes with $a \neq c$ appear in the requirement that the central idempotents commute with all elements from the tube algebra. These central idempotents correspond to minimally entangled states on the torus \cite{zhang2012quasiparticle}, and simultaneously provide a characterization of the anyonic excitations of the topological phase \cite{bultinck2017anyons}. These excitations are known to be classified by the monoidal center $Z(\mc{C})$ of the fusion category $\mc{C}$, and therefore the tube algebra and its idempotents provide an explicit construction of $Z(\mc{C})$. In the case of a modular fusion category, elements of $Z(\mc{C})$ can be labeled by tuples $(a,\bar{b})$ with $a,b$ simple objects in $\mc{C}$. When constructing CFT partition functions, which require a modular fusion category, the different idempotents $P_{a,\bar{b}}$ will correspond to the different bilinear combinations of characters $\chi_a \overline{\chi}_b$.

\subsection{MPO intertwiners and tube maps} \label{Sec:App:IntertwinersTubeMaps}
For a fixed string-net model $\mc{D}$, there are as many different PEPS representations as there are right $\mc{D}$-module categories $\mc{M}$. Importantly, these different representations can not be distinguished from each other by any local operator acting on the physical legs, but on the torus they might represent different ground states. This fact can be made very explicit by the existence of \textit{MPO intertwiners}, MPOs acting on the virtual legs of the PEPS that intertwine it from one representation to another. The simplest example is where we consider the PEPS representations based on the one hand on $\mc{D}$ as a right-$\mc{D}$ module category over itself (shaded green), and on the other hand a generic right $\mc{D}$-module category $\mc{M}$ (shaded blue). These MPO intertwiners should be freely movable through the lattice:
\begin{equation}
	\includegraphics[valign=c,page=34,scale=1]{figures/TN},
\end{equation}
and MPO symmetries should be able to end on them at a fusion/splitting tensor that can also be moved along the intertwiner
\begin{equation}
	\includegraphics[valign=c,page=35,scale=1]{figures/TN},
	\label{fusionEquation1}
\end{equation}
\begin{equation}
	\includegraphics[valign=c,page=36,scale=1]{figures/TN}.
\end{equation}
Additionally, these tensors satisfy various recoupling identities:
\begin{equation}
	\includegraphics[valign=c,page=43,scale=1]{figures/TN},
	\label{ladder_recouple}
\end{equation}
\begin{equation}
	\includegraphics[valign=c,page=44,scale=1]{figures/TN},
\end{equation}
\begin{equation}
	\includegraphics[valign=c,page=45,scale=1]{figures/TN}.
		\label{fusionEquation4}
\end{equation}
These equations are solved by the following explicit expressions for the MPO intertwiners:
\begin{equation}
		\includegraphics[scale=1,valign=c,page=37]{figures/TN} := \frac{\left(\F{3}^{C\alpha\gamma}_B\right)^{\beta,nk}_{A,jm}}{\sqrt{d_A d_\beta}}, \qquad
		\includegraphics[scale=1,valign=c,page=38]{figures/TN} := \frac{\left(\Fi{3}^{C\alpha\gamma}_B\right)^{\beta,nk}_{A,jm}}{\sqrt{d_A d_\beta}}, \label{intertwiner}
\end{equation}
and similarly defined fusion/splitting tensors \cite{lootens2020matrix}. On the torus, these MPO intertwiners can be used to construct a linear map between two different tube algebras. Consider the ground state $\ket{\mc{T}_{\alpha\beta\alpha}^{\gamma,jk}}$ of a string-net model $\mc{D}$ with the presence of the tube $\mc{T}_{\alpha\beta\alpha}^{\gamma,jk}$ where we take $\mc{D}$ as a right $\mc{D}$-module category over itself and consequently the MPOs are labeled by objects in $\mc{D}$. We now imagine inserting a closed bubble bounded by an MPO intertwiner $A$ of the PEPS in the representation determined by taking $\mc{M}$ as a right $\mc{D}$-module category:
\begin{equation}
	d_A \ket{\mc{T}_{\alpha\beta\alpha}^{\gamma,jk}} := \includegraphics[valign=c,page=46,scale=1]{figures/TN},
\end{equation}
where in the blue shaded region the MPOs are labeled by $\mc{C}$. By growing this bubble and fusing it with the tube $\mc{T}_{\alpha\beta\alpha}^{\gamma,jk}$ we obtain the following state:
\begin{equation}
	d_A \ket{\mc{T}_{\alpha\beta\alpha}^{\gamma,jk}} = \sum_{abc,mn} T_{\alpha\beta\gamma,jk;A}^{abc,mn} \ket{\mc{T}_{aba}^{c,mn}} = \sum_{abc,mn} T_{\alpha\beta\gamma,jk;A}^{abc,mn} \, \includegraphics[valign=c,page=47,scale=1]{figures/TN},
	\label{tubeMap}
\end{equation}
where the coefficients $T_{\alpha\beta\gamma,jk;A}^{abc,mn}$ can be determined using various $F$ moves. This means that the MPO intertwiners provide a way to go from the tube algebra based on MPOs $\mc{D}$ to the tube algebra based on MPOs $\mc{C}$. If the $(\mc{C},\mc{D})$-bimodule category $\mc{M}$ is invertible (which is always the case in this work), this map is an isomorphism between $Z(\mc{D})$ and $Z(\mc{C})$.

\subsection{Ladder algebra} \label{Sec:App:ladder}
Besides the tube algebra, which characterizes the topological sectors on the torus, we use a second kind of algebra in the main text when studying cylinder partition functions. This algebra is the \textit{ladder algebra}, where we define a \textit{ladder} as
\begin{equation}
	\mc{L}_{l_1} := \mc{L}_{AB,CD}^{a,jk} = \includegraphics[valign=c,page=33,scale=1]{figures/TN}.
	\label{ladderDefinition}
\end{equation}
Multiplication of these ladders is defined by stacking them vertically:
\begin{equation}
	\mc{L}_{l_1} \mc{L}_{l_2} = \sum_{l_3} \Lambda_{l_1,l_2}^{l_3} \mc{L}_{l_3}
\end{equation}
with a structure factor $\Lambda$ that can be worked out using $\F{1}$ transformations. Analogous to the tube algebra, we can find simple idempotents of the ladder algebra, to which only ladders with $A=C$ and $B=D$ will contribute on the cylinder. These simple idempotents have the distinguished property that when we fuse the closed rings labeled by $A$ and $B$ together and recoupling everything using $\F{1}$ transformations as in Eq. \eqref{ladder_recouple}, the result is a closed MPO $O_\alpha$, with $\alpha \in \mc{D}$. This is due to the fact that the ladder algebra provides a construction of the relative Deligne product $\mc{M}^\text{op} \, \boxtimes_\mc{C} \, \mc{M} = \mc{D}$ \cite{barter2019domain}, and by computing the simple idempotents of the ladder algebra we simply recover the simple objects $\alpha \in \mc{D}$.


\subsection{Reflection operator}
\label{App:reflection}

In the main text we defined the Klein bottle partition function as
\begin{equation}
	Z^\mathcal{K} = \tr\left(\mathcal{R}T^{L_y}\right),
\end{equation}
where $\mathcal{R}$ is the spatial reflection operator. To use $\mc{R}$ in combination with the MPO symmetries, we need to study the interplay between the two. One can convince oneself that the action of pulling $\mc{R}$ through an MPO is given by the following:
\begin{equation}
	\includegraphics[valign=c,page=1]{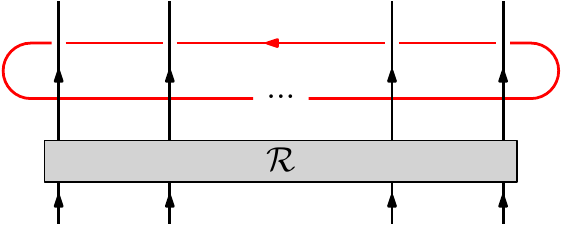}
	\quad = \quad \includegraphics[valign=c,page=2]{figures/ReflectionMPO},
\end{equation}
i.e. it is equivalent to transposing the individual MPO tensors along the virtual direction. For a unitary fusion category, using Eq. \eqref{virtual_transpose}, we can show the following relation:
\begin{equation}
	\includegraphics[valign=c,page=3]{figures/ReflectionMPO} = \left(\includegraphics[valign=c,page=4]{figures/ReflectionMPO}\right)^*,
	\label{mpo_reflect}
\end{equation}
meaning that pulling the operator $\mc{R}$ through an MPO amounts to a gauge transformation on the external MPO legs, and taking the complex conjugate. A similar argument applies to the transfer matrix itself, and since it is Hermitian, we get
\begin{equation}
	\includegraphics[valign=c,page=5]{figures/ReflectionMPO} = \left(\includegraphics[valign=c,page=6]{figures/ReflectionMPO}\right)^*.
	\label{transfer_reflect}
\end{equation}
In the presence of a twist in the transfer matrix, we have to consider the action of $\mc{R}$ on the MPO running perpendicular to $\mc{R}$, which leads to the following relation:
\begin{equation}
	\includegraphics[valign=c,page=7]{figures/ReflectionMPO} = \left(\includegraphics[valign=c,page=8]{figures/ReflectionMPO}\right)^*.
	\label{twist_reflect}
\end{equation}
Finally, in order to project the Klein bottle spectrum we need the central idempotents of the tube algebra. Since these are Hermitian, we get a similar relation:
\begin{equation}
	\includegraphics[valign=c,page=9]{figures/ReflectionMPO} = \left(\includegraphics[valign=c,page=10]{figures/ReflectionMPO}\right)^*.
	\label{idem_reflect}
\end{equation}
For the non-unitary Yang-Lee fusion category, the Hermitian conjugate is replaced by the transpose, so the complex conjugates in Eqs. \eqref{mpo_reflect}-\eqref{idem_reflect} are not needed. These calculations are similar as the calculation done in \cite{williamson2014matrix} and \cite{williamson2017symmetry} to prove their analogues for Hermitian conjugation, but now simply in the virtual direction.

\section{Fusion category data for specific models}
\label{App:data}

In this appendix, we give some data for the relevant fusion categories used in the main text and specifically in the result section. We focus mainly on the idempotent tables, expressing the topological superselection sectors (anyons) as a linear combination of the Ocneanu tube algebra elements (\ref{eq:idempotents}). This table forms the basis for our numerical simulations, singling out CFT characters in twisted transfer matrices. For the orbifold procedure, projecting the tetracritical Ising model on the Potts model, we give the explicit linear combinations of the tube elements, which is exactly what allows us to write down general twisted partition functions in terms of tetracritical Ising CFT characters.

\subsection{The Ising fusion category}

The Ising fusion category has a $\mathds{Z}_2$-grading on the objects $\{\bf{1},\psi\}\oplus\{\sigma\}$, with fusion rules $\psi \otimes \psi = \bf{1}$, $\sigma \otimes \sigma = \bf{1} + \psi$ and $\psi \otimes \sigma = \sigma \otimes \psi = \sigma$, and nontrivial $F$-symbols

\begin{align*}
[F^{\sigma\sigma\sigma}_{\sigma}]_{ij}=\frac{1}{\sqrt{2}}\begin{pmatrix}
1 & 1\\
1 & -1
\end{pmatrix},~ [F^{\sigma\psi\sigma}_{\psi}]_{\sigma}^{\sigma}=[F^{\psi\sigma\psi}_{\sigma}]_{\sigma}^{\sigma}=-1.
\label{IsingFsymbols}
\end{align*}

The corresponding quantum dimensions of the simple objects are $\{1,1,\sqrt{2}\}$, with total quantum dimension $D = 2$.
The category is modular and the idempotents can be labeled by the tensor product $\text{Ising} \boxtimes \overline{\text{Ising}}$ (Drinfeld center). The idempotents are given in Table \ref{IsingTable}.

\begin{center}
	\begin{table}[ht]
		\centering
		\resizebox{0.7\textwidth}{!}{
			\begin{tabular}{ |c|cccccccccc| }
				\hline
				&$\mathcal{T}_{\bf{1},1}^{\bf{1}}$&$\mathcal{T}_{\bf{1},\psi}^{\psi}$&$\mathcal{T}_{\psi,\psi}^{\bf{1}}$&$\mathcal{T}_{\psi,\bf{1}}^{\psi}$&$\mathcal{T}_{\sigma,\sigma}^{\bf{1}}$&$\mathcal{T}_{\sigma,\sigma}^{\psi}$&$\mathcal{T}_{\bf{1},\sigma}^{\sigma}$&$\mathcal{T}_{\psi,\sigma}^{\sigma}$&$\mathcal{T}_{\sigma,\bf{1}}^{\sigma}$&$\mathcal{T}_{\sigma,\psi}^{\sigma}$\\
				\hline
				$P^{\bf{1}\overline{\bf{1}}}$&1&1&&&&&$\sqrt{2}$&&&\\
				$P^{\psi\overline{\psi}}$&1&1&&&&&$-\sqrt{2}$&&&\\
				$P^{\psi\overline{\bf{1}}}$&&&1&-1&&&&$-i\sqrt{2}$&&\\
				$P^{\bf{1}\overline{\psi}}$&&&1&-1&&&&$i\sqrt{2}$&&\\
				$P^{\sigma\overline{\sigma}}$&2&-2&2&2&&&&&&\\
				$P^{\sigma\overline{\bf{1}}}$&&&&&1&$-i$&&&$e^{-\frac{\pi}{8}i}$&$e^{-\frac{3\pi}{8}i}$\\
				$P^{\sigma\overline{\psi}}$&&&&&1&$-i$&&&$e^{\frac{7\pi}{8}i}$&$e^{-\frac{5\pi}{8}i}$\\
				$P^{\bf{1}\overline{\sigma}}$&&&&&1&$i$&&&$e^{-\frac{\pi}{8}i}$&$e^{-\frac{3\pi}{8}i}$\\
				$P^{\psi\overline{\sigma}}$&&&&&1&$i$&&&$e^{-\frac{7\pi}{8}i}$&$e^{\frac{5\pi}{8}i}$\\
				\hline
		\end{tabular}}
		\caption{Idempotent table for the Ising model up to the normalization $1/D^2$.}
		\label{IsingTable}
	\end{table}
\end{center}

\subsection{The Yang-Lee fusion category}

The Yang-Lee model is obtained from the Fibonacci fusion category, which has two simple objects $\{\bf{1},\tau\}$ and nontrivial fusion rule ${\tau \times \tau = \bf{1} + \tau}$, by choosing the non-unitary solution of the pentagon equation. The non-trivial $F$-symbols are:

\begin{align}
[F^{\tau\tau\tau}_{\tau}]_{ij}=\frac{1}{\phi}\begin{pmatrix}
1 & \sqrt{\phi}\\
\sqrt{\phi} & -1
\end{pmatrix},
\end{align}

with $\phi = \frac{1}{2}(1-\sqrt{5})$ the negative solution of the golden ratio equation $\phi^2=1+\phi$. The corresponding quantum dimensions are $\{1,\phi\}$ and the total quantum dimension $D^2 = 2+\phi$. The idempotents (labeled by the double $\text{Fib} \boxtimes \overline{\text{Fib}}$) are given in table \ref{YangLeeTable}.

\begin{center}
	\begin{table}[ht]
		\centering
		\resizebox{0.5\textwidth}{!}{
			\begin{tabular}{ |c|ccccc| }
				\hline
				&$\mathcal{T}_{\bf{1},\bf{1}}^{\bf{1}}$&$\mathcal{T}_{\bf{1},\tau}^{\tau}$&$\mathcal{T}_{\tau,\tau}^{\bf{1}}$&$\mathcal{T}_{\tau,\bf{1}}^{\tau}$&$\mathcal{T}_{\tau,\tau}^{\tau}$\\
				\hline
				$P^{\bf{1}\overline{\bf{1}}}$&$1$&$\phi$&&&\\
				$P^{\tau\overline{\tau}}$&$\phi^2$&$-\phi$&$\phi$&$\phi$&$\frac{1}{\sqrt{\phi}}$\\
				$P^{\tau\overline{\bf{1}}}$&&&1&$e^{\frac{2\pi}{5}i}$&$\sqrt{\phi}e^{\frac{\pi}{5}i}$\\
				$P^{\bf{1}\overline{\tau}}$&&&1&$e^{-\frac{2\pi}{5}i}$&$\sqrt{\phi}e^{-\frac{\pi}{5}i}$\\
				\hline
		\end{tabular}}
		\caption{Idempotent table for the Yang-Lee model up to the normalization $1/D^2$.}
		\label{YangLeeTable}
	\end{table}
\end{center}

\subsection{The three-state Potts model categories}
\label{pottsData}

For the three-state Potts model as the $D_4$ RSOS model, we need three categories: a modular fusion category $\mc{D} = |\text{su}(2)_4|$, a right $|\text{su}(2)_4|$-module category $\mc{M}_{TY}$, and the fusion category $\mc{C}_{S_3}$. The modular fusion category $|\text{su}(2)_4|$ with 5 simple objects $\{\it{0,1/2,1,3/2,2}\}$ is part of the series of $|\text{su}(2)_k|$ modular fusion categories, the fusion rules are given by
\begin{equation}
	\begin{tabular}{|c|c c c c c|}
		\hline
		$\otimes$ & $\it{0}$ & $\it{1/2}$&$\it{1}$&$\it{3/2}$&$\it{2}$\\
		\hline
		$\it{0}$ & $\it{0}$ & $\it{1/2}$&$\it{1}$&$\it{3/2}$&$\it{2}$\\
		$\it{1/2}$ & $\it{1/2}$ & $\it{0+1}$&$\it{1/2+3/2}$&$\it{1+2}$&$\it{3/2}$\\
		$\it{1}$ & $\it{1}$ & $\it{1/2+3/2}$&$\it{0+1+2}$&$\it{1/2+3/2}$&$\it{1}$\\
		$\it{3/2}$ & $\it{3/2}$ & $\it{1+2}$&$\it{1/2+3/2}$&$\it{0+1}$&$\it{1/2}$\\
		$\it{2}$ & $\it{2}$ & $\it{3/2}$&$\it{1}$&$\it{1/2}$&$\it{0}$\\
		\hline
	\end{tabular}
\end{equation}
and the $F$-symbols can be found in, e.g., \cite{ardonne2010clebsch}, with the caveat that we take the solution of the pentagon equation that has all positive Frobenius-Schur indicators. In the context of the three-state Potts model, we refer to the $F$-symbols of $|\text{su}(2)_4|$ as $\F{4}$. The central idempotents of the tube algebra, which are elements of $Z(|\text{su}(2)_k|) = |\text{su}(2)_k| \boxtimes \overline{|\text{su}(2)_k|}$, can be found in Tables \ref{TetraTable1} - \ref{TetraTable4}.\\

The category $\mc{M}_\text{TY}$ is a right $|\text{su}(2)_4|$-module category and has 4 simple objects denoted as $\{A,\sigma,B,C\}$, where the action of $|\text{su}(2)_4|$ on $\mc{M}_\text{TY}$ is given by
\begin{equation}
\begin{tabular}{|c|c c c c c|}
\hline
$\ract$ & $\it{0}$ & $\it{1/2}$&$\it{1}$&$\it{3/2}$&$\it{2}$\\
\hline
$A$ & $A$ & $\sigma$&$B+C$&$\sigma$&$A$\\
$\sigma$ & $\sigma$ & $A+B+C$&$2 \sigma$&$A+B+C$&$\sigma$\\
$B$ & $B$ & $\sigma$&$A+C$&$\sigma$&$B$\\
$C$ & $C$ & $\sigma$&$A+B$&$\sigma$&$C$\\
\hline
\end{tabular}
\end{equation}
The module $F$-symbols of $\mc{M}_\text{TY}$ as a right $|\text{su}(2)_4|$-module category are denoted as $\F{3}$, and can be found in \cite{runkel2000structure}; additionally, they can be found in the supplementary material.\\

The fusion category $\mc{C}_{S_3}$ of topological defects has 8 simple objects with fusion rules
\begin{equation}
\begin{tabular}{|c|c c c c c c c c|}
\hline
$\otimes$ & $A^+$ & $\sigma^+$ & $B^+$ & $C^+$ & $A^-$ & $\sigma^-$ & $B^-$ & $C^-$\\
\hline
$A^+$ & $A^+$ & $\sigma^+$ & $B^+$ & $C^+$ & $A^-$ & $\sigma^-$ & $B^-$ & $C^-$\\
$\sigma^+$ & $\sigma^+$ & $A^+ + B^+ + C^+$ & $\sigma^+$ & $\sigma^+$ & $\sigma^-$ & $A^- + B^- + C^-$ & $\sigma^-$ & $\sigma^-$\\
$B^+$ & $B^+$ & $\sigma^+$ & $C^+$ & $A^+$ & $C^-$ & $\sigma^-$ & $A^-$ & $B^-$\\
$C^+$ & $C^+$ & $\sigma^+$ & $A^+$ & $B^+$ & $B^-$ & $\sigma^-$ & $C^-$ & $A^-$\\
$A^-$ & $A^-$ & $\sigma^-$ & $B^-$ & $C^-$ & $A^+$ & $\sigma^+$ & $B^+$ & $C^+$\\
$\sigma^-$ & $\sigma^-$ & $A^- + B^- + C^-$ & $\sigma^-$ & $\sigma^-$ & $\sigma^+$ & $A^+ + B^+ + C^+$ & $\sigma^+$ & $\sigma^+$\\
$B^-$ & $B^-$ & $\sigma^-$ & $C^-$ & $A^-$ & $C^+$ & $\sigma^+$ & $A^+$ & $B^+$\\
$C^-$ & $C^-$ & $\sigma^-$ & $A^-$ & $B^-$ & $B^+$ & $\sigma^+$ & $C^+$ & $A^+$\\
\hline
\end{tabular}
\end{equation}
The $F$-symbols of $\mc{C}_{S_3}$, which we denote as $\F{0}$, can be found in the supplementary material. The central idempotents of the tube algebra, which are elements of $Z(\mc{C}_{S_3})$, can be found in Tables \ref{PottsTableBlock1} - \ref{PottsTableBlock5}. Because $\mc{C}_{S_3}$ and $|\text{su}(2)_4|$ are Morita equivalent, their monoidal center is the same, and we can label elements of $Z(\mc{C}_{S_3})$ with labels in $|\text{su}(2)_k| \boxtimes \overline{|\text{su}(2)_k|}$ as well.\\

The category $\mc{M}_\text{TY}$ is also a left $\mc{C}_{S_3}$-module category, where the action of $\mc{C}_{S_3}$ on $\mc{M}_\text{TY}$ is given by
\begin{equation}
	\begin{tabular}{|c|c c c c|}
		\hline
		$\lact$ & $A$ & $\sigma$ & $B$ & $C$\\
		\hline
		$A^+$ & $A$ & $\sigma$ & $B$ & $C$\\
		$\sigma^+$ & $\sigma$ & $A+B+C$ & $\sigma$ & $\sigma$\\
		$B^+$ & $A$ & $\sigma$ & $C$ & $A$\\
		$C^+$ & $A$ & $\sigma$ & $A$ & $B$\\
		$A^-$ & $A$ & $\sigma$ & $C$ & $B$\\
		$\sigma^-$ & $\sigma$ & $A+B+C$ & $\sigma$ & $\sigma$\\
		$B^-$ & $A$ & $\sigma$ & $B$ & $A$\\
		$C^-$ & $A$ & $\sigma$ & $A$ & $C$\\
		\hline
	\end{tabular}
	\label{Table:N_CMM}
\end{equation}
The module $F$-symbols of $\mc{M}_\text{TY}$ as a left $\mc{C}_{S_3}$-module category are denoted as $\F{1}$ and can be found in the supplementary material.\\

A final set of $F$-symbols denoted as $\F{2}$ is obtained by requiring compatibility between $\mc{M}_\text{TY}$ as a right $|\text{su}(2)_4|$-module category and $\mc{M}_\text{TY}$ as a left $\mc{C}_{S_3}$-module category, and is provided in the supplementary material. \\

Note that the additional $F$-symbols $\F{0}$, $\F{1}$ and $\F{2}$ were not found in the literature and calculated numerically (symbolically). We leave a detailed explanation of such calculations to future work. 

\newpage
\begin{center}
	\begin{table*}[hbt!]
		\resizebox{\textwidth}{!}{
			\begin{tabular}{ |c|ccccccccccccccccccc| }
				\hline
				&$\mathcal{T}_{\it{0},\it{0}}^{\it{0}}$&$\mathcal{T}_{\it{0},\it{1}/\it{2}}^{\it{1}/\it{2}}$&$\mathcal{T}_{\it{0},\it{1}}^{\it{1}}$&$\mathcal{T}_{\it{0},\it{3}/\it{2}}^{\it{3}/\it{2}}$&$\mathcal{T}_{\it{0},\it{2}}^{\it{2}}$&$\mathcal{T}_{\it{1},\it{0}}^{\it{1}}$&$\mathcal{T}_{\it{1},\it{1}/\it{2}}^{\it{1}/\it{2}}$&$\mathcal{T}_{\it{1},\it{1}/\it{2}}^{\it{3}/\it{2}}$&$\mathcal{T}_{\it{1},\it{1}}^{\it{0}}$&$\mathcal{T}_{\it{1},\it{1}}^{\it{1}}$&$\mathcal{T}_{\it{1},\it{1}}^{\it{2}}$&$\mathcal{T}_{\it{1},\it{3}/\it{2}}^{\it{1}/\it{2}}$&$\mathcal{T}_{\it{1},\it{3}/\it{2}}^{\it{3}/\it{2}}$&$\mathcal{T}_{\it{1},\it{2}}^{\it{1}}$&$\mathcal{T}_{\it{2},\it{0}}^{\it{2}}$&$\mathcal{T}_{\it{2},\it{1}/\it{2}}^{\it{3}/\it{2}}$&$\mathcal{T}_{\it{2},\it{1}}^{\it{1}}$&$\mathcal{T}_{\it{2},\it{3}/\it{2}}^{\it{1}/\it{2}}$&$\mathcal{T}_{\it{2},\it{2}}^{\it{0}}$\\
				\hline
				$P^{\it{0}, \overline{\it{0}}}$&$1$&$\sqrt{3}$&$2$&$\sqrt{3}$&$1$&&&&&&&&&&&&&&\\
				$P^{\it{2}, \overline{\it{2}}}$&$1$&$-\sqrt{3}$&$2$&$-\sqrt{3}$&$1$&&&&&&&&&&&&&&\\
				$P^{\it{1/2}, \overline{\it{1/2}}}$&3&3&&$-3$&$-3$&3&$-\frac{3}{2}$&$\frac{3}{2}\sqrt{3}$&$\frac{3}{2}$&&$\frac{3}{2}$&$\frac{3}{2}\sqrt{3}$&$\frac{3}{2}$&3&&&&&\\
				$P^{\it{3/2}, \overline{\it{3/2}}}$&3&$-3$&&$3$&$-3$&3&$\frac{3}{2}$&$-\frac{3}{2}\sqrt{3}$&$\frac{3}{2}$&&$\frac{3}{2}$&$-\frac{3}{2}\sqrt{3}$&$-\frac{3}{2}$&3&&&&&\\
				$P^{\it{1}, \overline{\it{1}}}$&4&&$-4$&&4&4&&&2&4&$-2$&&&$-4$&4&&4&&4\\
				\hline
		\end{tabular}}
		\caption{Idempotent table (block 1) for $|\text{su(2)}_4|$.}
		\label{TetraTable1}
	\end{table*}
\end{center}

\begin{center}
	\begin{table*}[hbt!]
		\resizebox{\textwidth}{!}{
			\begin{tabular}{ |c|cccccccccccccccc| }
				\hline
				&$\mathcal{T}_{\it{1}/\it{2},\it{0}}^{\it{1}/\it{2}}$&$\mathcal{T}_{\it{1}/\it{2},\it{1}/\it{2}}^{\it{0}}$&$\mathcal{T}_{\it{1}/\it{2},\it{1}/\it{2}}^{\it{1}}$&$\mathcal{T}_{\it{1}/\it{2},\it{1}}^{\it{1}/\it{2}}$&$\mathcal{T}_{\it{1}/\it{2},\it{1}}^{\it{3}/\it{2}}$&$\mathcal{T}_{\it{1}/\it{2},\it{3}/\it{2}}^{\it{1}}$&$\mathcal{T}_{\it{1}/\it{2},\it{3}/\it{2}}^{\it{2}}$&$\mathcal{T}_{\it{1}/\it{2},\it{2}}^{\it{3}/\it{2}}$&$\mathcal{T}_{\it{3}/\it{2},\it{0}}^{\it{3}/\it{2}}$&$\mathcal{T}_{\it{3}/\it{2},\it{1}/\it{2}}^{\it{1}}$&$\mathcal{T}_{\it{3}/\it{2},\it{1}/\it{2}}^{\it{2}}$&$\mathcal{T}_{\it{3}/\it{2},\it{1}}^{\it{1}/\it{2}}$&$\mathcal{T}_{\it{3}/\it{2},\it{1}}^{\it{3}/\it{2}}$&$\mathcal{T}_{\it{3}/\it{2},\it{3}/\it{2}}^{\it{0}}$&$\mathcal{T}_{\it{3}/\it{2},\it{3}/\it{2}}^{\it{1}}$&$\mathcal{T}_{\it{3}/\it{2},\it{2}}^{\it{1}/\it{2}}$\\
				\hline
				$P^{\it{3/2}, \overline{\it{2}}}$&$\sqrt{3}e^{-\frac{\pi}{4}i}$&1&$2e^{\frac{2\pi}{3}i}$&$\sqrt{3}e^{-\frac{11\pi}{12}i}$&$\sqrt{3}e^{-\frac{5\pi}{12}i}$&$2e^{\frac{\pi}{6}i}$&$e^{i\frac{\pi}{2}i}$&$\sqrt{3}e^{-\frac{3\pi}{4}i}$&&&&&&&&\\
				$P^{\it{0}, \overline{\it{1/2}}}$&$\sqrt{3}e^{-\frac{3\pi}{4}i}$&1&$2e^{-\frac{2\pi}{3}i}$&$\sqrt{3}e^{-\frac{\pi}{12}i}$&$\sqrt{3}e^{-\frac{7\pi}{12}i}$&$2e^{\frac{-\pi}{6}i}$&$e^{\frac{-\pi}{2}i}$&$\sqrt{3}e^{-\frac{\pi}{4}i}$&&&&&&&&\\
				$P^{\it{2}, \overline{\it{3/2}}}$&$\sqrt{3}e^{\frac{\pi}{4}i}$&1&$2e^{-\frac{2\pi}{3}i}$&$\sqrt{3}e^{\frac{11\pi}{12}i}$&$\sqrt{3}e^{\frac{5\pi}{12}i}$&$2e^{-\frac{\pi}{6}i}$&$e^{-\frac{\pi}{2}i}$&$\sqrt{3}e^{\frac{3\pi}{4}i}$&&&&&&&&\\
				$P^{\it{1/2}, \overline{\it{0}}}$&$\sqrt{3}e^{\frac{3\pi}{4}i}$&1&$2e^{\frac{2\pi}{3}i}$&$\sqrt{3}e^{\frac{\pi}{12}i}$&$\sqrt{3}e^{\frac{7\pi}{12}i}$&$2e^{\frac{\pi}{6}i}$&$e^{\frac{\pi}{2}i}$&$\sqrt{3}e^{\frac{\pi}{4}i}$&&&&&&&&\\
				$P^{\it{1},\overline{\it{1/2}}}$&$2\sqrt{3}e^{\frac{7\pi}{12}i}$&2&$2e^{\frac{\pi}{3}i}$&$\sqrt{3}e^{\frac{\pi}{4}i}$&$\sqrt{3}e^{-\frac{\pi}{4}i}$&$2e^{\frac{5\pi}{6}i}$&$2e^{-\frac{\pi}{2}i}$&$2\sqrt{3}e^{-\frac{11\pi}{12}i}$&$2\sqrt{3}e^{\frac{7\pi}{12}i}$&$2e^{-\frac{\pi}{6}i}$&$2e^{\frac{\pi}{2}i}$&$\sqrt{3}e^{\frac{3\pi}{4}i}$&$\sqrt{3}e^{\frac{\pi}{4}i}$&$2$&$2e^{\frac{\pi}{3}i}$&$2\sqrt{3}e^{\frac{\pi}{12}i}$\\
				$P^{\it{1},\overline{\it{3/2}}}$&$2\sqrt{3}e^{-\frac{5\pi}{12}i}$&2&$2e^{\frac{\pi}{3}i}$&$\sqrt{3}e^{-\frac{3\pi}{4}i}$&$\sqrt{3}e^{\frac{3\pi}{4}i}$&$2e^{\frac{5\pi}{6}i}$&$2e^{-\frac{\pi}{2}i}$&$2\sqrt{3}e^{\frac{\pi}{12}i}$&$2\sqrt{3}e^{-\frac{5\pi}{12}i}$&$2e^{-\frac{\pi}{6}i}$&$2e^{\frac{\pi}{2}i}$&$\sqrt{3}e^{-\frac{\pi}{4}i}$&$\sqrt{3}e^{-\frac{3\pi}{4}i}$&$2$&$2e^{\frac{\pi}{3}i}$&$2\sqrt{3}e^{-\frac{11\pi}{12}i}$\\
				$P^{\it{1/2},\overline{\it{1}}}$&$2\sqrt{3}e^{-\frac{7\pi}{12}i}$&2&$2e^{-\frac{\pi}{3}i}$&$\sqrt{3}e^{-\frac{\pi}{4}i}$&$\sqrt{3}e^{\frac{\pi}{4}i}$&$2e^{-\frac{5\pi}{6}i}$&$2e^{\frac{\pi}{2}i}$&$2\sqrt{3}e^{\frac{11\pi}{12}i}$&$2\sqrt{3}e^{-\frac{7\pi}{12}i}$&$2e^{\frac{\pi}{6}i}$&$2e^{-\frac{\pi}{2}i}$&$\sqrt{3}e^{-\frac{3\pi}{4}i}$&$\sqrt{3}e^{-\frac{\pi}{4}i}$&$2$&$2e^{-\frac{\pi}{3}i}$&$2\sqrt{3}e^{-\frac{\pi}{12}i}$\\
				$P^{\it{3/2},\overline{\it{1}}}$&$2\sqrt{3}e^{\frac{5\pi}{12}i}$&2&$2e^{-\frac{\pi}{3}i}$&$\sqrt{3}e^{\frac{3\pi}{4}i}$&$\sqrt{3}e^{-\frac{3\pi}{4}i}$&$2e^{-\frac{5\pi}{6}i}$&$2e^{\frac{\pi}{2}i}$&$2\sqrt{3}e^{-\frac{\pi}{12}i}$&$2\sqrt{3}e^{\frac{5\pi}{12}i}$&$2e^{\frac{\pi}{6}i}$&$2e^{-\frac{\pi}{2}i}$&$\sqrt{3}e^{\frac{\pi}{4}i}$&$\sqrt{3}e^{\frac{3\pi}{4}i}$&$2$&$2e^{-\frac{\pi}{3}i}$&$2\sqrt{3}e^{\frac{11\pi}{12}i}$\\
				\hline
		\end{tabular}}
		\caption{Idempotent table (block 2) for $|\text{su(2)}_4|$.}
		\label{TetraTable2}
	\end{table*}
\end{center}

\begin{center}
	\begin{table*}[hbt!]
		\resizebox{\textwidth}{!}{
			\begin{tabular}{ |c|cccccccccccccc| }
				\hline
				&$\mathcal{T}_{\it{1},\it{0}}^{\it{1}}$&$\mathcal{T}_{\it{1},\it{1}/\it{2}}^{\it{1}/\it{2}}$&$\mathcal{T}_{\it{1},\it{1}/\it{2}}^{\it{3}/\it{2}}$&$\mathcal{T}_{\it{1},\it{1}}^{\it{0}}$&$\mathcal{T}_{\it{1},\it{1}}^{\it{1}}$&$\mathcal{T}_{\it{1},\it{1}}^{\it{2}}$&$\mathcal{T}_{\it{1},\it{3}/\it{2}}^{\it{1}/\it{2}}$&$\mathcal{T}_{\it{1},\it{3}/\it{2}}^{\it{3}/\it{2}}$&$\mathcal{T}_{\it{1},\it{2}}^{\it{1}}$&$\mathcal{T}_{\it{2},\it{0}}^{\it{2}}$&$\mathcal{T}_{\it{2},\it{1}/\it{2}}^{\it{3}/\it{2}}$&$\mathcal{T}_{\it{2},\it{1}}^{\it{1}}$&$\mathcal{T}_{\it{2},\it{3}/\it{2}}^{\it{1}/\it{2}}$&$\mathcal{T}_{\it{2},\it{2}}^{\it{0}}$\\
				\hline
				$P^{\it{1}, \overline{\it{0}}}$&$2e^{-\frac{2\pi}{3}i}$&$\sqrt{3}e^{\frac{2\pi}{3}i}$&$\sqrt{3}e^{-\frac{5\pi}{6}i}$&1&$2e^{\frac{2\pi}{3}i}$&$-1$&$\sqrt{3}e^{\frac{\pi}{6}i}$&$\sqrt{3}e^{\frac{2\pi}{3}i}$&$2e^{\frac{\pi}{3}i}$&&&&&\\
				$P^{\it{2}, \overline{\it{1}}}$&$2e^{\frac{2\pi}{3}i}$&$\sqrt{3}e^{\frac{\pi}{3}i}$&$\sqrt{3}e^{-\frac{\pi}{6}i}$&1&$2e^{-\frac{2\pi}{3}i}$&$-1$&$\sqrt{3}e^{\frac{5\pi}{6}i}$&$\sqrt{3}e^{\frac{2\pi}{3}i}$&$2e^{-\frac{\pi}{3}i}$&&&&&\\
				$P^{\it{0}, \overline{\it{1}}}$&$2e^{\frac{2\pi}{3}i}$&$\sqrt{3}e^{-\frac{2\pi}{3}i}$&$\sqrt{3}e^{\frac{5\pi}{6}i}$&1&$2e^{-\frac{2\pi}{3}i}$&$-1$&$\sqrt{3}e^{-\frac{\pi}{6}i}$&$\sqrt{3}e^{-\frac{2\pi}{3}i}$&$2e^{-\frac{\pi}{3}i}$&&&&&\\
				$P^{\it{1}, \overline{\it{2}}}$&$2e^{-\frac{2\pi}{3}i}$&$\sqrt{3}e^{-\frac{\pi}{3}i}$&$\sqrt{3}e^{\frac{\pi}{6}i}$&1&$2e^{\frac{2\pi}{3}i}$&$-1$&$\sqrt{3}e^{-\frac{5\pi}{6}i}$&$\sqrt{3}e^{-\frac{2\pi}{3}i}$&$2e^{\frac{\pi}{3}i}$&&&&&\\
				$P^{\it{3/2}, \overline{\it{1/2}}}$&$-3$&$\frac{3}{2}\sqrt{3}e^{\frac{\pi}{2}i}$&$\frac{3}{2}e^{\frac{\pi}{2}i}$&$\frac{3}{2}$&&$\frac{3}{2}$&$\frac{3}{2}e^{\frac{\pi}{2}i}$&$\frac{3}{2}\sqrt{3}e^{-\frac{\pi}{2}i}$&$-3$&$-3$&$3e^{\frac{\pi}{2}}$&&$3e^{\frac{\pi}{2}}$&3\\
				$P^{\it{1/2}, \overline{\it{3/2}}}$&$-3$&$\frac{3}{2}\sqrt{3}e^{-\frac{\pi}{2}i}$&$\frac{3}{2}e^{-\frac{\pi}{2}i}$&$\frac{3}{2}$&&$\frac{3}{2}$&$\frac{3}{2}e^{-\frac{\pi}{2}i}$&$\frac{3}{2}\sqrt{3}e^{\frac{\pi}{2}i}$&$-3$&$-3$&$3e^{-\frac{\pi}{2}}$&&$3e^{-\frac{\pi}{2}}$&3\\
				\hline
		\end{tabular}}
		\caption{Idempotent table (block 3) for $|\text{su(2)}_4|$.}
		\label{TetraTable3}
	\end{table*}
\end{center}

\begin{center}
	\begin{table*}[hbt!]
		\resizebox{\textwidth}{!}{
			\begin{tabular}{ |c|ccccccccccccc| }
				\hline
				&$\mathcal{T}_{\it{3}/\it{2},\it{0}}^{\it{3}/\it{2}}$&$\mathcal{T}_{\it{3}/\it{2},\it{1}/\it{2}}^{\it{1}}$&$\mathcal{T}_{\it{3}/\it{2},\it{1}/\it{2}}^{\it{2}}$&$\mathcal{T}_{\it{3}/\it{2},\it{1}}^{\it{1}/\it{2}}$&$\mathcal{T}_{\it{3}/\it{2},\it{1}}^{\it{3}/\it{2}}$&$\mathcal{T}_{\it{3}/\it{2},\it{3}/\it{2}}^{\it{0}}$&$\mathcal{T}_{\it{3}/\it{2},\it{3}/\it{2}}^{\it{1}}$&$\mathcal{T}_{\it{3}/\it{2},\it{2}}^{\it{1}/\it{2}}$&$\mathcal{T}_{\it{2},\it{0}}^{\it{2}}$&$\mathcal{T}_{\it{2},\it{1}/\it{2}}^{\it{3}/\it{2}}$&$\mathcal{T}_{\it{2},\it{1}}^{\it{1}}$&$\mathcal{T}_{\it{2},\it{3}/\it{2}}^{\it{1}/\it{2}}$&$\mathcal{T}_{\it{2},\it{2}}^{\it{0}}$\\
				\hline
				$P^{\it{2}, \overline{\it{1/2}}}$&$\sqrt{3}e^{-\frac{3\pi}{4}i}$&$2e^{\frac{5\pi}{6}i}$&$e^{\frac{\pi}{2}i}$&$\sqrt{3}e^{\frac{5\pi}{12}i}$&$\sqrt{3}e^{-\frac{\pi}{12}i}$&1&$2e^{-\frac{2\pi}{3}i}$&$\sqrt{3}e^{\frac{3\pi}{4}i}$&&&&&\\
				$P^{\it{1/2}, \overline{\it{2}}}$&$\sqrt{3}e^{\frac{3\pi}{4}i}$&$2e^{-\frac{5\pi}{6}i}$&$e^{-\frac{\pi}{2}i}$&$\sqrt{3}e^{-\frac{5\pi}{12}i}$&$\sqrt{3}e^{\frac{\pi}{12}i}$&1&$2e^{\frac{2\pi}{3}i}$&$\sqrt{3}e^{-\frac{3\pi}{4}i}$&&&&&\\
				$P^{\it{0}, \overline{\it{3/2}}}$&$\sqrt{3}e^{\frac{\pi}{4}i}$&$2e^{\frac{5\pi}{6}i}$&$e^{\frac{\pi}{2}i}$&$\sqrt{3}e^{-\frac{7\pi}{12}i}$&$\sqrt{3}e^{\frac{11\pi}{12}i}$&1&$2e^{-\frac{2\pi}{3}i}$&$\sqrt{3}e^{-\frac{\pi}{4}i}$&&&&&\\
				$P^{\it{3/2}, \overline{\it{0}}}$&$\sqrt{3}e^{-\frac{\pi}{4}i}$&$2e^{-\frac{5\pi}{6}i}$&$e^{-\frac{\pi}{2}i}$&$\sqrt{3}e^{\frac{7\pi}{12}i}$&$\sqrt{3}e^{-\frac{11\pi}{12}i}$&1&$2e^{\frac{2\pi}{3}i}$&$\sqrt{3}e^{\frac{\pi}{4}i}$&&&&&\\
				$P^{\it{2}, \overline{\it{0}}}$&&&&&&&&&1&$\sqrt{3}e^{-\frac{\pi}{2}}$&$-2$&$\sqrt{3}e^{\frac{\pi}{2}}$&1\\
				$P^{\it{0}, \overline{\it{2}}}$&&&&&&&&&1&$\sqrt{3}e^{\frac{\pi}{2}}$&$-2$&$\sqrt{3}e^{-\frac{\pi}{2}}$&1\\
				\hline
		\end{tabular}}
		\caption{Idempotent table (block 4) for $|\text{su(2)}_4|$.}
		\label{TetraTable4}
	\end{table*}
\end{center}

\newpage
\begin{table*}[hbt!]
	\center
	\scalebox{0.8}{
		\begin{tabular}{ |c|cccccccc| }
			\hline
			&$\mathcal{T}_{A^+,A^+}^{A^+}$&$\mathcal{T}_{A^+,\sigma^+}^{\sigma^+}$&$\mathcal{T}_{A^+,B^+}^{B^+}$&$\mathcal{T}_{A^+,C^+}^{C^+}$&$\mathcal{T}_{A^+,A^-}^{A^-}$&$\mathcal{T}_{A^+,\sigma^-}^{\sigma^-}$&$\mathcal{T}_{A^+,B^-}^{B^-}$&$\mathcal{T}_{A^+,C^-}^{C^-}$\\
			\hline
			$P^{\it{0}, \overline{\it{0}}}$&$1$&$\sqrt{3}$&$1$&$1$&$1$&$\sqrt{3}$&$1$&$1$\\
			$P^{\it{2}, \overline{\it{0}}}$&$1$&$\sqrt{3}$&$1$&$1$&$-1$&$-\sqrt{3}$&$-1$&$-1$\\
			$P^{\it{0}, \overline{\it{2}}}$&$1$&$-\sqrt{3}$&$1$&$1$&$-1$&$\sqrt{3}$&$-1$&$-1$\\
			$P^{\it{2}, \overline{\it{2}}}$&$1$&$-\sqrt{3}$&$1$&$1$&$1$&$-\sqrt{3}$&$1$&$1$\\
			$P^{\it{1},\overline{\it{1}}}$&$8$&&$-4$&$-4$&&&&\\
			\hline
	\end{tabular}}		
	\caption{Idempotent table (block 1) for $\mc{C}_{S_3}$.}
	\label{PottsTableBlock1}
\end{table*}
\begin{center}
	\begin{table*}[hbt!]
		\resizebox{\textwidth}{!}{
			\begin{tabular}{ |c|cccccccccccc| }
				\hline
				&$\mathcal{T}_{\sigma^+,A^+}^{\sigma^+}$&$\mathcal{T}_{\sigma^+,\sigma^+}^{A^+}$&$\mathcal{T}_{\sigma^+,\sigma^+}^{B^+}$&$\mathcal{T}_{\sigma^+,\sigma^+}^{C^+}$&$\mathcal{T}_{\sigma^+,B^+}^{\sigma^+}$&$\mathcal{T}_{\sigma^+,C^+}^{\sigma^+}$&$\mathcal{T}_{\sigma^+,A^-}^{\sigma^-}$&$\mathcal{T}_{\sigma^+,\sigma^-}^{A^-}$&$\mathcal{T}_{\sigma^+,\sigma^-}^{B^-}$&$\mathcal{T}_{\sigma^+,\sigma^-}^{C^-}$&$\mathcal{T}_{\sigma^+,B^-}^{\sigma^-}$&$\mathcal{T}_{\sigma^+,C^-}^{\sigma^-}$\\
				\hline
				$P^{\it{2},\overline{\it{1/2}}}$&$\sqrt{3}e^{-\frac{3\pi}{4}i}$&$1$&$e^{-\frac{2\pi}{3}i}$&$e^{-\frac{2\pi}{3}i}$&$\sqrt{3}e^{-\frac{\pi}{12}i}$&$\sqrt{3}e^{-\frac{\pi}{12}i}$&$\sqrt{3}e^{\frac{7\pi}{12}i}$&$-1$&$e^{\frac{\pi}{3}i}$&$e^{\frac{\pi}{3}i}$&$\sqrt{3}e^{-\frac{3\pi}{4}i}$&$\sqrt{3}e^{-\frac{3\pi}{4}i}$\\
				$P^{\it{0},\overline{\it{3/2}}}$&$\sqrt{3}e^{\frac{\pi}{4}i}$&$1$&$e^{-\frac{2\pi}{3}i}$&$e^{-\frac{2\pi}{3}i}$&$\sqrt{3}e^{\frac{11\pi}{12}i}$&$\sqrt{3}e^{\frac{11\pi}{12}i}$&$\sqrt{3}e^{-\frac{5\pi}{12}i}$&$-1$&$e^{\frac{\pi}{3}i}$&$e^{\frac{\pi}{3}i}$&$\sqrt{3}e^{\frac{\pi}{4}i}$&$\sqrt{3}e^{\frac{\pi}{4}i}$\\
				$P^{\it{0},\overline{\it{1/2}}}$&$\sqrt{3}e^{-\frac{3\pi}{4}i}$&$1$&$e^{-\frac{2\pi}{3}i}$&$e^{-\frac{2\pi}{3}i}$&$\sqrt{3}e^{-\frac{\pi}{12}i}$&$\sqrt{3}e^{-\frac{\pi}{12}i}$&$\sqrt{3}e^{-\frac{5\pi}{12}i}$&$1$&$e^{-\frac{2\pi}{3}i}$&$e^{-\frac{2\pi}{3}i}$&$\sqrt{3}e^{\frac{\pi}{4}i}$&$\sqrt{3}e^{\frac{\pi}{4}i}$\\
				$P^{\it{2},\overline{\it{3/2}}}$&$\sqrt{3}e^{\frac{\pi}{4}i}$&$1$&$e^{-\frac{2\pi}{3}i}$&$e^{-\frac{2\pi}{3}i}$&$\sqrt{3}e^{\frac{11\pi}{12}i}$&$\sqrt{3}e^{\frac{11\pi}{12}i}$&$\sqrt{3}e^{\frac{7\pi}{12}i}$&$1$&$e^{-\frac{2\pi}{3}i}$&$e^{-\frac{2\pi}{3}i}$&$\sqrt{3}e^{-\frac{3\pi}{4}i}$&$\sqrt{3}e^{-\frac{3\pi}{4}i}$\\
				$P^{\it{1},\overline{\it{3/2}}}$&$4\sqrt{3}e^{-\frac{5\pi}{12}i}$&$4$&$2e^{\frac{\pi}{3}i}$&$2e^{\frac{\pi}{3}i}$&$2\sqrt{3}e^{-\frac{3\pi}{4}i}$&$2\sqrt{3}e^{-\frac{3\pi}{4}i}$&&&&&&\\
				$P^{\it{1},\overline{\it{1/2}}}$&$4\sqrt{3}e^{\frac{7\pi}{12}i}$&$4$&$2e^{\frac{\pi}{3}i}$&$2e^{\frac{\pi}{3}i}$&$2\sqrt{3}e^{\frac{\pi}{4}i}$&$2\sqrt{3}e^{\frac{\pi}{4}i}$&&&&&&\\
				\hline
		\end{tabular}}
		\caption{Idempotent table (block 2) for $\mc{C}_{S_3}$.}
		\label{PottsTableBlock2}
	\end{table*}
\end{center}
\begin{center}
	\begin{table*}[hbt!]
		\scalebox{0.8}{
			\begin{tabular}{ |c|cccccccccc| }
				\hline
				&$\mathcal{T}_{B^+,A^+}^{C^+}$&$\mathcal{T}_{B^+,\sigma^+}^{\sigma^+}$&$\mathcal{T}_{B^+,B^+}^{A^+}$&$\mathcal{T}_{B^+,C^+}^{B^+}$&$\mathcal{T}_{B^+,\sigma^-}^{\sigma^-}$&$\mathcal{T}_{C^+,A^+}^{B^+}$&$\mathcal{T}_{C^+,\sigma^+}^{\sigma^+}$&$\mathcal{T}_{C^+,B^+}^{C^+}$&$\mathcal{T}_{C^+,C^+}^{A^+}$&$\mathcal{T}_{C^+,\sigma^-}^{\sigma^-}$\\
				\hline
				$P^{\it{1},\overline{\it{1}}}$&$4$&&$4$&$4$&&$4$&&$4$&$4$&\\
				$P^{\it{2},\overline{\it{1}}}$&$2e^{\frac{2\pi}{3}i}$&&$2$&$2e^{-\frac{2\pi}{3}i}$&$2\sqrt{3}e^{\frac{\pi}{3}i}$&$2e^{\frac{2\pi}{3}i}$&&$2e^{-\frac{2\pi}{3}i}$&$2$&$2\sqrt{3}e^{\frac{\pi}{3}i}$\\
				$P^{\it{1},\overline{\it{2}}}$&$2e^{-\frac{2\pi}{3}i}$&$2\sqrt{3}e^{-\frac{\pi}{3}i}$&$2$&$2e^{\frac{2\pi}{3}i}$&&$2e^{-\frac{2\pi}{3}i}$&$2\sqrt{3}e^{-\frac{\pi}{3}i}$&$2e^{\frac{2\pi}{3}i}$&$2$&\\
				$P^{\it{1},\overline{\it{0}}}$&$2e^{-\frac{2\pi}{3}i}$&$2\sqrt{3}e^{\frac{2\pi}{3}i}$&$2$&$2e^{\frac{2\pi}{3}i}$&&$2e^{-\frac{2\pi}{3}i}$&$2\sqrt{3}e^{\frac{2\pi}{3}i}$&$2e^{\frac{2\pi}{3}i}$&$2$&\\
				$P^{\it{0},\overline{\it{1}}}$&$2e^{\frac{2\pi}{3}i}$&&$2$&$2e^{-\frac{2\pi}{3}i}$&$2\sqrt{3}e^{-2\frac{\pi}{3}i}$&$2e^{\frac{2\pi}{3}i}$&&$2e^{-\frac{2\pi}{3}i}$&$2$&$2\sqrt{3}e^{-\frac{2\pi}{3}i}$\\
				\hline
		\end{tabular}}
		\caption{Idempotent table (block 3) for $\mc{C}_{S_3}$.}
		\label{PottsTableBlock3}
	\end{table*}
\end{center}
\begin{center}
	\begin{table*}[hbt!]
		\resizebox{\textwidth}{!}{
			\begin{tabular}{ |c|cccccccccccc| }
				\hline
				&$\mathcal{T}_{A^-,A^+}^{A^-}$&$\mathcal{T}_{A^-,\sigma^+}^{\sigma^-}$&$\mathcal{T}_{A^-,A^-}^{A^+}$&$\mathcal{T}_{A^-,\sigma^-}^{\sigma^+}$&$\mathcal{T}_{B^-,A^+}^{B^-}$&$\mathcal{T}_{B^-,\sigma^+}^{\sigma^-}$&$\mathcal{T}_{B^-,\sigma^-}^{\sigma^+}$&$\mathcal{T}_{B^-,B^-}^{A^+}$&$\mathcal{T}_{C^-,A^+}^{C^-}$&$\mathcal{T}_{C^-,\sigma^+}^{\sigma^-}$&$\mathcal{T}_{C^-,\sigma^-}^{\sigma^+}$&$\mathcal{T}_{C^-,C^-}^{A^+}$\\
				\hline
				$P^{\it{1/2},\overline{\it{1/2}}}$&$3$&$3$&$3$&$3$&$3$&$3e^{\frac{\pi}{3}i}$&$3e^{\frac{2\pi}{3}i}$&$3$&$3$&$3e^{\frac{\pi}{3}i}$&$3e^{\frac{2\pi}{3}i}$&$3$\\
				$P^{\it{3/2},\overline{\it{3/2}}}$&$3$&$-3$&$3$&$-3$&$3$&$3e^{-\frac{2\pi}{3}i}$&$3e^{-\frac{\pi}{3}i}$&$3$&$3$&$3e^{-\frac{2\pi}{3}i}$&$3e^{-\frac{\pi}{3}i}$&$3$\\
				$P^{\it{1/2},\overline{\it{3/2}}}$&$-3$&$3$&$3$&$-3$&$-3$&$3e^{\frac{\pi}{3}i}$&$3e^{-\frac{\pi}{3}i}$&$3$&$-3$&$3e^{\frac{\pi}{3}i}$&$3e^{-\frac{2\pi}{3}i}$&$3$\\
				$P^{\it{3/2},\overline{\it{1/2}}}$&$-3$&$-3$&$3$&$3$&$-3$&$3e^{-\frac{2\pi}{3}i}$&$3e^{\frac{2\pi}{3}i}$&$3$&$-3$&$3e^{-\frac{2\pi}{3}i}$&$3e^{\frac{2\pi}{3}i}$&$3$\\
				\hline
		\end{tabular}}
		\caption{Idempotent table (block 4) for $\mc{C}_{S_3}$.}
		\label{PottsTableBlock4}
	\end{table*}
\end{center}
\begin{center}
	\begin{table*}[hbt!]
		\resizebox{\textwidth}{!}{
			\begin{tabular}{ |c|cccccccccccc| }
				\hline
				&$\mathcal{T}_{\sigma^-,A^+}^{\sigma^-}$&$\mathcal{T}_{\sigma^-,\sigma^+}^{A^-}$&$\mathcal{T}_{\sigma^-,\sigma^+}^{B^-}$&$\mathcal{T}_{\sigma^-,\sigma^+}^{C^-}$&$\mathcal{T}_{\sigma^-,B^+}^{\sigma^-}$&$\mathcal{T}_{\sigma^-,C^+}^{\sigma^-}$&$\mathcal{T}_{\sigma^-,A^-}^{\sigma^+}$&$\mathcal{T}_{\sigma^-,\sigma^-}^{A^+}$&$\mathcal{T}_{\sigma^-,\sigma^-}^{B^+}$&$\mathcal{T}_{\sigma^-,\sigma^-}^{C^+}$&$\mathcal{T}_{\sigma^-,B^-}^{\sigma^+}$&$\mathcal{T}_{\sigma^-,C^-}^{\sigma^+}$\\
				\hline
				$P^{\it{1/2},\overline{\it{0}}}$&$\sqrt{3}e^{\frac{3\pi}{4}i}$&$1$&$e^{-\frac{\pi}{3}i}$&$e^{-\frac{\pi}{3}i}$&$\sqrt{3}e^{\frac{\pi}{12}i}$&$\sqrt{3}e^{\frac{\pi}{12}i}$&$\sqrt{3}e^{\frac{5\pi}{12}i}$&$1$&$e^{\frac{2\pi}{3}i}$&$e^{\frac{2\pi}{3}i}$&$\sqrt{3}e^{-\frac{\pi}{4}i}$&$\sqrt{3}e^{-\frac{\pi}{4}i}$\\
				$P^{\it{3/2},\overline{\it{2}}}$&$\sqrt{3}e^{-\frac{\pi}{4}i}$&$1$&$e^{-\frac{\pi}{3}i}$&$e^{-\frac{\pi}{3}i}$&$\sqrt{3}e^{-\frac{11\pi}{12}i}$&$\sqrt{3}e^{-\frac{11\pi}{12}i}$&$\sqrt{3}e^{-\frac{7\pi}{12}i}$&$1$&$e^{\frac{2\pi}{3}i}$&$e^{\frac{2\pi}{3}i}$&$\sqrt{3}e^{3\frac{\pi}{4}i}$&$\sqrt{3}e^{\frac{3\pi}{4}i}$\\
				$P^{\it{3/2},\overline{\it{0}}}$&$\sqrt{3}e^{-\frac{\pi}{4}i}$&$-1$&$e^{\frac{2\pi}{3}i}$&$e^{\frac{2\pi}{3}i}$&$\sqrt{3}e^{-\frac{11\pi}{12}i}$&$\sqrt{3}e^{-\frac{11\pi}{12}i}$&$\sqrt{3}e^{\frac{5\pi}{12}i}$&$1$&$e^{\frac{2\pi}{3}i}$&$e^{\frac{2\pi}{3}i}$&$\sqrt{3}e^{-\frac{\pi}{4}i}$&$\sqrt{3}e^{-\frac{\pi}{4}i}$\\
				$P^{\it{1/2},\overline{\it{2}}}$&$\sqrt{3}e^{\frac{3\pi}{4}i}$&$-1$&$e^{\frac{2\pi}{3}i}$&$e^{\frac{2\pi}{3}i}$&$\sqrt{3}e^{\frac{\pi}{12}i}$&$\sqrt{3}e^{\frac{\pi}{12}i}$&$\sqrt{3}e^{-\frac{7\pi}{12}i}$&$1$&$e^{\frac{2\pi}{3}i}$&$e^{\frac{2\pi}{3}i}$&$\sqrt{3}e^{\frac{3\pi}{4}i}$&$\sqrt{3}e^{\frac{3\pi}{4}i}$\\
				$P^{\it{1/2},\overline{\it{1}}}$&$4\sqrt{3}e^{-\frac{7\pi}{12}i}$&&&&$2\sqrt{3}e^{-\frac{\pi}{4}i}$&$2\sqrt{3}e^{-\frac{\pi}{4}i}$&&$4$&$2e^{-\frac{\pi}{3}i}$&$2e^{-\frac{\pi}{3}i}$&&\\
				$P^{\it{3/2},\overline{\it{1}}}$&$4\sqrt{3}e^{\frac{5\pi}{12}i}$&&&&$2\sqrt{3}e^{\frac{3\pi}{4}i}$&$2\sqrt{3}e^{\frac{3\pi}{4}i}$&&$4$&$2e^{-\frac{\pi}{3}i}$&$2e^{-\frac{\pi}{3}i}$&&\\
				\hline
		\end{tabular}}
		\caption{Idempotent table (block 5) for $\mc{C}_{S_3}$.}
		\label{PottsTableBlock5}
	\end{table*}
\end{center}

\clearpage
\bibliographystyle{SciPost_bibstyle}
\bibliography{Potts}

\end{document}